\documentclass[twocolumn]{aa}

\usepackage{aalongtable}
\usepackage{graphicx}
\usepackage{natbib}
\bibpunct{(}{)}{;}{a}{}{,}
\usepackage{txfonts}

\begin{document}

\title{The BeppoSAX catalog of GRB X-ray afterglow observations}

\offprints{M. De Pasquale}

\author{M. De Pasquale\inst{1}\fnmsep\thanks{Present address : Mullard Space Science Laboratory, Holmbury St. Mary, Dorking, Surrey RH5 6NT, United Kingdom}, L. Piro\inst{1}, B. Gendre\inst{1}, L. Amati\inst{2}, L.A. Antonelli\inst{3}, E. Costa\inst{1}, M. Feroci\inst{1},\\ F. Frontera\inst{4}, L. Nicastro\inst{6}, P. Soffitta\inst{1}, \and J. in't Zand\inst{5}}

\institute{
INAF Rome, via fosso del cavaliere 100, 00133, Roma, Italy\\
\email{\hspace{-1.5cm}mdp@mssl.ucl.ac.uk, luigi.piro@rm.iasf.cnr.it,
bruce.gendre@rm.iasf.cnr.it, enrico.costa@rm.iasf.cnr.it,
marco.feroci@rm.iasf.cnr.it, paolo.soffitta@rm.iasf.cnr.it} \and
INAF - Istituto di Astrofisica Spaziale e Fisica Cosmica, via P. Gobetti 101, I-40219, Bologna, Italy \\
\email{amati@bo.iasf.cnr.it} \and Rome Astronomical Observatory, via
di Frascati 33, 00044 Rome\\
\email{a.antonelli@mporzio.astro.it}
\and
Università di Ferrara, Via Paradiso 12, 44100 Ferrara, Italy\\
\email{frontera@fe.infn.it}
\and
Space Research Organization of the Netherlands, Sorbonnelaan 2, 3584 CA Utrecht, Netherlands.\\
\email{j.heise@sron.nl, jeanz@sron.nl}
\and
Istituto Astrofisica Spaziale e Fisica Cosmica, Sezione di Palermo, INAF, Via U. La Malfa 153, 90146 Palermo, Italy.\\
\email{nicastro@pa.iasf.cnr.it}
 }

\date{Received --; accepted --}

\abstract{ We present the X-ray afterglow catalog of BeppoSAX from
the launch of the satellite to the end of the mission. Thirty-three
X-ray afterglows were securely identified based on their fading
behavior out of 39 observations. We have extracted the continuum
parameters (decay index, spectral index, flux, absorption) for all
available afterglows.  We point out a possible correlation between
the X-ray afterglow luminosity and the energy emitted during the
prompt $\gamma$-ray event. We do not detect a significant jet
signature within the afterglows, implying a lower limit on the
beaming angle, neither a standard energy release when X-ray fluxes
are corrected for beaming. Our data support the hypothesis that the
burst should be surrounded by an interstellar medium rather than a
wind environment, and that this environment should be dense. This
may be explained by a termination shock located near the burst
progenitor. We finally point out that some dark bursts may be
explained by an intrinsic faintness of the event, while others may
be strongly absorbed.

 \keywords{ X-rays:general--
             Gamma-rays:bursts -- Catalogs
             }

}

\titlerunning{BeppoSAX GRB X-ray afterglows catalog}
\authorrunning{De Pasquale et al.}

\maketitle

 \section{Introduction}

 Discovered in the early 70's \citep{kle73}, Gamma-Ray Bursts (GRBs)
have been a mysterious phenomenon for 25 years. The lack of any
optical counterpart prevented observers from determining the
distance - galactic or extragalactic - and therefore the amount of
energy involved, which was uncertain within 10 orders of magnitude.
A lot of different models were at that time able to explain the
observed prompt gamma-ray emission.

 The situation changed dramatically with the first fast and precise
localization of GRB, that was obtained by the BeppoSAX satellite
\citep{piro95, boe97} in 1997. This satellite was combining a
gamma-ray burst monitor (that provided the burst trigger) with X-ray
cameras (that were able to asses a precise position and to carry out
follow-up observations). This observational strategy led to the
discovery of the X-ray \citep{cos97}, optical \citep{van97} and
radio \citep{fra97} afterglows. The spectroscopy of the optical
counterpart of the burst also allowed the distance of these events
to be firmly established as cosmologic \citep{met97}.

 With the end of the BeppoSAX mission (April 2002) and its
reentry, a page of the GRB afterglow study was turned, but the
observations remain within the archives. To prepare the future, we
have initiated a complete re-analysis of all X-ray observations
done. In this first paper, we present the legacy of BeppoSAX : its
X-ray afterglow catalog, focusing on the continuum properties. We
will also compare our results with those of previous studies on GRB
X-ray afterglows \citep{fro05,piro04}.
 In two forthcoming papers \citep[][Gendre et al. in
preparation]{gen05}, we will discuss GRB afterglow observations made
by XMM-Newton and Chandra, and a systematic study of line emission
in the X-ray afterglow spectra.

 This article is organized as follows. In Sec. \ref{sec_analyse} we
present the data analysis and the results. We discuss these results
in Sec. \ref{sec_discu} in the light of the fireball model. We
investigate the so called {\it Dark Burst} phenomenon in Sec.
\ref{sec_dark}, before concluding.

\section{ Data reduction and analysis }
\label{sec_analyse}

 BeppoSAX detected and localized simultaneously in the Gamma Ray Burst
Monitor \citep[GRBM,][]{fro97} and Wide Field Cameras
\citep[WFC,][]{jag97} 51 GRBs within its six year long lifetime
\citep{fro04}. These bursts have been included in our analysis
sample. We note that this set is biased against X-ray rich GRBs and
especially X-ray flashes \citep{hei03}, i.e. bursts with weak or
absent signal in the GRBM and normal counterpart in the WFC. In our
sample, we also included GRB991106, GRB020410 and GRB020427,
although they gave no detection in the GRBM\footnote{In the case of
GRB020410, GRBM was actually switched off at the time of the burst.
A $\gamma$-ray signal was detected by Konus\citep{nic04}}, due to
the fact that a subsequent observation with BeppoSAX was performed
after the localization with the WFC. Data on these bursts are
reported in Tables \ref{table1} and \ref{table2}. We have not
included the bursts discovered after an archive re-analysis.
 Overall, it was possible to follow up 36 burst with the narrow
field instruments. One other afterglow observation (\object{GRB
000926}) was carried out following external triggers. Finally, in
the case of \object{GRB 980703}, BeppoSAX detected the burst while
it was outside failed the WFC field of view, and the follow up
observation was performed on the basis of a localization by the RXTE
All Sky Monitor. In this paper we present the data gathered by the
Narrow Field Instruments (NFI) Low Energy Concentrator Spectrometer
\citep[LECS, 0.1 - 10 keV,][]{par97} and Medium Energy Concentrator
Spectrometer \citep[MECS, 1.6 - 10 keV,][]{boe97}. The first of this
sample (\object{GRB 960720}) was followed up late, while 38 had fast
(within 1 day\footnote{2 days for GRB000926}) follow up
observations. We analyzed 37 of these fast follow-up, excluding
\object{GRB 990705} due to its high contamination of a nearby X-ray
source.

 A typical observation starts $\sim 8-9$ hours after the burst and
its duration is about $1\times 10^{5}$ seconds for MECS and $7
\times 10^{4} $ for LECS. The net exposure lasts $\sim 1/2$ of the
observation for MECS and $1/4$ for LECS.

\subsection{Afterglow identification and temporal analysis}

\begin{figure*}
\centering
\hspace{-2cm}
\includegraphics{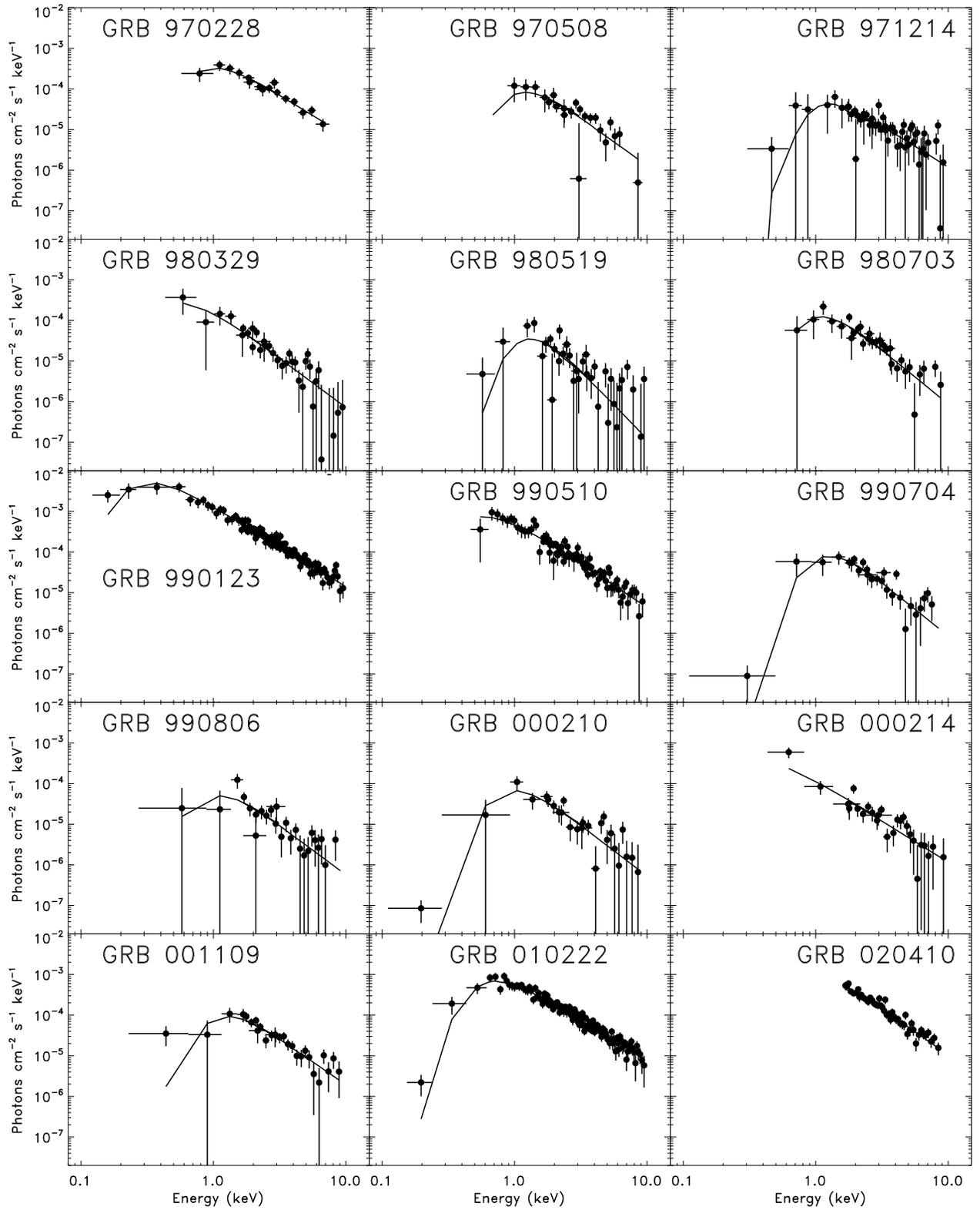}
\caption{\label{fig2} X-ray spectra of the afterglows observed by BeppoSAX.}
\end{figure*}

 The first step of data analysis is source detection, in order to
find the afterglow. For this purpose, we used the MECS data, because
this instrument has a sensitivity higher than that of the LECS. We
extracted the image, ran the detection tool {\it Ximage}
4.3\footnote{see
\textit{http://heasarc.gsfc.nasa.gov/docs/xanadu/ximage/ximage.html
}} on this image and selected all the sources with at least a
$3\sigma$ significance located inside the WFC error box. In the
special cases of \object{GRB 980703} and \object{GRB 000926} we used
the IPN error box \citep{hur00} and ASM error box \citep{lev98}
respectively as these bursts were outside the WFC field of view. The
afterglow was
recognized by its fading behavior. The light curves were generated
from counts extracted within a circle area centered on the source
with a radius of 4 arcminutes. We chose this value because $\gtrsim90\%$
of the source energy is within this region \citep{fio99}. We
also selected counts between 1.6 and 10 keV interval, which is the
optimal range of work for the MECS.

 The associated background was extracted using an annulus centered
at the same position than the source extraction region, with inner
and outer radii of 4.5 and 10 arcminutes respectively. To take into
account the effects of effective area variation and the MECS
support, we renormalized the counts extracted in the annuli by a
factor determined by comparing the counts in the same regions of the
library background fields.

 We used the local background rather than the library background
for light curves in order to take into account any possible time
fluctuation. We developed an IDL script to construct and fit light
curves. This algorithm can calculate adequate errors even in the
case of few counts per bin, by using a Poissonian statistics.
However, if possible, the width of temporal bins was chosen wide
enough to have at least 15-20 counts/bin (background subtracted) at
least, in order to apply a proper Gaussian fit (see below). When
available, subsequent TOOs were also used to better constraint the
light curve behavior.

 The light curves were fitted using a simple power law, using the
Levenberg-Marquardt method to minimize the $\chi^2$ statistic. We
detected 31 sources with a positive decaying index (in the
following, we used the convention $F_{X} \varpropto t^{-\delta}$,
thus a decaying source has a positive decay index) at the $90\%$
confidence level. These sources were identified as the X-ray
afterglow of each burst\footnote{In the cases of GRB000926 and
GRB020427, we have used data gathered by the {\rm Chandra} X-ray
observatory to constrain the decay index \citep[see][]{piro01,
gen05}. For GRB011121, we have used the last WFC data points
\citep[see][]{piro05}}. For three of these sources (\object{GRB
971227}, \object{GRB 990217} and \object{GRB 000529}) the value of
the decay index is greater than zero but not well constrained. We
report in Table \ref{table3} the decay index we obtained for all
these 31 sources (henceforth, all errors reported are at $1\sigma$,
while upper limits are quoted at the 90\% confidence level, unless
otherwise specified).

 In three cases (\object{GRB 970111}, \object{GRB 991106} and
\object{GRB 000615}), we detected within the WFC error box only one
source that did not display any significant fading behavior. We
refer to these as {\it candidate} afterglows. We have calculated the
probability to observe a serendipitous source at the observed flux
level within the WFC error box for these 3 bursts, adopting the Log
N - Log S distribution for BeppoSAX released by \citet{gio00}. The
probability are $\cong 0.027$ for \object{GRB 000615} and
$\cong0.05$ for \object{GRB 970111} and \object{GRB 991106}. The
probability that all of these 3 sources are not afterglows is $\sim
10^{-4}$. We note, however, that these probabilities have been
calculated for extragalactic sources; for low Galactic latitude
events, like GRB991106 ($b\simeq-3\degr$), the value may differ
significantly. \citet{cor02} indicated that {GRB 991106} could in
fact be a Galactic type-I X-ray burster.

 In two cases (\object{GRB 010220} and \object{GRB020321}) we did
not detect any source with $3\sigma$ significance within the WFC
error box. We report in Table \ref{table3} the $3\sigma$ detection
upper limits.

 Some observations deserve special comments. \object{GRB 990907} was
observed for $\sim1000$ seconds only and no decaying behavior can be
detected within the light curve of the source found inside the WFC
error box. However, given the high flux of this source
($\sim10^{-12}$ erg cm$^{-2}$ sec$^{-1}$ in the 1.6-10 keV band),
the probability to have observed a serendipitous source was
$\sim10^{-3}$. We have thus assumed that this source was indeed the
X-ray afterglow of \object{GRB 990907}. In the case of \object{GRB
980425}, we analyzed the source S1 coincident with \object{SN1998bw}
\citep{pia99}. We do not include it in the following discussion as
the detected X-ray emission could be strongly affected by
\object{SN1998bw}.

We present the light curves in Fig. \ref{fig1}.

\subsection{Spectral analysis}

 The X-ray afterglow spectra have been accumulated from the LECS and
MECS during the first TOO only, for those afterglows with more than
150 photons in the MECS (background subtracted). 15 GRBs passed this
criterion; their spectra are presented in Fig. \ref{fig2}.

 We have generally collected LECS counts within a circle centered on
the source with radius $r=8$ arcminutes, which again encircles $>90\%$
of source energy. We operated with LECS data in the range 0.1-4.0
keV, where the response matrix is more accurate. As for MECS, we
collected counts with the same criteria we applied for the time
analysis.
 For spectral analysis, we used the library spectral backgrounds for both
LECS and MECS as they have a very good signal-to-noise ratio, due to
long exposition\footnote{In the case of GRB970111 and GRB970402,
better results were obtained by using local background}. However,
the library backgrounds have been taken at high Galactic latitudes,
with an average Galactic absorption around 2-3 $\times10^{20}$.
Several afterglows in our sample have been observed in fields with
an absorption much higher than this value. For these bursts, the
local background would differ from the library one at low energy
(e.g. below 0.3 keV). The use of a library background from 0.1 keV
would result in an underestimate of the low-energy signal and
consequently a too high estimate of the intrinsic absorbing column
of the burst. Therefore, to evade this problem, we have taken the
minimum energy for LECS to be 0.4 keV if the Galactic column density
was $N_{H}\ge 5\times 10^{20}$ cm$^{-2}$.
 Similarly to the time analysis, the spectral analysis was
performed by requiring at least $15-20$ counts/bin.
 The standard model to fit the spectral data consists of a
constant, a Galactic absorption, an extragalactic absorption (i.e.
{\em in situ}) and a power law. The constant has been included
because of the differences in the LECS and MECS instruments. Its
value is obtained in each case by fitting LECS and MECS data in the
1.6 - 4 keV interval (to avoid absorption effects) with a simple
power law model.

 In our work, we have calculated the 1.6 - 10 keV flux of X-ray
afterglows 11 hours after the burst trigger. We have chosen this
time to avoid effects of changes in the decaying slope. The average
count rate in the MECS has been associated with the average flux
given by the spectrum. Successively, we have taken the count rate at
11 hours, which is given by the light curves, to compute the flux at
that time. In most cases, observations include it. In a few cases
(e.g. \object{GRB 000926}) the flux has been extrapolated.

 For those afterglows with $<150$ counts, we used a canonical model
with an power law energy index of $\alpha=1.2$ (which is typical of
X-ray afterglow spectra) to convert the count rate 11 hours after
the trigger to the corresponding flux.

\begin{table*}
\centering \caption{\label{table3} Properties of the X-ray
afterglows detected by BeppoSAX. We indicate the absorbed flux
extrapolated or interpolated to 11 hours after the burst, the
temporal decay and the energy spectral index, and the excess
absorption around the burst (assuming a distance of z=1 when the
host galaxy redshift was unknown.}

\begin{tabular}{lccccc}
\hline\hline
GRB name            & $1.6 - 10$ keV Flux   &  Decay                 & Spectral             & Density               \\
                    & ($10^{-13}$           &  index                 &  index               & column                \\
                    &erg cm$^{-2}$ s$^{-1}$)&  $\delta$              &  $\alpha$            & ($10^{22}$ cm$^{-2}$) \\
\hline
\object{GRB 970111} & $0.75 \pm 0.47$       & $2.8^{-3.7}$           &         ---          &         ---            \\
\object{GRB 970228} & $20.8 \pm 2.7$        & $1.32^{+0.15}_{-0.20}$ &$1.04^{+0.21}_{-0.27}$&  $<1.12 $              \\
\object{GRB 970402} & $1.35 \pm 0.73$       & $1.11^{+1.5}_{-0.76}$  &         ---          &     ---                \\
\object{GRB 970508} & $5.72 \pm 0.90$       & $0.80^{+0.18}_{-0.15}$ &$1.40^{+0.32}_{-0.27}$& $2.63^{+2.5}_{-1.37}$  \\
\object{GRB 971214} & $6.36 \pm 0.91$       & $1.00 \pm 0.22$        &$1.08^{+0.40}_{-0.23}$& $<53$                  \\
\object{GRB 971227} &      ---              & $>0.4$                 &          ---         &     ---                \\
\object{GRB 980329} & $6.00 \pm 0.56$       & $1.42^{+0.62}_{-0.48}$ &$1.44^{+0.32}_{-0.26}$&$<3.07$                 \\
\object{GRB 980425} & $2.82 \pm 0.59$       & $0.10\pm0.06$          &          ---         &           ---          \\
\object{GRB 980515} & $5.6  \pm 2.2 $       & $>0.51$                &          ---         &           ---          \\
\object{GRB 980519} & $3.9^{+1.2}_{-1.1}$   & $2.18^{+0.89}_{-0.65}$ &$2.43^{+0.97}_{-0.65}$&$5.1^{+6.0}_{-3.8}$     \\
\object{GRB 980613} & $2.6^{+1.2}_{-1.1}$   & $1.49^{+1.9}_{-0.86}$  &          ---         &          ---           \\
\object{GRB 980703} & $14.0^{+7.0}_{-3.2}$  & $1.10^{+0.36}_{-0.28}$ & $1.71 \pm0.29 $      &$2.6^{+2.0}_{-1.3}$     \\
\object{GRB 981226} & $2.8^{2.1}_{1.3}$     & $0.66^{+0.68}_{-0.44}$ &          ---         &       ---              \\
\object{GRB 990123} & $54.2 \pm 1.7$        & $1.45\pm0.06$          & $0.99\pm0.05$        &$0.10^{+0.08}_{-0.06}$  \\
\object{GRB 990217} & $2.8^{+5.1}_{-1.4}$   & $>0$                   &          ---         &     ---                \\
\object{GRB 990510} & $34.7 \pm 2.1 $       & $1.4\pm0.1$            & $1.17\pm 0.09$       &$<0.93$                 \\
\object{GRB 990627} & $3.3^{+1.6}_{-1.5}$   & $1.32^{+1.7}_{-0.92}$  &          ---         &     ---                \\
\object{GRB 990704} & $5.87 \pm 0.84$       & $0.88^{+0.28}_{-0.20}$ &$1.68^{+0.45}_{-0.38}$&$4.1^{+3.4}_{-2.3}$     \\
\object{GRB 990806} & $3.20 \pm 0.87$       & $0.9^{+0.47}_{-0.42}$  &$1.31^{+0.57}_{-0.43}$&$<13.15$                \\
\object{GRB 990907} & $10.6 \pm 4.0 $       &         ---            &          ---         &           ---          \\
\object{GRB 991014} & $5.4^{+1.9}_{-1.5}$   & $1.10^{+0.50}_{-0.32}$ &          ---         &           ---          \\
\object{GRB 991106} & $1.26 \pm 0.47 $      & $1.1^{+2.5}_{-2.1}$    &          ---         &           ---          \\
\object{GRB 000210} & $3.10^{+0.90}_{-0.96}$& $1.41^{+0.98}_{-0.77}$ & $1.54^{+0.31}_{-4}$  &$2.1^{+2.0}_{-1.3}$     \\
\object{GRB 000214} & $6.2^{+2.1}_{-1.8}$   & $0.68\pm0.41$          & $1.04\pm0.27$        &  $<0.36$               \\
\object{GRB 000528} & $3.0^{+4.1}_{-1.4}$   & $0.8^{+0.5}_{-1.5}$    &          ---         &   ---                  \\
\object{GRB 000529} & $1.6  \pm 1.2$        & $>0$                   &          ---         &   ---                  \\
\object{GRB 000615} & $1.28 \pm 0.38 $      & $-0.23^{+1.4}_{-0.94}$ &          ---         &   ---                  \\
\object{GRB 000926} & $32.6^{+15.7}_{-8.7}$ & $1.79^{+0.21}_{-0.16}$ &          ---         &   ---                  \\
\object{GRB 001109} & $23.2^{+5.8}_{-4.5}$  & $1.47^{+0.22}_{-0.27}$ &$1.29^{+0.27}_{-0.26}$&$3.4^{+2.3}_{-1.7}$     \\
\object{GRB 010214} & $3.06^{+0.71}_{-0.64}$& $1.90^{+0.90}_{-0.53}$ &          ---         &           ---          \\
\object{GRB 010220} & $<1.43$               &         ---            &          ---         &           ---          \\
\object{GRB 010222} & $70.6 \pm 3.4$        & $1.35\pm0.06$          & $1\pm0.06$           &$1.27^{+0.33}_{-0.31}$  \\
\object{GRB 011121} & $13.6 \pm 1.5$        & $1.30\pm0.03$          &          ---         &   ---                  \\
\object{GRB 020321} & $<3.4$                &          ---           &          ---         &   ---                  \\
\object{GRB 020322} & $3.8 \pm 0.8$         & $0.84^{+0.46}_{-0.35}$ &          ---         &   ---                  \\
\object{GRB 020410} & $77.8^{+6.3}_{-6.9}$  & $0.92\pm0.12$          &  $1.3\pm0.19$        &      $<4.8$            \\
\object{GRB 020427} & $4.8 \pm 1.7$         & $1.3^{+0.10}_{-0.12}$ &          ---         &   ---                  \\
\hline
\end{tabular}
\end{table*}

All the results of our X-ray afterglow analysis are summarized in
Table \ref{table3}. In Table \ref{table3b}, we report results of the
previous analysis on single \textit{BeppoSAX} GRBs, mostly taken by
a review of \citet{fro04}. We can see a general agreement of the
previous results with ours.

 In order to increase the statistical significance of the sample
of X-ray afterglows with known redshift, we included in our
successive analysis GRB011211. For this burst, which was observed by
\textrm{XMM-Newton}, we assumed a flux of $1.7\pm0.04$ 10$^{-13}$
erg cm$^{-2}$ s$^{-1}$, a spectral and decay index of
$\alpha=1.3\pm0.06$ and $\delta=2.1\pm0.2$ respectively (Gendre et
al. 2005).

\begin{table*}
\centering \caption{\label{table3b} Results reported from previous
analysis of X-ray afterglows detected by BeppoSAX. In this table we
indicate the temporal decay, the energy index, the fitted value of
$n_H$ compared to the galactic column density and the associated
reference. A label 'W' close to the GRB name indicates that the
decay index was obtained by means of WFC and NFI data.}
\begin{tabular}{ccccccc}
\hline\hline
 GRB                & Temporal               & Energy             & n$_H$ /n$_H^G$              & 2--10 keV flux          & Ref. \\
 name               & index$^a$              & index              &                             &  at 10$^5$ s   $^{a}$   &  \\
                    &$\delta$                &$\alpha$            & ($\times10^{21}$ cm$^{-1}$) & (erg cm$^{-2}$s$^{-1}$) &  \\
\hline
\object{GRB 970111} & $>$1.5                 & ---                & ---                         & $<1.0\times10^{-13}$    & \citet{fer98}\\
\object{GRB 970228} & 1.3 $\pm$ 0.2          & 1.1$\pm$0.3        & 3.5$_{-2.3}^{+3.3}$ / 1.6   & $\sim6.8\times10^{-13}$ & \citet{cos97,fro98}\\
\object{GRB 970402} & $1.45 \pm 0.15$        & 1.7$\pm$0.6        & $<$20 / 2.0                 & $\sim4.5\times10^{-14}$ & \citet{nic98v}\\
\object{GRB 970508} & 1.1$\pm$0.1$^b$        & 1.5$\pm$0.55       & 6.0$_{-3.3}^{+7.9}$ / 0.5   & $3.5\times10^{-13}$     & \citet{piro98b,piro99}\\
\object{GRB 971214} & $\sim1.2$              & 0.6$\pm$0.2        & 1.0$_{-1.0}^{+2.3}$ / 0.6   & ---                     & \citet{dal00}\\
\object{GRB 971227} & 1.12$^{+0.08}_{-0.05}$(W) & [1.1]              & [0.13] / 0.13               & $\sim1.4\times10^{-13}$ & \citet{ant99}\\
\object{GRB 980329} & 1.3$\pm$0.03 (W)       & 1.4$\pm$0.4        & 10$\pm$4 / 0.9              &  $2.0\times10^{-13}$    & \citet{zan98}\\
\object{GRB 980425} & 0.16$\pm$0.04          & 1.0$\pm$0.18       & [0.39] / 0.39               & $\sim4.0\times10^{-13}$ & \citet{pia00}\\
\object{GRB 980519} & 1.83$\pm$0.30          & 1.8$^{+0.6}_{-0.5}$& 3--20 / 1.73                & $8.0\times10^{-14}$     & \citet{nic98}\\
\object{GRB 980613} & 1.19$\pm$0.17(W)       & ----               & ---                         & $\sim2.3\times10^{-13}$ & \citet{sof02}\\
\object{GRB 980703} & $>$0.91                & 1.51$\pm$0.32      &36$_{-13}^{+22}$$^{\mathrm c}$ / 0.34  & $4.5\times10^{-13}$     &\citet{vre99} \\
\object{GRB 981226} & 1.3$^{+0.5}_{-0.4}$    & 0.92$\pm$0.47      & [0.18] / 0.18 &  $\sim2.0\times10^{-13}$& \citet{fro00b}\\
\object{GRB 990123}$^d$ & 1.46$\pm$0.04  & 0.94$\pm$0.08 & 0.9$^{+15} _{0.9}$ / 0.21   & $1.25\times10^{-12}$    & \citet{mai05}\\
\object{GRB 990510} & 1.42$\pm$0.07          & 1.03$\pm$0.08      & 2.1$\pm$0.6/0.94            & $9.6\times10^{-13}$     & \citet{kuu00}\\
\object{GRB 990704} & 0.83$\pm$0.16          & 0.7$^{+0.4}_{-0.2}$& [0.3] / 0.3                 & $\sim3.3\times10^{-13}$ & \citet{fer01}\\
\object{GRB 990705} & 1.58$\pm$0.06          & ---                & -                           & $<1.2\times10^{-13}$    & \citet{fro05}\\
\object{GRB 990806} & 1.15$\pm$0.03(W)       &1.16$^{+0.3}_{-0.37}$ & [0.35] / 0.35            & $\sim2.0\times10^{-13}$ & \citet{mon01}\\
\object{GRB 991014} & $>$0.4                 & 0.53$\pm$0.25      & [2.5]/ 2.5                  & $\sim3.0\times10^{-13}$ & \citet{zan00b}\\
\object{GRB 000210} & $1.38\pm0.03$(W)       & $0.75\pm0.3$       & $<4\times10^{21}$           & $\sim2\times 10^{-13}$  &\citet{piro02}\\
\object{GRB 000214} & 0.8$\pm$0.3            & 1.0$\pm$0.18       &0.7$_{-0.7}^{+7.5}$/ 0.55   & $\sim3.5\times10^{-13}$ & \citet{ant00} \\
\object{GRB 000926} & 1.89$^{+0.16 e}_{-0.19}$& $0.9\pm0.42$ &4/0.27$^e,^f$                      & $9.0\times10^{-13}$     & \citet{piro01}\\
\object{GRB 001109} & 1.18$\pm$0.05          & 1.4$\pm$0.3        & 8.7$\pm$0.4$ /0.42$         & $\sim8.0\times10^{-13}$ & \citet{ama02b}\\
\object{GRB 010214} &  2.1$^{+0.6}_{-1.0}$   &0.3$^{+0.8}_{-0.6}$ & [0.27] / 0.27               & ---                     & \citet{gui03}\\
\object{GRB 010222} & 1.33$\pm$0.04          & 0.97$\pm$0.05      & 1.5$\pm$0.3/ 0.16 & $2.4\times10^{-12}$ & \citet{zan01}\\
\object{GRB 011121} & $1.29\pm0.03$(W)       & $1.6\pm0.5$        & $<100/$                     & $\sim 10^{-13}$         & \citet{piro05}\\
\object{GRB 020321} & ---                    & ---                &   ---                       & $<3\times10^{-13}$      & \citet{zand03}\\
\object{GRB 020410} & $0.81\pm0.07$          & $1.05\pm0.08$      &      ---                    & $\sim3.5\times10^{-12}$ & \citet{nic04}\\
\object{GRB 020427} & $1.3^{+0.13} _{0.09}$  & $1^{+2.2} _{-1.1}$ &  0.29/0.29                  & $\sim10^{-13}$          & \citet{ama04}\\
\hline
\end{tabular}

$^a$ All upper limits are 3$\sigma$ except for GRB990705 which are 2$\sigma$.\\
$^b$ from 6$\times$10$^{4}$ s to 5.8$\times$10$^{5}$ s \\
$^c$ n$_H$ value corrected for redshift. \\
$^d$ Spectral data of the first 20,000 s. The time decaying index includes the whole NFI TOO.\\
$^e$ \emph{SAX} plus \emph{CHANDRA} data \citep{piro01}. \\
$^f$ Corrected for redshift \citep{piro01}. This n$_H^z$ value was added to the Galactic column density n$_H^G$.\\

\end{table*}

\section{Results and Discussion}
\label{sec_discu}

\subsection{General properties of X-ray afterglows}
\label{section31}

\begin{figure}
\centering
\includegraphics[width=8cm]{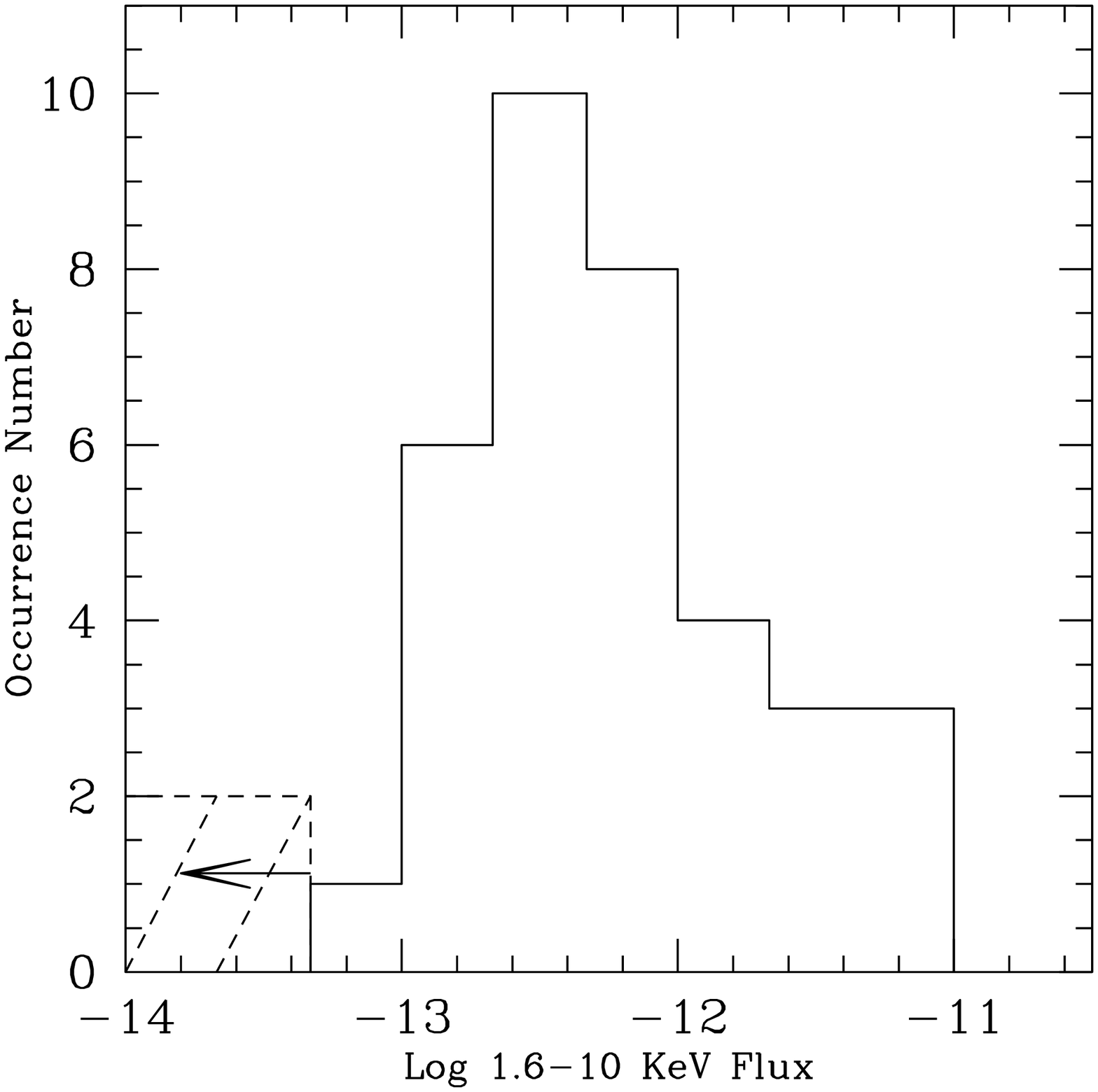}
\caption{The distribution of 1.6-10 keV fluxes in the BeppoSAX GRB
afterglow sample. All fluxes are indicated 11 hours after the burst.
Upper limits have been set to $10^{-13}$ for clarity. \label{fig3}}
\end{figure}

\begin{figure*}
\centering
\includegraphics[width=8cm]{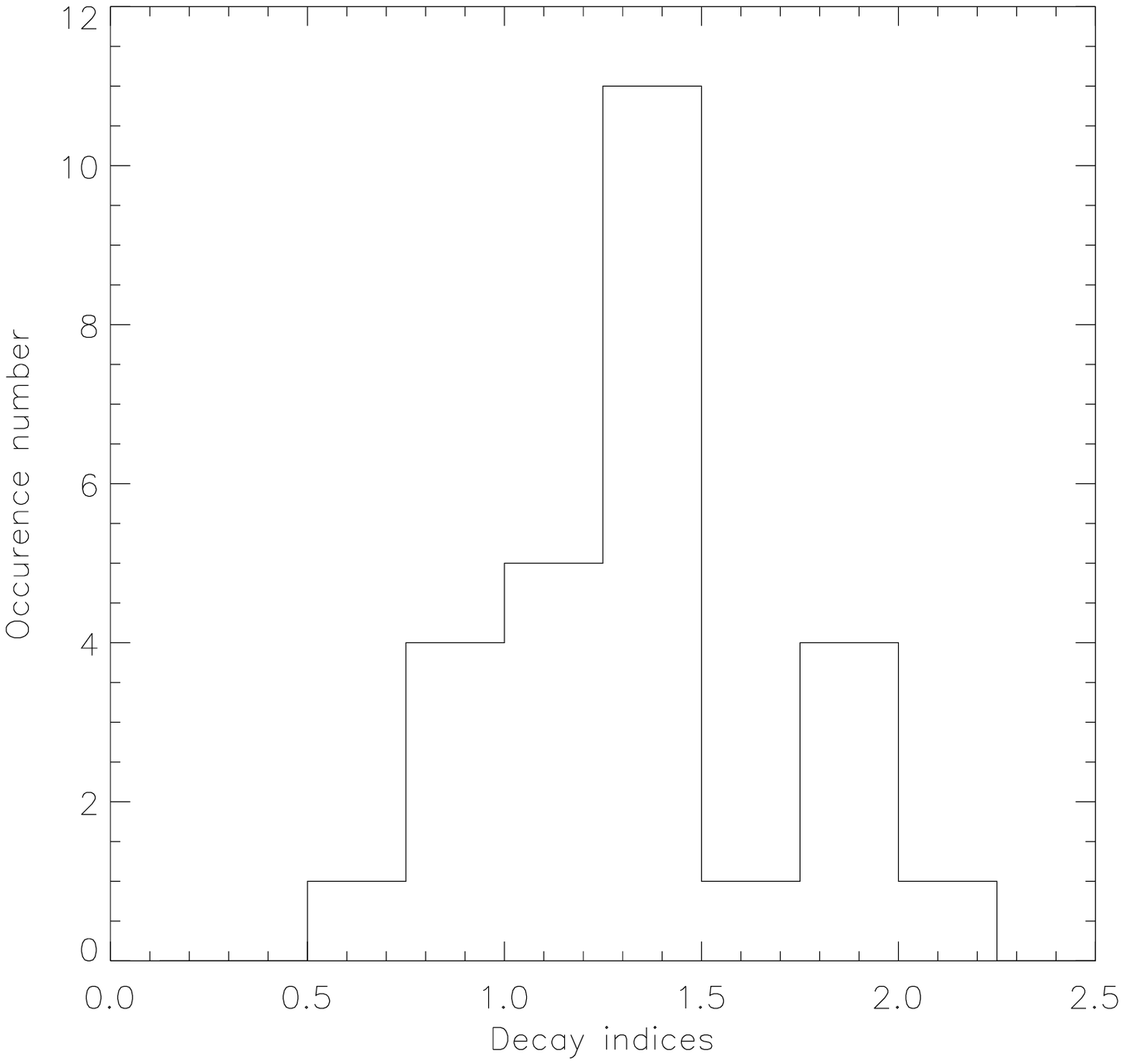}
\includegraphics[width=8cm]{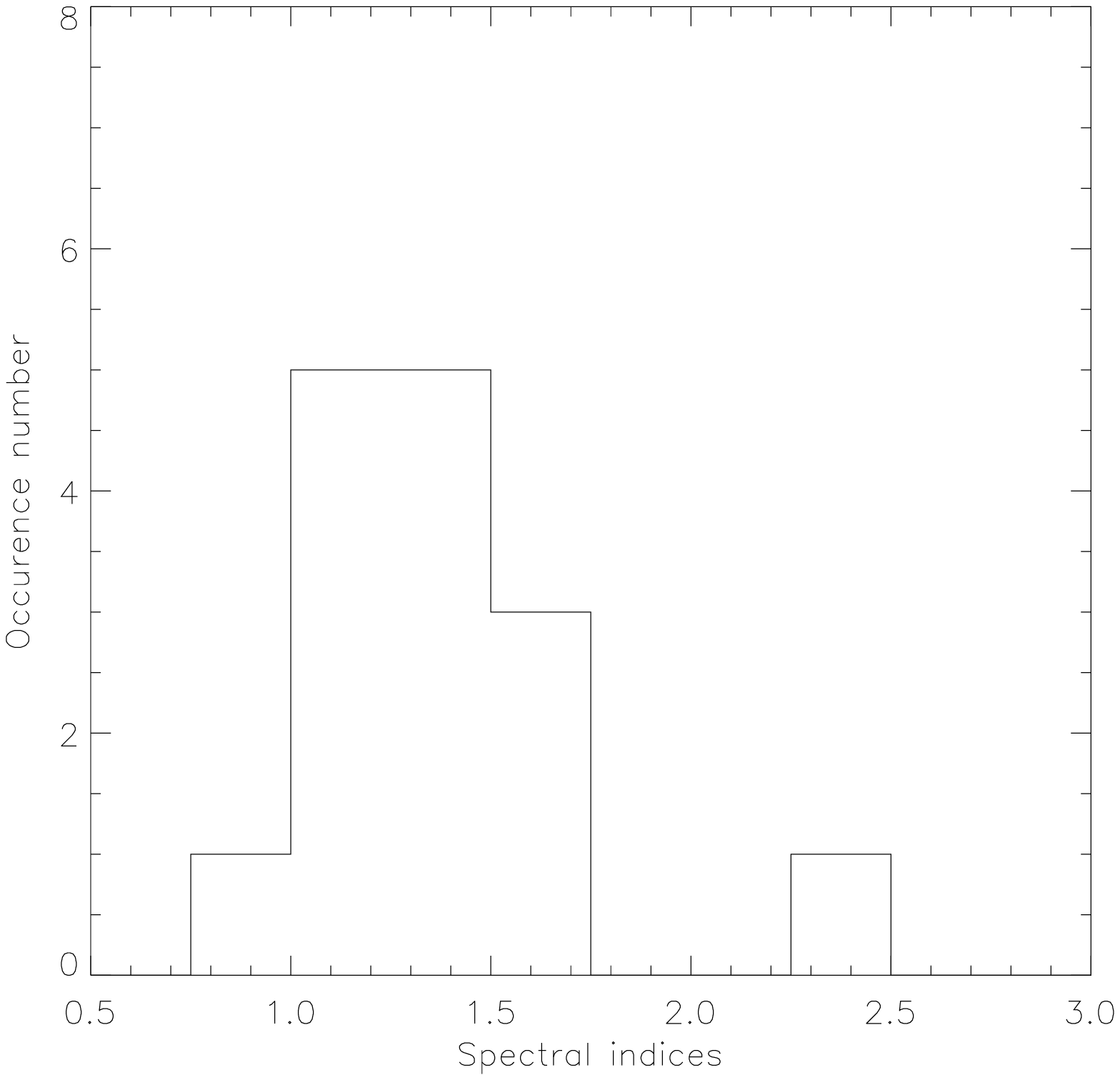}
\caption{Left. Distribution of the spectral indexes of the afterglow of the BeppoSAX bursts. Right. Distribution of the decay indexes of the afterglow of the BeppoSAX bursts. \label{fig:alphadelta}}
\end{figure*}

 We detect an X-ray afterglow in 31 of 36 cases. This constitutes
$86\%$ of the sample. If all doubtful sources are considered as
afterglows, then the fraction of X-ray afterglows increases up to
$94\%$.

 In Fig. \ref{fig3} we present the distribution of the X-ray
afterglow flux F$_{X}$ in the 1.6-10 keV band. It spans
approximately 2 orders of magnitude. GRB 020410 afterglow is the
object with the highest flux, $\sim8\times10^{-12}$ erg cm$^{-2}$
s$^{-1}$, while the weakest is $970402$, $\sim10^{-13}$ erg
cm$^{-2}$ s$^{-1}$.
 The fit of this distribution with a Gaussian provides a
logarithmic mean and width of $<F_{X}>-12.2\pm 0.1$ and $\sigma_{Fx}
= 0.5$ respectively. One may wonder if some faint afterglows could
be missed due to the detection limit (either due to a low luminosity
or to a large distance). In this case, the true distribution could
be broader than that we measure. However, the fact that we detect
X-ray afterglows in $\sim90\%$ of follow-up observations indicates
that this is not the case.

 We have also estimated the distribution of the spectral and decay
indexes (Fig.\ref{fig:alphadelta}). The values we have obtained for
those parameters are the result of the convolution of the intrinsic
distribution with the measurement error. Under the assumption that
both are Gaussian, it is possible to deconvolve the two
distributions. We have adopted a maximum likelihood method \citep[see][]{dep03, mac88} to gather the best
estimates of the parent distribution in the BeppoSAX sample. We have
obtained from the spectral index distribution a mean value of
$\alpha=1.2\pm0.1$ with a width of $0.13_{-0.05}^{+0.11}$, and from
the decay index distribution a mean value of $\delta=1.3\pm0.1$ with
a width of $0.3 \pm 0.1$. These values depend on the value of $p$,
the energy power law index of the electrons which radiate by
synchrotron emission within the fireball, and the state of the
fireball itself (fast/slow cooling, position of the cooling
frequency, beaming, surrounding medium). In section 3.4 we will show
that the average properties of the afterglow are consistent with a
cooling frequency below the X-ray range. In this case, following
Sari et al. 1998, we can determine an average value for
$p=2.4\pm0.2$.

\begin{table}
\centering \caption{\label{table4}X-ray luminosity (assuming
isotropy, $L_{X} ^{iso}$, and after beaming correction, $L_{X}
^{corr}$), Energy emitted during the prompt $\gamma$-ray event
(assuming isotropy, $E_{\gamma} ^{iso}$) in units of $10^{51}$ erg,
and beaming angle for BeppoSAX GRBs with a measured beaming angle
(extracted from literature). }
\begin{tabular}{cccccc}
  \hline\hline
  GRB name & $L_{X} ^{iso}$           & $E_{\gamma} ^{iso}$ & $\theta$ & $L_{X} ^{corr}$   \\
           & $10^{44}$ erg s$^{-1}$ & $10^{51}$ erg       &  rad & $10^{44}$ erg s$^{-1}$ \\
  \hline
\object{GRB 970228} & 28.6  & 9.9    & $>0.32$  & $>1.46$   \\
\object{GRB 970508} & 16.1  & 3.5    & 0.391    &  1.23     \\
\object{GRB 971214} & 147   & 125    & $>0.1$   & $>0.74$   \\
\object{GRB 980613} & 7.21  & 4.26   & $>0.226$ & $>0.2$    \\
\object{GRB 980703} & 37.4  & 74.1   & 0.2      & 0.75      \\
\object{GRB 990123} & 373   & 692    & 0.089    & 1.48      \\
\object{GRB 990510} & 269.7 & 144.5  & 0.054    & 0.39      \\
\object{GRB 990705} &       & 79.4   & 0.096    &           \\
\object{GRB 990712} &       & 3.32   & $>0.777$ &           \\
\object{GRB 000210} & 6.96  & 130    & $>0.139$ & $>0.07$   \\
\object{GRB 000214} & 3.4   & 3.17   & $>0.115$ & $>0.023$  \\
\object{GRB 000926} & 335   & 155    & 0.140    & 2.14      \\
\object{GRB 010222} & 377   & 375    & 0.08     & 13.1      \\
\object{GRB 011121} & 5.1   & 3.74   & 0.145    & 0.05      \\
\object{GRB 011211} & 20    & 68.8   & 0.115    & 0.12      \\
\hline
\end{tabular}
\end{table}


\subsection{General properties of the prompt emission and selection effects}

\begin{figure}
\centering
\includegraphics[width=8cm]{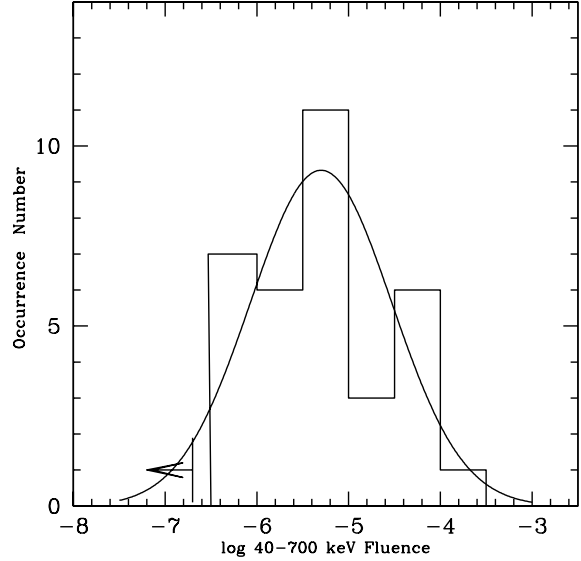}
\caption{The 40-700 keV fluence distribution of the BeppoSAX GRB
sample. Data are extracted from the literature.\label{fig3bis}}
\end{figure}

We list in Table \ref{table2} the properties of the prompt emission
of GRB detected by BeppoSAX, extracted from the literature. Figure
\ref{fig3bis} displays the distribution of the $\gamma$-ray fluence
of the BeppoSAX sample. The fit with a Gaussian provides a mean
logarithmic fluence of $S_{\gamma}=-5.31$ and a width of
distribution $\sigma_{S\gamma}=0.77$\footnote{GRB 980425 has not
been included in this calculation and in the successive ones for its
peculiarity.}.

\begin{figure*}
\centering
\includegraphics[width=8cm]{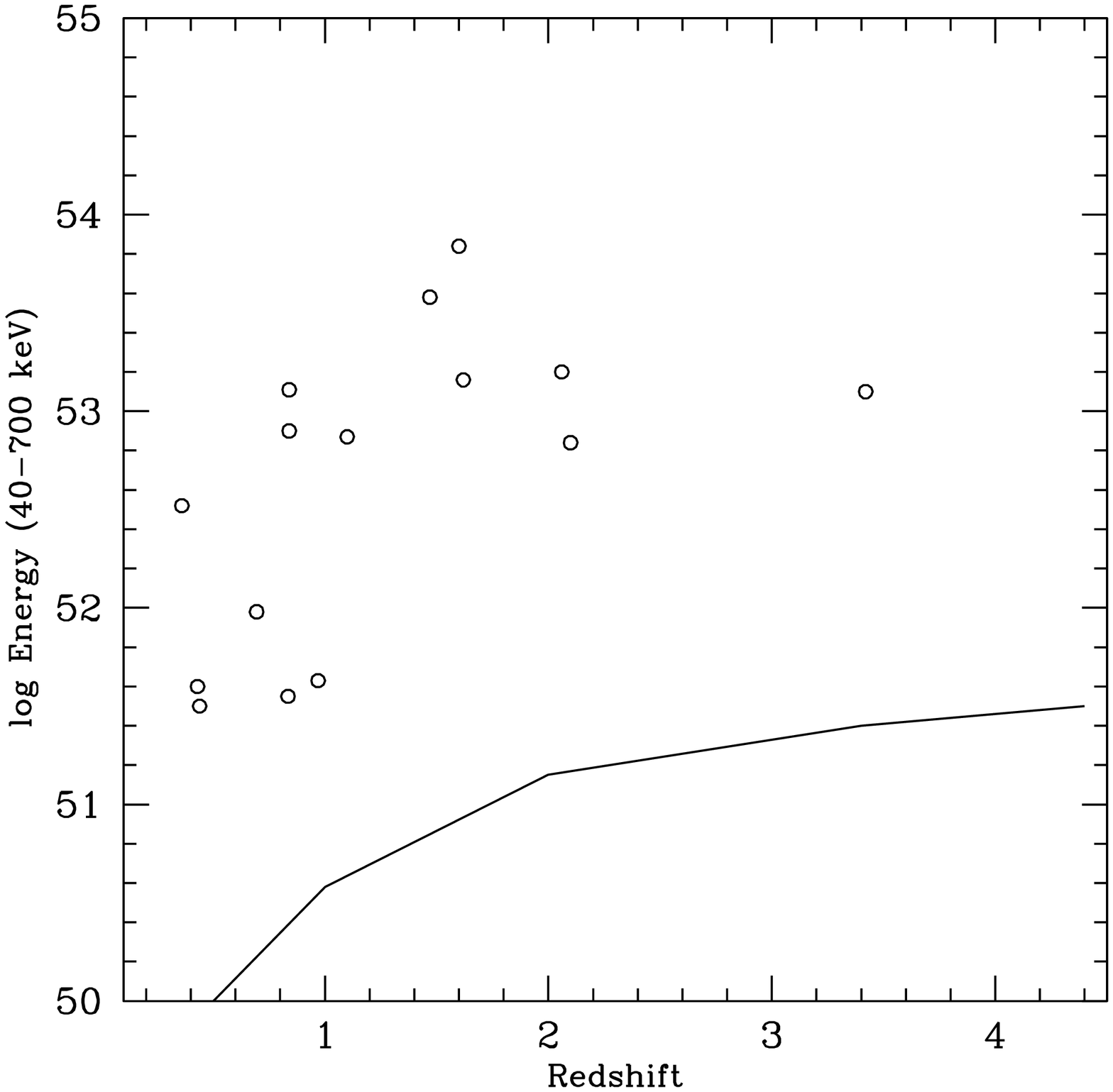}
\includegraphics[width=8cm]{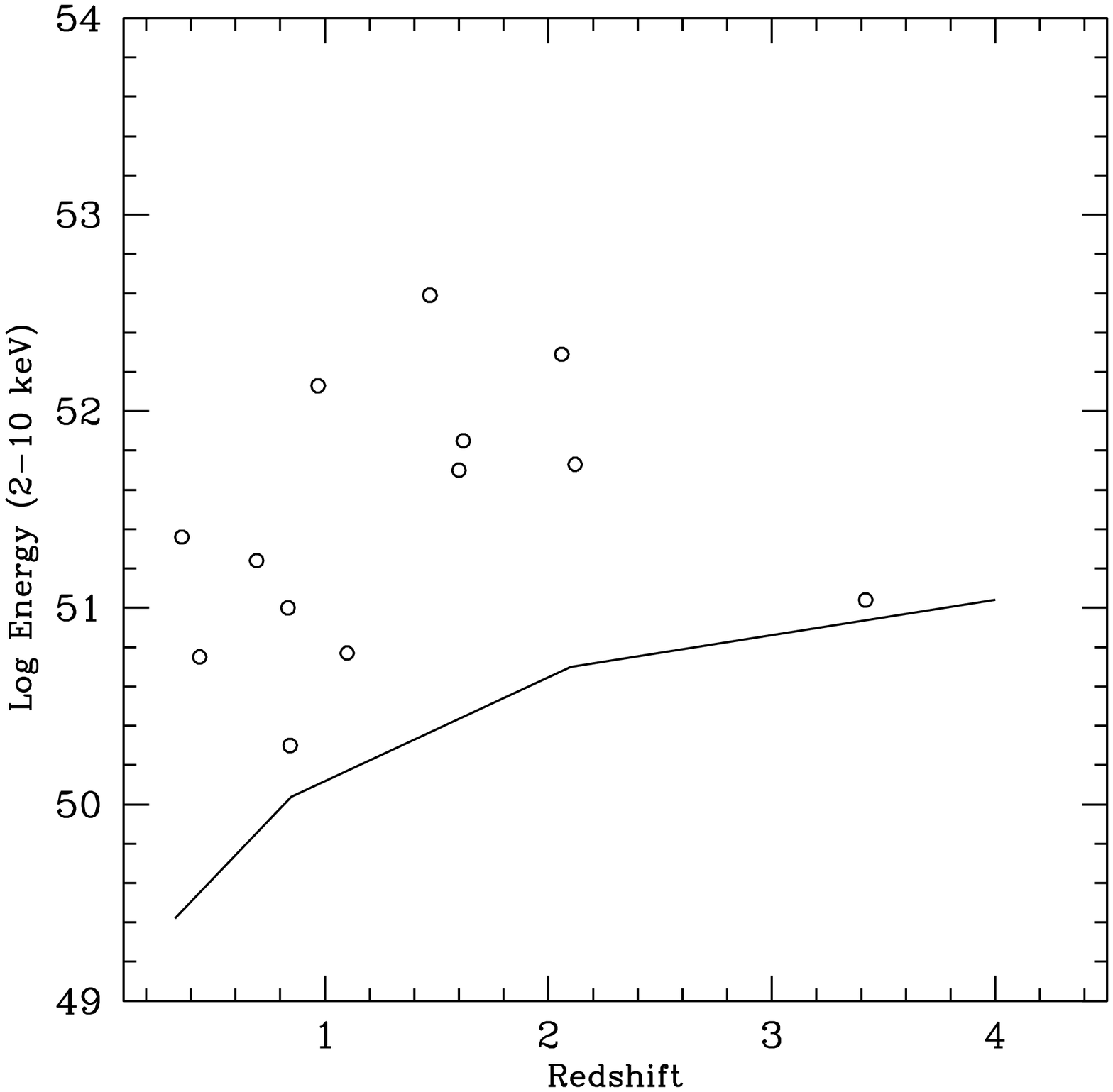}
\caption{Left. Energy emitted during the prompt phase in the 40-700
keV band in the burst rest frame. The solid line represents the
detection threshold of the GRBM discussed in the text. Right.
Energy emitted during the prompt phase in the 2-10 keV band in the
burst rest frame. The solid line represent the detection threshold
of the WFC discussed in the text.\label{fig:selection}}
\end{figure*}

 An important question regards the possible selection effects on the
flux of the prompt phase. In Fig. \ref{fig:selection} we present the
isotropic gamma-ray energy and X-ray energy for events of known
redshift, emitted in the 40-700 and 2-10 keV band respectively in
the GRB cosmological rest frames. They have been calculated by using
the k-correction of Bloom et al. 2001, with cosmological parameters
$H_{0}$=65 km s$^{-1}$ Mpc$^{-1}$, $\Omega_{\Lambda}$=0.7,
$\Omega_{M}$=0.3.

 The continuous lines indicate the detection thresholds as function
of the redshift, \textit{for a typical GRB.} Note that these are
indicative values because the sensitivity depends on the exposed
area as function of the off-set angle and the duration of the event.
The minimum energy required for a detection have been calculated
taking the fluence detection thresholds of the two instruments,
around $S=10^{-7}$ erg cm$^{-2}$ for the GRBM and $S=8\times
10^{-8}$ erg cm$^{-2}$ for the WFC. In the case of the WFC this
corresponds to about 200 mCrab in 20 seconds. From the figures it is
evident that the gamma-ray energies are well above the GRBM
threshold. On the contrary the sample is limited by the WFC
detection threshold, roughly corresponding to a isotropic energy in
the 2-10 keV range of $\sim 10^{50}$ erg  at z=1 and $\sim 10^{51}$
erg at z=4.

 We note, however, that this may not be true for X-ray
rich GRBs and X-ray Flashes \citep{hei03}: the $\gamma$-ray emission
of these objects is weak or absent. In these cases, only the WFC
could detect distant events.

\subsection{Correlation between Afterglow Luminosity and Gamma-Ray Energy.}
\label{sec_energetic}


 We note that the width of the $\gamma$-ray fluence distribution is
not very different from that of the X-ray afterglow flux
distribution (see Fig. \ref{fig3} and Fig. \ref{fig3bis})
 A few authors, e.g.\citet{kup00}, have proposed that the energy
emission from the fireball surface need not be isotropic, but that
large spatial variations of $dE_{\gamma}/d\Omega$ in the fireball
could exist. During the prompt emission phase, the radiation is
highly beamed, due to very high Lorentz factor of the ejecta. These
circumstances would lead to a large spread of $\gamma$-ray fluences.
In the afterglow phase, X-rays are less beamed due to the lower
Lorentz factor, and hence the fluctuations are averaged over a
larger region. Therefore, X-ray flux afterglow distribution would be
less broad than the $\gamma$-ray fluence. As we do not observe such
a difference in the two distribution widths, we cannot support the
hypothesis of \citet{kup00}.

\begin{figure}
\centering
\includegraphics[width=9cm]{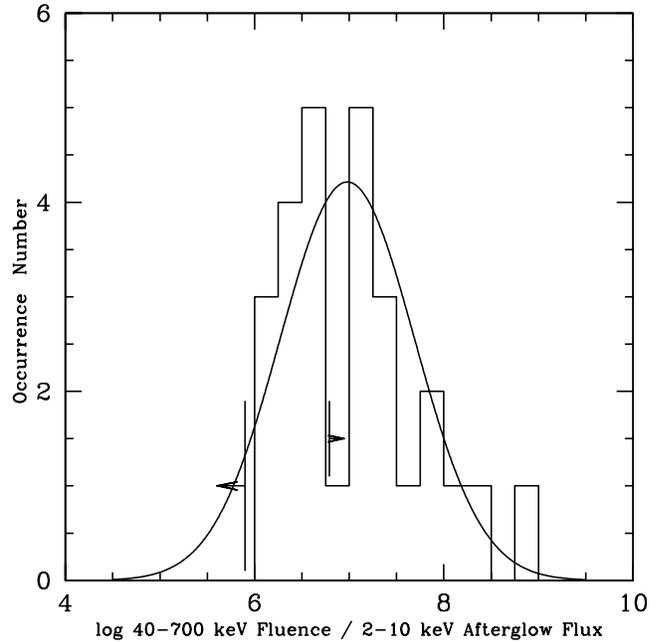}
\caption{Distribution of the logarithmic ratios of the prompt
$\gamma$-ray fluence versus the X-ray afterglow flux for the
BeppoSAX GRB sample.
 \label{fig4}}
\end{figure}

 The distribution of S$_{\gamma}$ - F$_{X}$ ratio is not very broad
($\sigma =0.71$), suggesting a correlation between the X-ray
afterglow luminosity and the gamma-ray energy  (see Fig.
\ref{fig4}). For the sample of burst with known redshift we have
then derived L$_{X}$  by the formula \citep{lam00} :

\begin{equation}
 F(\nu,t) = \frac{L_{\nu}(\nu,t)} {4\pi D^2(z) (1+z)^{1+\alpha-\delta}}
\end{equation}

Luminosity is obtained in the 1.6-10 keV energy band and at 11 hours
after the burst in the rest frame. We have adopted the average
values of $\alpha$, $\delta$ reported in the previous section. The
cosmological parameters used are the same as for the computation of
the emitted energy (see Sec. \ref{section31}) \footnote{As for
GRB000214, $z=0.44$ was adopted.}.

\begin{figure}
\centering
\includegraphics[width=9cm]{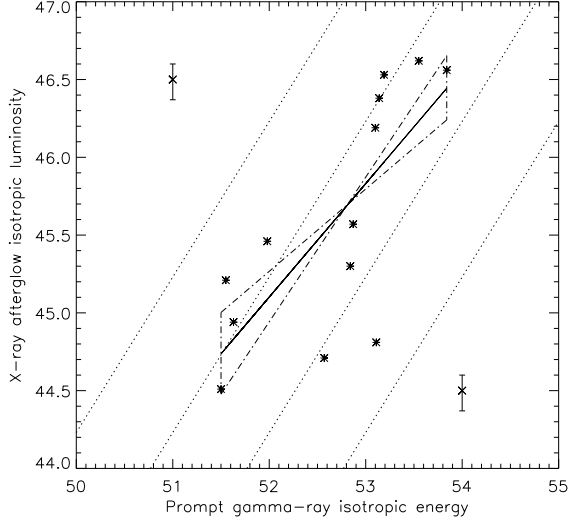}
\caption{\label{fig6} 1.6-10 keV Afterglow Luminosity vs 40-700 keV
Energy of the prompt emission. The fit between these two quantities,
discussed in the text, is also shown together with its confidence interval (dot-dashed box). The correlation coefficient is
$r=0.74$. The dotted lines represent the Eqn. \ref{eqn_max} is case
of $\epsilon_{e}^{1.3}/\epsilon_{\gamma} = 10$ (upper line), 1 and
0.1 (middle lines) and 0.01 (lower line)}
\end{figure}

 In Fig.\ref{fig6} we plot L$_{X}$ vs E$_{\gamma}$. The correlation
coefficient is r=0.74 and the probability of chance correlation is
0.008. It is worth noting that some indication of correlation
between prompt and afterglow luminosity is also found in a small set
of \textit{Swift} bursts \citep{chi05}.


 Assuming that the observed X-ray frequency $\nu_{X}$ is above the cooling
frequency $\nu_{c}$, the measurement of X-ray luminosity at a fixed
time after the burst gives an estimate of isotropic kinetic energy
of the fireball $E_{K,A}$ \citep{fre01} :

\begin{equation}
\label{equation_2}
E_{K,A} = C \epsilon_{e} ^{\frac{-4p+4}{p+2}} \nu t L_{X}
\end{equation}

In that equation, $C$ is a parameter which depends very
weakly on the fraction of energy carried by the magnetic field
$\epsilon_{B}$, the luminosity distance, the flux density, the time
$t$ and the frequency of observation $\nu$. $C$ has a stronger
dependence on the value of $p$, however henceforth we will make the
simplifying assumption that the value of this parameter is the same
for all bursts examined. For our purposes, the value of $C$ can thus
be considered constant. We also note that Eqn. \ref{equation_2} does not depend on the
value of the density of the circumbust medium, so it holds either in
the case of expansion in interstellar medium, with constant density,
or in the case of medium affected by wind of the progenitor star,
with a typical density profile decreasing from the center of the
explosion.

 Using $p=2.4$, the value determined from the data, a luminosity distance
of $3\times10^{28}$ cm, time and frequency of observation of $40000$
sec and $2.4\times 10^{17}$ Hz, a flux density of $0.3 \mu$Jy,
$\epsilon_{B}$=0.01, Eqn. \ref{equation_2} becomes :

\begin{equation}
E_{K,A} = 5.8\times 10^{6} \epsilon _{e} ^{-1.3} L_{X}
\end{equation}

 In the case of gamma-ray emission, we have to consider an unknown
coefficient of conversion of relativistic energy of the fireball
into gamma-ray energy \citep{pira01}.

 \begin{equation}
 E_{\gamma} = \epsilon_{\gamma} E_{K,P}
 \end{equation}

 where $E_{K,P}$ is the isotropic relativistic energy of the fireball in the
prompt phase. We may suppose $E_{K,P}\backsimeq E_{K,A}$, because
$\epsilon_{\gamma}$ cannot be too close to unity otherwise there
will not be an afterglow \citep{kob97, pira01}. We assume that
radiative losses also are negligible. From the previous equations we
derive:

\begin{equation}
\label{eqn_max}
  L_{X} = 1.73 \times 10^{-7} \epsilon_{e}^{1.3} \epsilon_{\gamma}^{-1} E_{\gamma}
\end{equation}

 We plot in Fig. \ref{fig6} this relationship (dotted lines),
assuming $\epsilon_{e}^{1.3}/\epsilon_{\gamma}$ equal to 0.01, 0.1,
1 and 10 respectively.
 As one can see, the correlation we have found implies that the ratio
$\epsilon_{e}^{1.3}/\epsilon_{\gamma}$ does not strongly vary from
burst to burst. Assuming that $\epsilon_e$ is not too close to zero
\citep[a common value observed is $\sim$ 0.3][]{fre01}, this implies
that $\epsilon_{e}$ is approximately proportional to
$\epsilon_{\gamma}$. Thus, the fraction of fireball energy carried
by relativistic electrons in the external shock and emitted in the
afterglow is roughly proportional to the fraction of the fireball
relativistic energy converted into $\gamma$-rays during the prompt
phase.

\subsection {Jet collimation}
\label{sec_beaming}
\begin{figure}
\centering \resizebox{8.5cm}{!}{\includegraphics[bb=0 92 592
800]{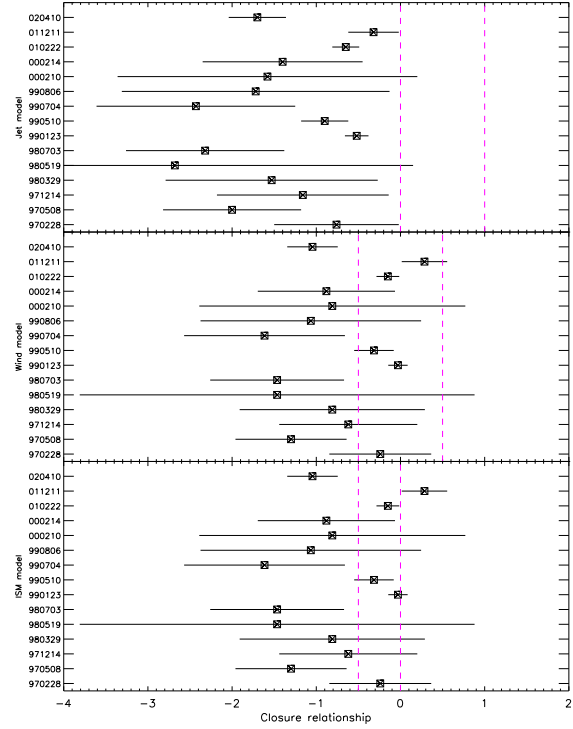}}
\caption{The closure relationships for all burst with constraints on
both the spectral and temporal decay indexes. We indicate the
closure relationships for the three cases (Jet expansion, Wind
model, ISM model) in the three panels. Vertical lines indicate the
theoretical expected values.\label{fig7}}
\end{figure}

 According to \citet{sar98, che99, rho97}, the decay index and
the spectral index values are linked together by closure relationships
that depend on the burst geometry and environment. We present the closure
relationships for each burst in Fig. \ref{fig7}, and focus first on the
burst geometry (shown in the top panel of Fig. \ref{fig7}).

As one can see, the jet signature is ruled out in most of the cases from our
analysis. This is also evident when we calculate the mean value for the
closure relationship. For a jet signature, this is :

\begin{eqnarray}
\label{eq1}
\delta - 2 \alpha - 1 = -2.1 \pm 0.22 & & \nu_x < \nu_c \\
\label{eq1bis}
\delta - 2 \alpha \phantom{-1} = -1.1\pm 0.22 & & \nu_x > \nu_c
\end{eqnarray}

In Eq. \ref{eq1} and \ref{eq1bis} we should expect a value of 0,
clearly excluded by the data. This implies that the beaming angle
may be large. We can set a lower limit on its value ($\theta$).
According to \citet{sar99}, we have :

\begin{equation}
\label{eq2}
\theta = 0.057 \left(\frac{n_{-1}}{E_{\gamma,i,53}} \right)^{1/8}
t_{\theta,day}^{3/8} \left(\frac{\epsilon_{\gamma}}{0.2}\right)^{1/8}
\left(\frac{1+z}{2}\right)^{-3/8}
\end{equation}

  In Eqn. \ref{eq2}, E$_{\gamma,i,53}$ represents the isotropic energy
emitted in $\gamma$-rays by the fireball in units of $10^{53}$ erg, $n_{-1}$ is
the density in 0.1 particle cm$^{-3}$ unit, $\epsilon_{\gamma}$ is the efficiency of
conversion of explosion energy into $\gamma$-rays, and $t_{\theta,day}$
the date when the break of light curve, due to the beamed emission,
appears (expressed in days).

  BeppoSAX TOOs are mostly carried out within 2 days after the GRB. Because
decay and spectral slopes are not consistent with a collimated
outflow, we can derive $t_{\theta,day} > 2$.  Assuming a typical
E$_{\gamma,i,53}$=$1$, $\epsilon_{\gamma}=0.2$, $n_{-1}=100$ \citep{ber03} we
obtain a limit of $\theta \gtrsim 0.1 $ rad, which in turn give us a
lower limit on the beaming factor $f_{b} \approxeq 0.005$. This
result is of the same order of magnitude of that claimed by
\citet{fra01}. We note that the majority of beaming angles, mostly
inferred by breaks in optical light curves, are consistent with this
result. Only GRB 990510 and GRB010222 seem to represent exceptions
(see table \ref{table4}).

  A density of $n_{-1}=100$ is typical of the interstellar medium.
On the other hand, several authors proposed that GRBs are originated
by massive stars \citep[e.g.][]{woo93}. In such a case, these stars
should produce the GRB within their original forming region, which
are usually very dense. If we assume $n_{-1}=10^{4}$, which is
typical of Giant Molecular Clouds, the beaming angle limit increase
to $\theta \gtrsim 0.24$ rad, which corresponds to a beaming factor
limit of $f_{b} \approxeq 0.03$.

\begin{figure}
\centering
\includegraphics[width=8cm]{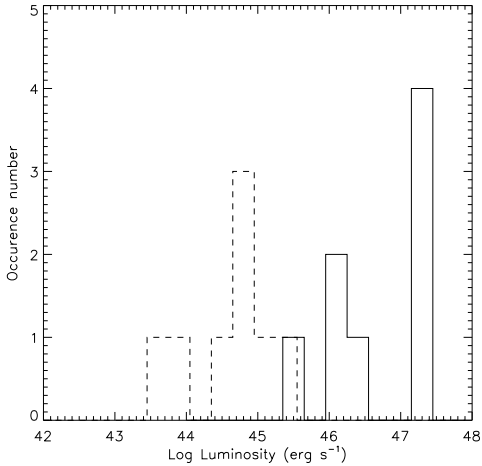}
\caption{Afterglow Luminosity of BeppoSAX GRBs with known redshift.
Solid line : before correction for beaming. Dashed line : after
correction for beaming.\label{fig8}}
\end{figure}

\citet{ber03} claimed that the distribution of X-ray afterglow
luminosity appears to converge significantly toward a common value
after beaming correction. We have tested this hypothesis with our
sample, using the beaming angle values reported in the literature
\citep[see Table \ref{table4}; most of them are extracted from the
article by][]{ber03}. The isotropic luminosity is corrected for
beaming by applying a multiplicative factor depending on the beaming
angle \citep[see][for details]{ber03}. Before beaming correction,
the luminosity distribution displays a logarithmic width of 0.8 (see
Fig. \ref{fig8}), with a mean value of $7.2 \times 10^{45}$ erg
s$^{-1}$. After the beaming correction, the distribution width
shrinks to a value of 0.4, very similar to the 0.3 value
\citet{ber03} obtained. The mean luminosity decreases to $9.5 \times
10^{43}$ erg s$^{-1}$ (Fig. \ref{fig8}).

 One may note that the beaming angle was calculated assuming a density of
10 cm$^{-3}$ when it was unknown. This may have strong consequences. As
an example, \citet{zan01} has reported a density value of $10^6$
cm$^{-3}$ for \object{GRB 010222}. When using this value, rather than
that reported by \citet{ber03}, the beaming angle increases up to 0.26
rad. This leads the beaming corrected luminosity distribution width to
increase to a value of 0.7, clearly not supporting anymore the
hypothesis of a standard energy release in the afterglow. Thus, such
claims should be accepted with caution, depending on the assumptions made
on the density values.

\subsection{The density profile of the environment}
\label{sec_environ}

 Figure \ref{fig7} displays also the closure relationships for an
expansion into a wind environment (the WIND case, middle panel) and
a constant density medium (the ISM case, bottom panel). These
closure relationships present a degeneration when $\nu_{c}<\nu_{X}$,
which prevents us from drawing any conclusion. One can see from Fig.
\ref{fig7} that most of the bursts are in that situation. The
uncertainties of other bursts do not allow us to draw any conclusion
for most of them using only the X-ray data. This is also shown by
the mean closure relationships reported in Table \ref{table6}: the
two medium cases can fit the mean value if the cooling frequency is
below the X-rays, while none of them can fit the mean value in the
opposite case.

\begin{table}[h!]
\centering
\caption{\label{table6} Mean closure relationship from our sample. We indicate the wind and ISM closure relationships, depending of the cooling frequency position.}
 \small
 \begin{tabular}{ccc}
   \hline\hline
    & ISM  & Wind \\
   \hline
   $\nu_{X}<\nu_{c}$ & $\delta -1.5\alpha = -0.5\pm0.2 $&  $\delta -1.5\alpha -0.5 =
-1\pm0.2$ \\
   $\nu_{c}<\nu_{X}$ & $\delta -1.5\alpha + 0.5 = 0 \pm0.2 $  & $\delta - 1.5\alpha +0.5 =
0\pm 0.2 $ \\
   \hline
 \end{tabular}
\end{table}

 To get rid of this degeneration, we need to use also the optical
observations. From the fireball model, the X-ray decay index is
larger than the optical one if the cooling frequency is between the
optical and X-ray bands and if the fireball is expanding into a
constant density medium \citep{sar98}. The difference between the
optical and X-ray decay index is $-0.25$. If the fireball expands
into a wind environment (also assuming the cooling frequency to be
between the optical and X-ray bands), then it is the optical decay
index which is larger than the X-ray decay index. The difference
between the optical and X-ray decay index is now 0.25. Assuming that
the cooling frequency is indeed between the optical and the X-ray
bands, we can remove the degeneration.

\begin{table}
\centering
\caption{\label{table5}Optical decay indexes and comparison with the X-ray band
decay indexes.}
\begin{tabular}{cccc}
  \hline\hline
  GRB & $\delta_{O}$ & $\delta_{X} - \delta_{O}$ & Reference \\
  \hline
\object{GRB 970228} & $1.21\pm0.02$ & 0.11  & 1 \\
\object{GRB 970508} & $0.15\pm0.02$ & 0.65  & 2 \\
\object{GRB 971214} & $1.20\pm0.02$ & -0.20 & 3 \\
\object{GRB 980329} & $1.28\pm0.19$ & 0.14  & 4 \\
\object{GRB 980613} & $0.8\pm0.5$   & 0.69  & 5 \\
\object{GRB 980703} & $1.22\pm0.35$ & -0.12 & 6 \\
\object{GRB 990123} & $1.10\pm0.35$ & 0.34  & 7 \\
\object{GRB 990510} & $0.8\pm0.2$   & 0.64  & 8 \\
\object{GRB 010222} & $1.32\pm0.03$ & 0.03  & 9 \\
\object{GRB 011121} & $1.63\pm0.61$ & -0.33 & 10\\
\object{GRB 011211} & $0.95\pm0.2$  & 1.15  & 11\\
\object{GRB 020322} & $0.5\pm0.25$ & 0.34  & 12, 13\\
  \hline
\end{tabular}

References : 1: \citet{mas98} 2: \citet[][the index shown is
relative to the BeppoSAX observation interval]{gal98} 3:
\citet{die98} 4: \citet{rei99} 5: \citet{hjo02} 6: \citet{blo99} 7:
\citet{kul99} 8: \citet{har99} 9 : \citet{mas01} 10: \citet{pri02}
11: \citet{jac03} 12: \citet{blo02} 13: \citet{gre02}
\end{table}

\begin{figure}[h!]
\centering
  \includegraphics[width=8cm]{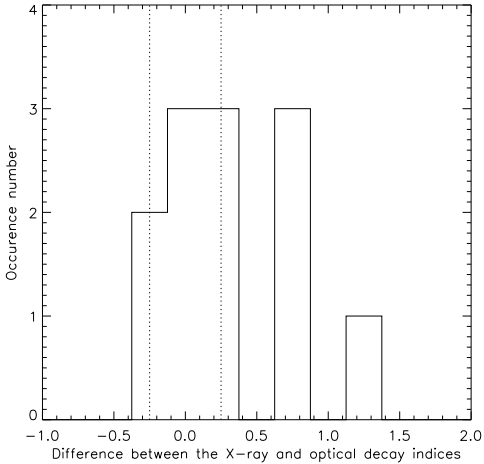}
  \caption{\label{fig9}Difference of the X-ray and optical decay indexes of BeppoSAX sample. Right line:
$\delta_{X}=\delta{o}+0.25$ (as expected for an ISM environment). Left line: $\delta_{X}=\delta{o}-0.25$ (as expected for a wind environment). }
\end{figure}

In Table \ref{table5} we show the optical vs X-ray band decay indexes
(results taken from the literature). We excluded GRB 980519 and
GRB000926 from our set because in their case the jet phase started
slightly after the beginning of BeppoSAX observations (Jaunsen et
al. 2001, Fynbo et al. 2001), therefore we may have their decaying
behavior largely affected by the change of slope.

 For the remaining GRBs with both X-ray and optical afterglows detected,
the average value of the decay index is $\delta_{O}=1\pm0.2$ in the
optical and $\delta_{X}=1.3\pm0.2$ in the X-rays. The difference
between these two values is $0.3 \pm 0.3$. A constant density medium
surrounding the burst is thus favored, but a wind environment is not
ruled out. This is also visible in Fig. \ref{fig9}, where we plot
the $\delta_{X} - \delta_{O}$ value for each single burst. For a
majority of them, the value 0.25 is preferred, thus implying also
that we observe a constant density medium surrounding the burst, for
some others, we observe indeed a wind medium. This is tricky, as one
should expect, if the long GRB progenitor is indeed a massive star
\citep[as the GRB-supernova association claimed for several GRBs
implies, see e.g.][]{sta03, hjo03}, the surrounding medium to be the
wind arising from the star for all bursts \citep{che99}.
\citet{ram01} suggested the existence of a termination shock that
could maintain the wind close to the star \citep[see also][]{che04}.
This would explain our observations. In such a case, this implies
that the termination shock has been crossed before the observations
(thus early after the burst), which should then imply a dense
surrounding medium. This is supported by the large absorption
observed around the bursts (see Table \ref{table3}): such a high
density column may be due to a compact and dense layer around the
burst. This is also supported by the observation of \object{GRB
010222}. For this burst, the surrounding medium is indeed the
interstellar medium (see Fig. \ref{fig7}). \citet{zan01} has
proposed this burst to be surrounded by a very dense ($10^6$
cm$^{-3}$) medium or affected by a jet effect. We can discard the
jet effects (see Fig. \ref{fig7}), and thus confirm the proposed
explanation. Such a medium, with a large density, would be very
efficient to maintain the termination shock nearby the GRB
progenitor.

 Finally, we would like to underline the fact that inferences
drawn from our afterglow analysis are in general agreement with
those of the reviews of \citet{fro05} and \citet{piro04}. This is
not very surprising, however, because of the wide consistency of
\citet{fro05} results with ours, while \citet{piro04} used a large
part of the same GRB X-ray afterglow set and basically the same data
analysis to derive his conclusions.

 \section{Dark GRBs}
\label{sec_dark}
 About 90 \% of the GRBs detected by BeppoSAX present an X-ray
afterglow. On the other hand, only 16 GRBs present an optical
afterglow. Taking into account the late follow up of \object{GRB
960720} and the absence of optical observations of \object{GRB
980515} and \object{GRB 020427}, this implies that only 42 \% of the
GRBs detected by BeppoSAX have an identified optical afterglow. This
led to the definition of the so called {\it Dark} bursts
\citep{dep03}. Several authors \citep[e.g.][]{fyn01, fox03, rol05}
pointed out that this definition can in fact hide an instrumental
bias (as this does not take into account the date of the optical
follow up and the decay rate of the optical afterglow). In fact, the
non detection of the optical afterglow can be due to several
reasons: a late follow up, a steep decay, an intrinsic faintness, a
large dust extinction and a distant burst. While the first two
possibilities are instrumental bias, the last three give information
about the burst.

 For those bursts with a rapid optical follow up and a non detection
of the optical follow up, it has been shown that on average the
optical flux should be $2$ magnitude lower than bursts with an
optical afterglow in order to explain the non detection of the
optical source \citep{laz02}. Another study made with a sample of 31
BeppoSAX GRB afterglows indicated that the X-ray afterglow fluxes of
dark GRBs are, on average, $4.8$ times weaker than those of normal
bursts \citep{dep03}. The probability that this flux distribution
comes from a single population of burst is 0.002, i.e.a $3\sigma$
rejection. Using the whole BeppoSAX sample, this probability does
not change significantly.

\begin{figure}
\centering
\includegraphics[width=8cm]{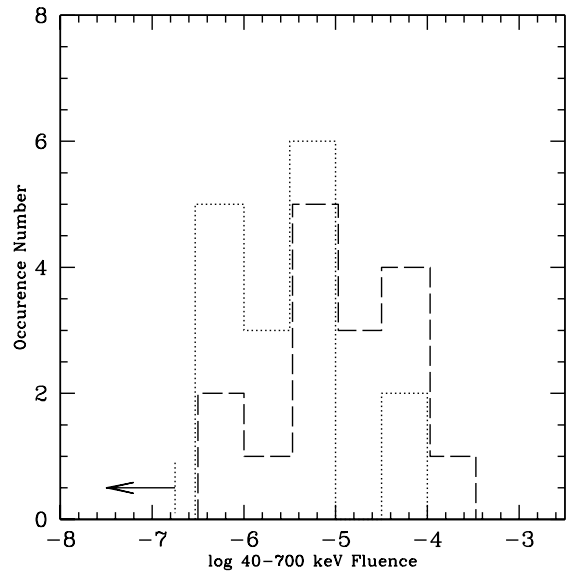}
\caption{\label{fig10}Comparison of the $\gamma$-ray fluences of
dark (dotted line) and optically bright (dashed line) GRBs.}
\end{figure}

 The results exposed in Sec. \ref{sec_energetic} imply that this
X-ray faintness should extend to the prompt phase, and thus that
dark GRBs should present a fainter $\gamma$-ray fluence. We have
tested this hypothesis and present the result in Fig. \ref{fig10}.
As one can see, there is indeed a trend for the dark burst (dotted
line) to have a low  $\gamma$-ray fluence compared to GRBs with
optical transient (OT GRBs). The ratio between the average dark GRB
fluence and OT GRBs fluence is 5.7, similar to the value of the
ratio of X-ray fluxes and the expected value derived from the
correlation observed in Sec. \ref{sec_energetic}. The probability
that optically bright GRBs and dark GRBs fluence distributions
derive from an unique population of burst is 0.01. It thus seems
that faintness is an intrinsic property of dark GRBs at all
wavelengths.

 The above statements can explain the non detection of the optical
afterglow. But they imply that {\it the whole afterglow} is affected
by this effect (i.e. the faintness is observed in all the
observation bands). On the contrary, extinct optical afterglow and
distant bursts should also feature a faintness that is wavelength
dependent (due to dust-to gas laws in the first case and due to the
Lymann-$\alpha$ forest redshifted in the optical band in the second
case). To discriminate all these effects and to validate their
interpretation, \citet{dep03} also carried out a comparison of the
X-ray and optical fluxes. They found that 75 \% of dark bursts were
compatible with a global faintness, and thus that these bursts were
dark because searches were not fast or deep enough.

 For the remaining GRBs, the optical-to-X-ray flux ratio is at
least a factor 5-10 lower than the average value observed in normal
GRBs. In terms of spectral index, these events have optical to X-ray
spectral index $\alpha_{OX}\lesssim0.6$, whereas for OT GRBs the
average value is $\simeq0.75$. These facts strongly suggest that for
these bursts the spectrum is depleted in the optical band.
\citet{jac04}, using a similar method and comparing their results
with the fireball model expectations, indicated that at least 10 \%
of their sample was not compatible with the fireball model and thus
were {\it truly dark} GRBs. It is worth noting that the
\textit{Swift} mission \citep{geh05}, recently begun, has
already confirmed that a considerable fraction of GRBs has tight
upper limits for the optical emission (Roming et al. 2005, in preparation) We can
thus indicate that about 10-20 \% of GRBs is characterized by an
optical afterglow emission fainter than that expected from the X-ray
afterglow flux. These bursts could be distant (z$>5$) or extinct
bursts.

 Two dark bursts have been associated with host galaxies at
z $<5$ \citep{djo01, piro02}. We also note (see Table \ref{table3})
that the X-ray absorption around some bursts is important and could
be responsible of an important optical extinction \citep[see
e.g.][]{str04}. Thus, for some of these events, the likely
explanation of the darkness is an optical depletion by dust in star
forming region. This in turn supports the massive star progenitor
hypothesis for long GRBs, as these massive stars are likely to
explode in their original star forming region. On the other hand,
this does not rule out the distance explanation for some dark bursts
with no known host. In fact, it is likely that the dark burst
population is the sum of these three (faint, distant and extinct)
populations. In principle, these cases could be disentangled by
other measurements such as column density, prompt E$_{peak}$, X-ray
flux. However, it is important to be cautious, because a few X-ray
flashes (see Heise et al. 2001) could have the values of these
parameters consistent with those of very high redshift GRBs, even if
they are not actually placed at $z>5$.

\section{Conclusions}
\label{sec_conclu}
 We have presented the BeppoSAX X-ray afterglow catalog. Thirty-nine
BeppoSAX afterglow observations were carried out on a sample of 52
detected GRBs. Thirty-one X-ray afterglows were securely identified
due to their fading behavior. Three other observations led to the
detection of only one source within the prompt positional error box.
Thus, X-ray afterglows are present in $\sim90$\% of the
observations.

 We derived the main properties - flux, decay index, spectral index,
absorption - for 15 afterglows, and give constraints on decay slope
and flux for the remaining. The width of the prompt fluence and
X-ray afterglow flux distributions are similar, suggesting no strong
spatial variation of the energy emission within the beamed fireball.
We pointed out a likely correlation between the X-ray afterglow
luminosity and the energy emitted during the prompt $\gamma$-ray
event. Such a correlation suggests that the fraction of fireball
energy carried by relativistic electrons in the external shock and
emitted in the afterglow is roughly proportional to the fraction of
the fireball relativistic energy converted into $\gamma$-ray during
the prompt phase.

 We do not detect significant jet signature within the afterglow
observations, implying a lower limit on the beaming angle of
$\sim0.1$. Moreover, we note that the hypothesis of standard energy
release in the afterglow as claimed by \citet{ber03} may be consistent with our sample,
but it strongly depends on the assumptions made about the density of
the surrounding medium.

 The average value of the spectral index of
the electron energy distribution, inferred by our time and spectral
analysis, is $p=2.4\pm0.2$.

 Our data support the fact that GRBs should be typically
surrounded by a medium with a constant density rather than a wind
environment, and that this medium should be dense. This may be
explained by a termination shock located near the burst progenitor.
 We finally pointed out that some bursts without optical counterpart
may be explained by an intrinsic faintness of the event, while
others can be strongly absorbed.

 A first
comparison with the bursts observed by XMM-Newton and Chandra are
presented in \citet{gen05}. In a forthcoming paper (Gendre et al.,
in preparation), we will search the spectra for metal lines and
other deviations from the continuum properties.

\begin{acknowledgements}
The BeppoSAX satellite was a joint program of Italian (ASI) and
Dutch (NIVR) space agencies.
BG acknowledges a support by the EU FP5 RTN 'Gamma ray bursts: an enigma and a tool'.
\end{acknowledgements}

\clearpage

\begin{longtable}{lcccccl}
\caption{\label{table1} GRBs localized and/or observed by BeppoSAX.
We indicate the position the first TOO start and end times, the sum
of the Good Time Interval (GTI), and the date of the subsequent
TOOs. A 'WFC' following the position means that this GRB was
localized only the the WFC, 'NFI' a localization obtained by NFI. An
external trigger of a BeppoSAX TOO is indicated by giving in
parenthesis the satellite that localized the burst, but the
localization displayed has been provided by NFI. We also indicate in
the table if an optical afterglow was detected together with the
distance and other information obtained from the
optical study.}\\
\hline\hline
GRB name & Position          & Localization & First TOO & \hspace{0.1cm} Sum of    & Other TOOs & Optical afterglow\\
         & (Right Ascention, &              & start-end & \hspace{0.1cm} GTI$^{a}$ & start-end  & detection (redshift)\\
         &   Declination)    &              & (hours)   & \hspace{0.1cm}(ksec)     & (hours)     & \\
\hline
\endfirsthead
\caption{continued.}\\
\hline\hline
GRB name            & Position          & Localization & First TOO & Sum of  & Other TOOs & Optical afterglow\\
                    & (Right Ascension, &              & start-end &  GTI$^{a}$ & start-end  & detection (redshift)\\
                    &    Declination)   &              & (hours)   & (ksec) & (hours)     & \\
\hline
\endhead
\hline
\endfoot
\object{GRB 960720} & $17^h30^m37^s \phantom{00} +49\degr05\arcmin48\arcsec$ & WFC   &  3715-3765.2 &  49.1  &     ---     & N \\
\object{GRB 970111} & $15^h28^m10^s \phantom{00} +19\degr36\arcmin17\arcsec$ & NFI   &  16-46.5     &  56    &     ---     & N  \\
\object{GRB 970228} & $05^h01^m47^s \phantom{00} +11\degr46\arcmin41\arcsec$ & NFI   &  8-16.7      &  14.3  & 89.6 - 98.8 & Y (z=0.695) \\
\object{GRB 970402} & $14^h50^m03^s \phantom{00} -69\degr20\arcmin06\arcsec$ & NFI   &  8-19        &  23.6  & 40.9-58.5   & N \\
\object{GRB 970508} & $06^h53^m49^s \phantom{00} +79\degr16\arcmin20\arcsec$ & NFI   &  6-21.6      &  35.5  & 66-74   & Y (z=0.835) \\
                    &                                                        &       &              &        & 136.3-160 & \\
\object{GRB 971214} & $11^h56^m25^s \phantom{00} +65\degr12\arcmin43\arcsec$ &  NFI  &  6.5-60.7    &  101   &    ---      & Y (z=3.42)\\
\object{GRB 971227} & $12^h57^m15^s \phantom{00} +59\degr23\arcmin26\arcsec$ &  NFI  &  12-31.2     &  37    &     ---     & N \\
\object{GRB 980109} & $00^h25^m56^s \phantom{00} -63\degr01\arcmin24\arcsec$ &  WFC  &    ---       &  ---   &   ---       & N \\
\object{GRB 980326} & $08^h36^m26^s \phantom{00} -18\degr53\arcmin00\arcsec$ &  WFC  &     ---      &   ---  &     ---     & Y \\
\object{GRB 980329} & $07^h02^m37^s \phantom{00} +38\degr50\arcmin46\arcsec$ &  NFI  &  7-48.6      &  63.8  &     ---     & Y \\
\object{GRB 980425} & $19^h35^m02^s \phantom{00} -52\degr50\arcmin16\arcsec$ &  NFI  &  10.2-52.4   &  52.1  & 161-185     & SN (z=0.0085)\\
                    &                                                        &       &              &        & Nov 10.75-12& \\
\object{GRB 980515} & $21^h16^m44^s \phantom{00} -67\degr13\arcmin05\arcsec$ &  NFI  &  10-47.2     &  49.1  & 218-265     & No study \\
\object{GRB 980519} & $23^h22^m17^s \phantom{00} +77\degr15\arcmin53\arcsec$ &  NFI  &  9.7-35.2    &  78    &     ---     & Y \\
\object{GRB 980613} & $10^h18^m04^s \phantom{00} +71\degr33\arcmin58\arcsec$ &  NFI  &  8.6-35.3    &  61.5  &     ---     & Y (z=1.1)\\
\object{GRB 980703} & $23^h59^m07^s \phantom{00} +08\degr35\arcmin06\arcsec$ &(RXTE) &  22.3-45.6   & 39.2   & 110.3-132.6 & Y (z=0.97)\\
\object{GRB 981226} & $23^h29^m37^s \phantom{00} -23\degr55\arcmin45\arcsec$ &  NFI  &  6.5-61      &  89      & 172-191     & N \\
\object{GRB 990123} & $15^h25^m31^s \phantom{00} +44\degr45\arcmin52\arcsec$ &  NFI  &  5.8-53.9    &  81.9  &     ---     & Y (z=1.62)\\
\object{GRB 990217} & $03^h02^m45^s \phantom{00} -53\degr06\arcmin11\arcsec$ &  NFI  &  6-44      &  56.4    &     ---     & N \\
\object{GRB 990510} & $13^h38^m03^s \phantom{00} -80\degr29\arcmin44\arcsec$ &  NFI  &  8-44.4       &  67.9  &     ---     & Y (z=1.6)\\
\object{GRB 990625} & $00^h26^m34^s \phantom{00} -32\degr12\arcmin00\arcsec$ &  WFC  & ---          & ---    & ---         & No study\\
\object{GRB 990627} & $01^h48^m23^s \phantom{00} -77\degr05\arcmin22\arcsec$ &  NFI  &  8-39.7      & 30   &     ---     & N\\
\object{GRB 990704} & $12^h19^m28^s \phantom{00} -03\degr50\arcmin00\arcsec$ &  NFI  &  7.5-29.5    & 37     &  169.8-195  & N\\
\object{GRB 990705} & $05^h09^m52^s \phantom{00} -72\degr08\arcmin02\arcsec$ &  WFC  &  11-33.8     & 77.8     &     ---     & Y (z=0.86)\\
\object{GRB 990712} & $22^h31^m49^s \phantom{00} -73\degr24\arcmin24\arcsec$ &  WFC  &   ---        &  ---   &     ---     & Y (z=0.43)\\
\object{GRB 990806} & $03^h10^m36^s \phantom{00} -68\degr07\arcmin13\arcsec$ &  NFI  &  8-48.9      & 77.9   &     ---     & N \\
\object{GRB 990907} & $07^h31^m07^s \phantom{00} -69\degr27\arcmin24\arcsec$ &  NFI  &  11-11.4     & 1.1    &     ---     & N \\
\object{GRB 990908} & $06^h52^m53^s \phantom{00} -74\degr59\arcmin17\arcsec$ &  WFC  &   ---        &  ---   &     ---     & N \\
\object{GRB 991014} & $06^h51^m02^s \phantom{00} +11\degr35\arcmin37\arcsec$ &  NFI  &  13-33.9     & 36.1   & 258-285.8   & N \\
\object{GRB 991105} & $12^h03^m29^s \phantom{00} -67\degr45\arcmin25\arcsec$ &  WFC  &    ---       &  ---   &   ---       & N \\
\object{GRB 991106} & $22^h24^m43^s \phantom{00} +54\degr23\arcmin22\arcsec$ &  NFI  &  8-26.8      & 31.6   &     ---     & N \\
\object{GRB 000210} & $01^h59^m17^s \phantom{00} -40\degr39\arcmin17\arcsec$ &  NFI  &  7.2-40.2    & 44.4   &     ---     & N (z=0.835)\\
\object{GRB 000214} & $18^h54^m28^s \phantom{00} -66\degr27\arcmin59\arcsec$ &  NFI  &  12-41.5     & 50.8   &     ---     &N (z=0.37-0.47)\\
\object{GRB 000528} & $10^h45^m09^s \phantom{00} -33\degr59\arcmin01\arcsec$ &  NFI  &  12-27.3     & 26.6   &     78.8-99  & N     \\
\object{GRB 000529} & $00^h09^m27^s \phantom{00} -61\degr31\arcmin43\arcsec$ &  NFI  &  7.4-50.5    & 34.8   &     ---     & N    \\
\object{GRB 000615} & $15^h32^m42^s \phantom{00} +73\degr47\arcmin23\arcsec$ &  NFI  &  10-41.6     &  44.6   &     ---     & N \\
\object{GRB 000620} & $07^h35^m29^s \phantom{00} +69\degr11\arcmin56\arcsec$ &  WFC  &    ---       &  ---   &   ---       & N  \\
\object{GRB 000926} & $17^h04^m06^s \phantom{00} +51\degr47\arcmin37\arcsec$ & (IPN) &  48.9-61     & 19.6    &   ---       & Y (z=2.066)\\
\object{GRB 001011} & $18^h23^m04^s \phantom{00} +50\degr53\arcmin56\arcsec$ &  WFC  &    ---       &  ---   &   ---       & N  \\
\object{GRB 001109} & $18^h30^m08^s \phantom{00} +55\degr18\arcmin14\arcsec$ &  NFI  & 16-37.8      & 33.2   &   70-106    & N  \\
\object{GRB 010213} & $17^h09^m22^s \phantom{00} +39\degr15\arcmin36\arcsec$ &  WFC  &    ---       &  ---   &   ---       & no study \\
\object{GRB 010214} & $17^h40^m56^s \phantom{00} +48\degr34\arcmin52\arcsec$ &  NFI  & 6-51.8       & 83   &     ---     & N \\
\object{GRB 010220} & $02^h36^m59^s \phantom{00} +61\degr45\arcmin57\arcsec$ &  WFC  & 15-36        & 17.2   &    ---      & N \\
\object{GRB 010222} & $14^h52^m12^s \phantom{00} +43\degr01\arcmin00\arcsec$ &  NFI  & 8-64         & 88.3     &     ---     & Y (z=1.48)\\
\object{GRB 010304} & $21^h06^m22^s \phantom{00} +53\degr12\arcmin36\arcsec$ &  WFC  &    ---       &   ---  &   ---       & no study\\
\object{GRB 010412} & $19^h39^m39^s \phantom{00} +13\degr37\arcmin05\arcsec$ &  WFC  &    ---       &   ---  &   ---       & N\\
\object{GRB 010501} & $19^h06^m50^s \phantom{00} -70\degr10\arcmin48\arcsec$ &  WFC  &    ---       &   ---  &     ---     & no study\\
\object{GRB 010518} & $10^h46^m43^s \phantom{00} -57\degr47\arcmin37\arcsec$ &  WFC  &    ---       &   ---  &     ---     & no study\\
\object{GRB 011121} & $11^h34^m29^s \phantom{00} -76\degr01\arcmin52\arcsec$ &  NFI  &  21.9-65     &  32.5  & 86.7-120    & Y (z=0.36)\\
\object{GRB 011211} & $11^h15^m16^s \phantom{00} -21\degr55\arcmin44\arcsec$ &  WFC  &    ---       &   ---  &     ---     & Y (z=2.14)\\
\object{GRB 020321} & $16^h13^m05^s \phantom{00} -83\degr42\arcmin35\arcsec$ &  WFC  &  6-10.8      &  6.1   &     ---     & N \\
\object{GRB 020322} & $18^h00^m58^s \phantom{00} +81\degr06\arcmin41\arcsec$ &  NFI  &  6-12.4      &  12.3  &  26.8-33.2     & Y \\
\object{GRB 020409} & $08^h45^m14^s \phantom{00} +66\degr41\arcmin16\arcsec$ &  WFC  &    ---       &   ---  &     ---     & N \\
\object{GRB 020410} & $22^h06^m27^s \phantom{00} -83\degr49\arcmin28\arcsec$ &  NFI  &  20-27.5     &  22.8  &  54.3-59.6  & Y \\
\object{GRB 020427} & $22^h09^m21^s \phantom{00} -65\degr19\arcmin42\arcsec$ &  NFI  &  11-14.3     &  6.8   &  60.2-66    & N \\
\hline
\end{longtable}

$a$ First TOO.

\clearpage
\begin{longtable}{lccccc}
\caption{\label{table2}Properties of the prompt emission of BeppoSAX
Gamma Ray Bursts reported in Table \ref{table1}. We indicate the
duration and fluence both in X-ray (2.0-10.0 keV band) and
$\gamma$-ray (40.0-700) keV band. A X following the source name
denotes an X-ray rich GRB or an X-ray flash}\\
 \hline\hline
GRB name & $\gamma$-ray    & X-ray  & $\gamma$-ray & X-ray    & Ref.\\
         & duration &  duration     &  fluence     &  fluence &  \\
         & (T, s) & (T, s) & $10^{-7}$ erg  cm$^{-2}$ & $10^{-7}$ erg  cm$^{-2}$\\
\hline
\endfirsthead
\caption{continued.}\\
\hline\hline
GRB name & $\gamma$-ray    & X-ray  & $\gamma$-ray & X-ray    & Ref.\\
         & duration &  duration     &  fluence     &  fluence &  \\
         & (T, s) & (T, s) & $10^{-7}$ erg  cm$^{-2}$ & $10^{-7}$ erg  cm$^{-2}$\\
\hline
\endhead
\hline
\endfoot
\object{GRB 960720} & 8    & 17   &    $26\pm3$           &    $0.8\pm0.2$          &    1, 2, 3    \\
\object{GRB 970111} & 43   & 60   &  $430\pm30$           &  $16\pm1$               &    4, 2, 3     \\
\object{GRB 970228} & 80   & 80   &  $64.5$               &  $15.4$                 &    5      \\
\object{GRB 970402} & 150  & 150  &  $82\pm9$             &  $4.7\pm1.5$            &    2     \\
\object{GRB 970508} & 15   & 29   &  $14.5$               &  $5.3$                  &    5      \\
\object{GRB 971214} & 35   & 35   &  $64.9$               &  $2.34$                 &    5,3      \\
\object{GRB 971227} & 7    & 7    &  $6.6\pm0.7$          &  1                      &    6,3     \\
\object{GRB 980109} & 20   & 20   &  $32.3\pm3$           &     ---                 &    3,7     \\
\object{GRB 980326} & 9    & 9    &  $7.5\pm1.5$          &  $2\pm0.3$              &    5, 3      \\
\object{GRB 980329} & 58   & 68   &  $650\pm50$           &  $9.7\pm0.7$            &    5,      \\
\object{GRB 980425} & 31   & 40   &  $28.5\pm5$           &  $7.8\pm0.2$            &    2, 3     \\
\object{GRB 980515} & 15   & 20   &  $23\pm3$             &  -                      &    7, 3     \\
\object{GRB 980519} & 30   & 190  &  $81\pm5$             &  18                     &    8,9, 3  \\
\object{GRB 980613} & 50   & 50   &  $9.9$                &    $2.3$                &    5, 3      \\
\object{GRB 981226X}& 20   & 260  &  $4\pm1$              &  $5.7\pm1$              &    10,3       \\
\object{GRB 980703} & 90   & ---  & $300\pm100$           &      ---                &    11     \\
\object{GRB 990123} & 100  & 100  & $1790$                &     $22.9$              &    5, 3      \\
\object{GRB 990217} & 25   & 25   & $12.7\pm1.5$          &      ---                &    7, 3     \\
\object{GRB 990510} & 75   & 80   & $181$                 &  $17.9$                 &    3, 5       \\
\object{GRB 990625} & 11   & 11   &     ---               &     ---                 &    3        \\
\object{GRB 990627} & 28   & 60   &     ---               &   $\sim15$              &    3,12       \\
\object{GRB 990704X}& 23   & 40   & $10\pm1$              & $15\pm0.8$              &    13, 3     \\
\object{GRB 990705} & 42   & 45   & $423$                 & $22.5$                  &    5, 3      \\
\object{GRB 990712} & 30   & 30   & $65\pm3$              & $28.6$                  &    5, 3      \\
\object{GRB 990806} & 30   & 30   & $\sim42$              & $\sim2.5$               &    14, 3     \\
\object{GRB 990907} & 1    & 220  &     ---               &     ---                 &    3        \\
\object{GRB 990908} & 50   & 130  &     ---               &     ---                 &    3       \\
\object{GRB 991014} & 3    & 10   & $9\pm1$               & 1                       &    15,16, 3    \\
\object{GRB 991105} & 13   & 40   &     ---               &      ---                &    3         \\
\object{GRB 991106}$^{\mathrm a}$ & ---  & 5  & $<1.2$ $^{\mathrm b}$ &                    &  17      \\
\object{GRB 000210} & 10   & 115  & $610\pm20$            & $\sim15$                &    18, 3    \\
\object{GRB 000214} & 115  & 100  & $61.7$                & $11.6$                  &    5, 3      \\
\object{GRB 000528} & 80   & 120  & $14.4\pm0.4$          &      ---                &    19, 20    \\
\object{GRB 000529} & 14   & 30   &     ---               &     ---                 &    3         \\
\object{GRB 000615X}& 12   & 120  & $9.8\pm0.9$           & $17\pm1$                &    21, 3    \\
\object{GRB 000620} & 15   & 20   &    ---                &    ---                  &    3       \\
\object{GRB 001011} & 31   & 60   &    ---                &    ---                  &    3          \\
\object{GRB 001109} & 60   & 65   & $49.7\pm1.9$          & $6.4\pm0.33$            &    22, 3     \\
\object{GRB 010213} & 23   & 25   &    ---                &    ---                  &    3          \\
\object{GRB 010214} & 15   & 30   & $45\pm0.8$            & $2\pm0.3$               &    23    \\
\object{GRB 010220} & 40   & 150  &    ---                &   ---                   &    3      \\
\object{GRB 010222} & 170  & 280  & $753$                 & $95$                    &    5, 3       \\
\object{GRB 010304} & 15   & 24   &    ---                &   ---                   &    3     \\
\object{GRB 010501} & 37   & 41   &    ---                &   ---                   &    3      \\
\object{GRB 010412} & 74   & 90   &    ---                &   ---                   &    3         \\
\object{GRB 010518} & 25   & 30   &    ---                &   ---                   &    3         \\
\object{GRB 011121} & 105  & 100  & $1000\pm20$           & $140\pm3$               &    24, 3 \\
\object{GRB 011211} & 400  & 400  & $37\pm4$              & $11\pm1$                &    24, 3 \\
\object{GRB 020321} & 70   & 90   & 30                    & 0.9                     &    25, 3        \\
\object{GRB 020322} & 15   & 50   & ---                   & ---                     &    3         \\
\object{GRB 020409} & 40   & 60   & ---                   & ---                     &    3        \\
\object{GRB 020410} & 1800 &$>$1290& $\sim290$            & $>47$                   &    26, 3     \\
\object{GRB 020427X}&  --- & 60   &      $<2.9$           &  $3.7\pm0.3$            &    27, 3 \\
\hline

\end{longtable}

$a$ Perhaps not a GRB. See Cornelisse et al. 2000.

$b$ Conservative $3\sigma$ upper limit based on GCN 448

References :
1: \citet{piro98}, 
2: \citet{fro00a}, 
3: \citet{fro04}, 
4: \citet{fer98}, 
5: \citet{ama02a}, 
6. \citet{ant99}, 
7: \citet{ama99}, 
8 : \citet{nic98}, 
9 : \citet{zan99}, 
10: \citet{fro00b}, 
11: \citet{ama98}, 
12: \citet{mul99b}, 
13: \citet{fer01}, 
14: \citet{mon01}, 
15: \citet{tas99}, 
16: \citet{zan00b}, 
17: \citet{gan99}, 
18: \citet{piro02}, 
19: \citet{gui00}, 
20: \citet{zan00a}, 
21: \citet{nic01}, 
22: \citet{gui03}, 
23: \citet{gui03}, 
24: \citet{piro05}, 
25: \citet{zand03}, 
26: \citet{nic04}, 
27: \citet{ama04}. 

Note 1:  When not available, values of 2-10 keV fluences have been
calculated from the 2-26 keV fluences and assuming the spectral
parameters reported in the references.

Note 2: The X-ray and $\gamma$ fluences reported by \citet{ama02a}
have been obtained by reporting at z$=0$ the parameters of the WFC
and GRBM spectra fit (see table 2 of the same article).

\clearpage

\begin{figure*}
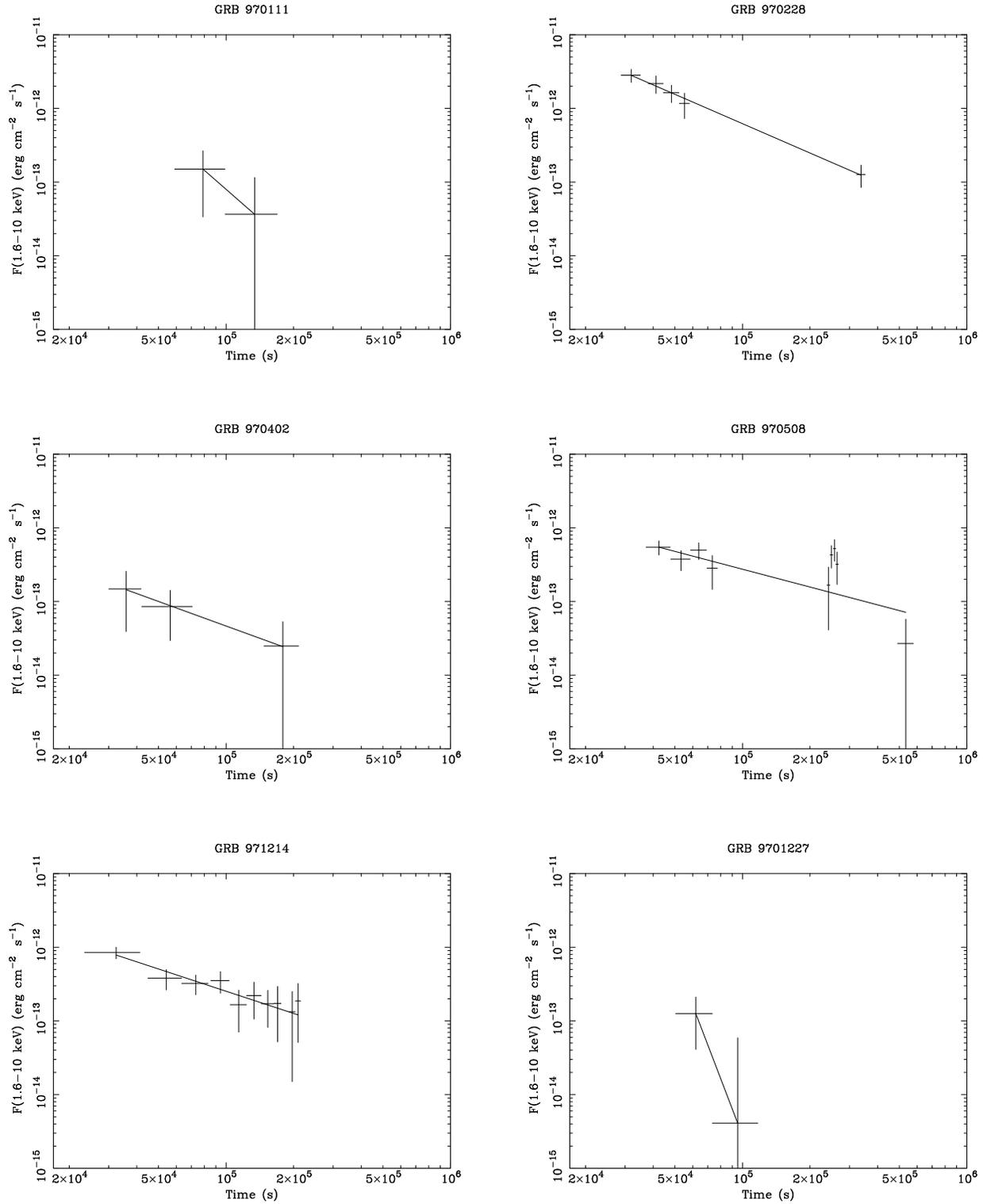

\centering
\vspace{1cm}
\includegraphics[width=6cm,angle=270]{grb970111_lc.ps}\hspace{1cm}
\includegraphics[width=6cm,angle=270]{grb970228_lc.ps}\vspace{1cm}

\includegraphics[width=6cm,angle=270]{grb970402_lc.ps}\hspace{1cm}
\includegraphics[width=6cm,angle=270]{grb970508_lc.ps}\vspace{1cm}

\includegraphics[width=6cm,angle=270]{grb971214_lc.ps}\hspace{1cm}
\includegraphics[width=6cm,angle=270]{grb971227_lc.ps}\vspace{1cm}

\caption{\label{fig1}X-ray lightcurves of the afterglows observed by
beppoSAX in the 1.6-10 keV band.}
\end{figure*}

\setcounter{figure}{11}
\begin{figure*}
\centering
\vspace{1cm}
\includegraphics[width=6cm,angle=270]{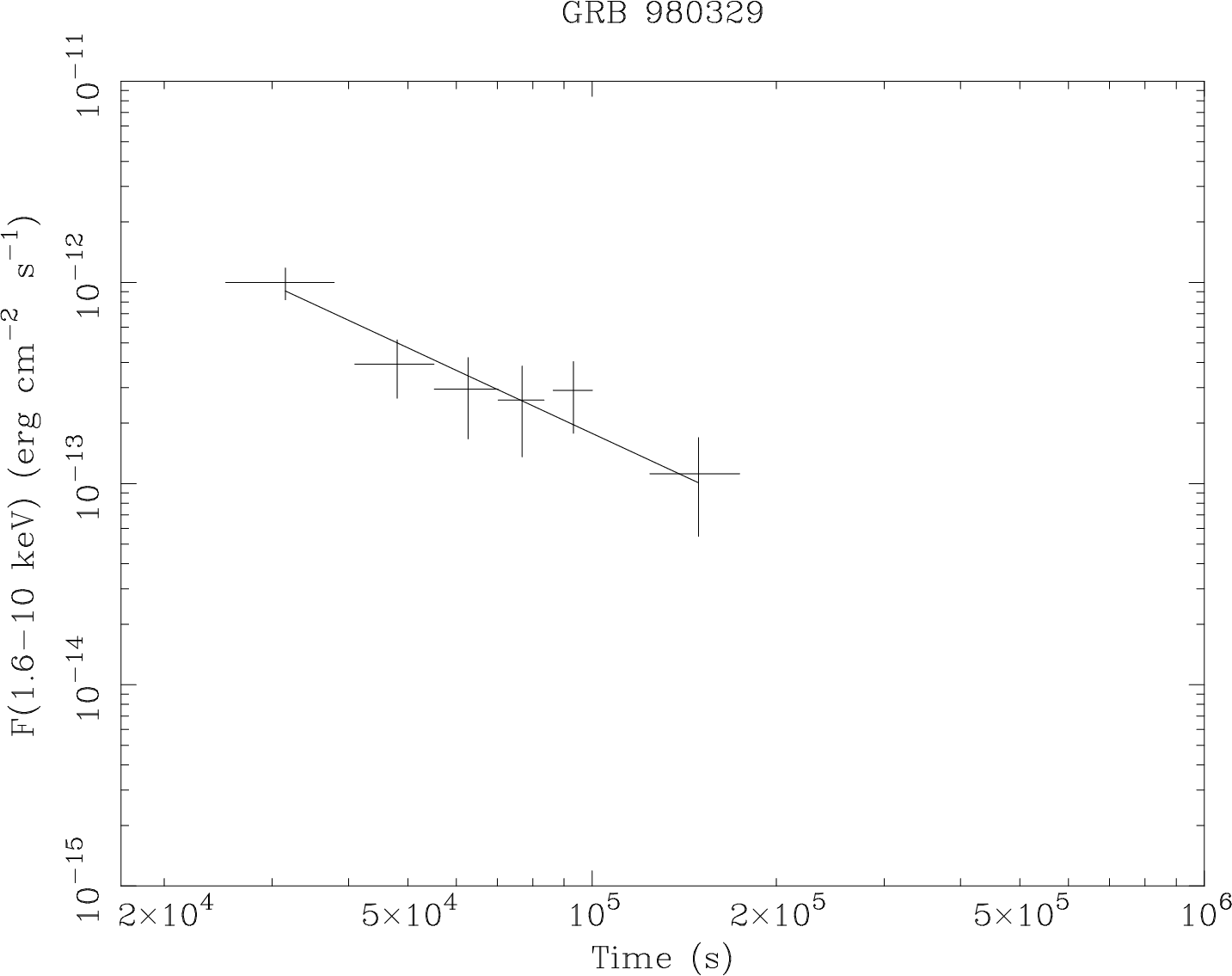}\hspace{1cm}
\includegraphics[width=6cm,angle=270]{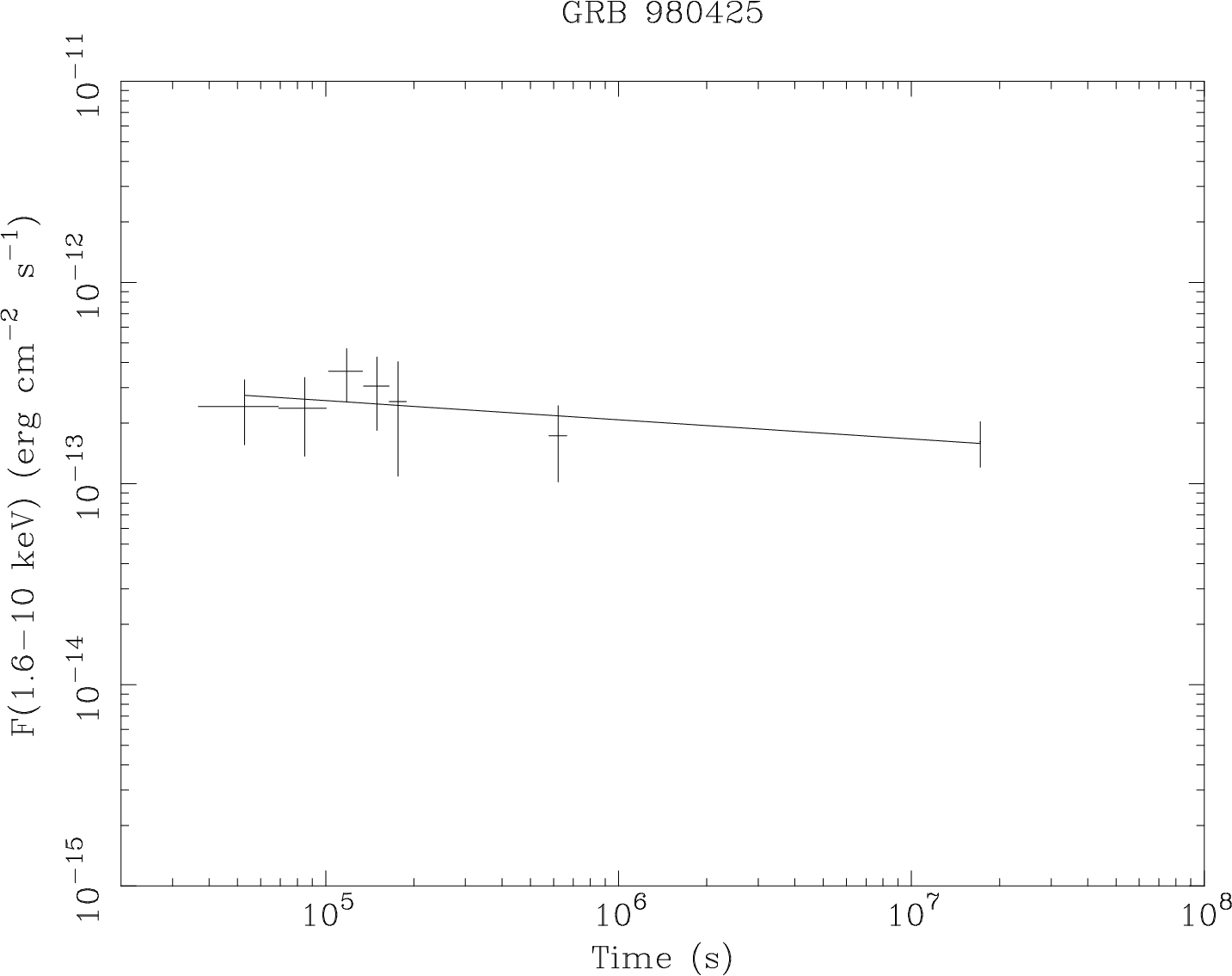}\vspace{1cm}

\includegraphics[width=6cm,angle=270]{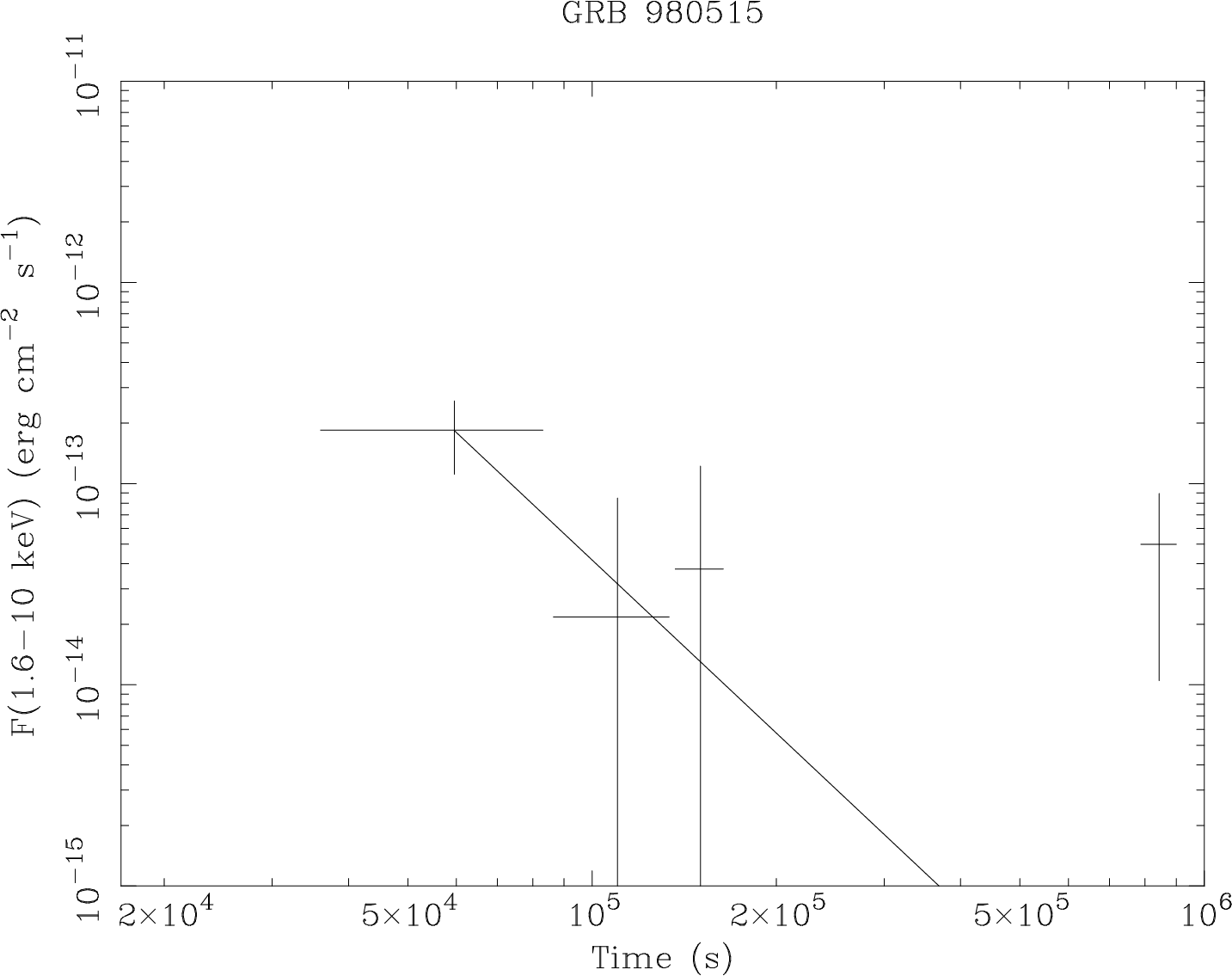}\hspace{1cm}
\includegraphics[width=6cm,angle=270]{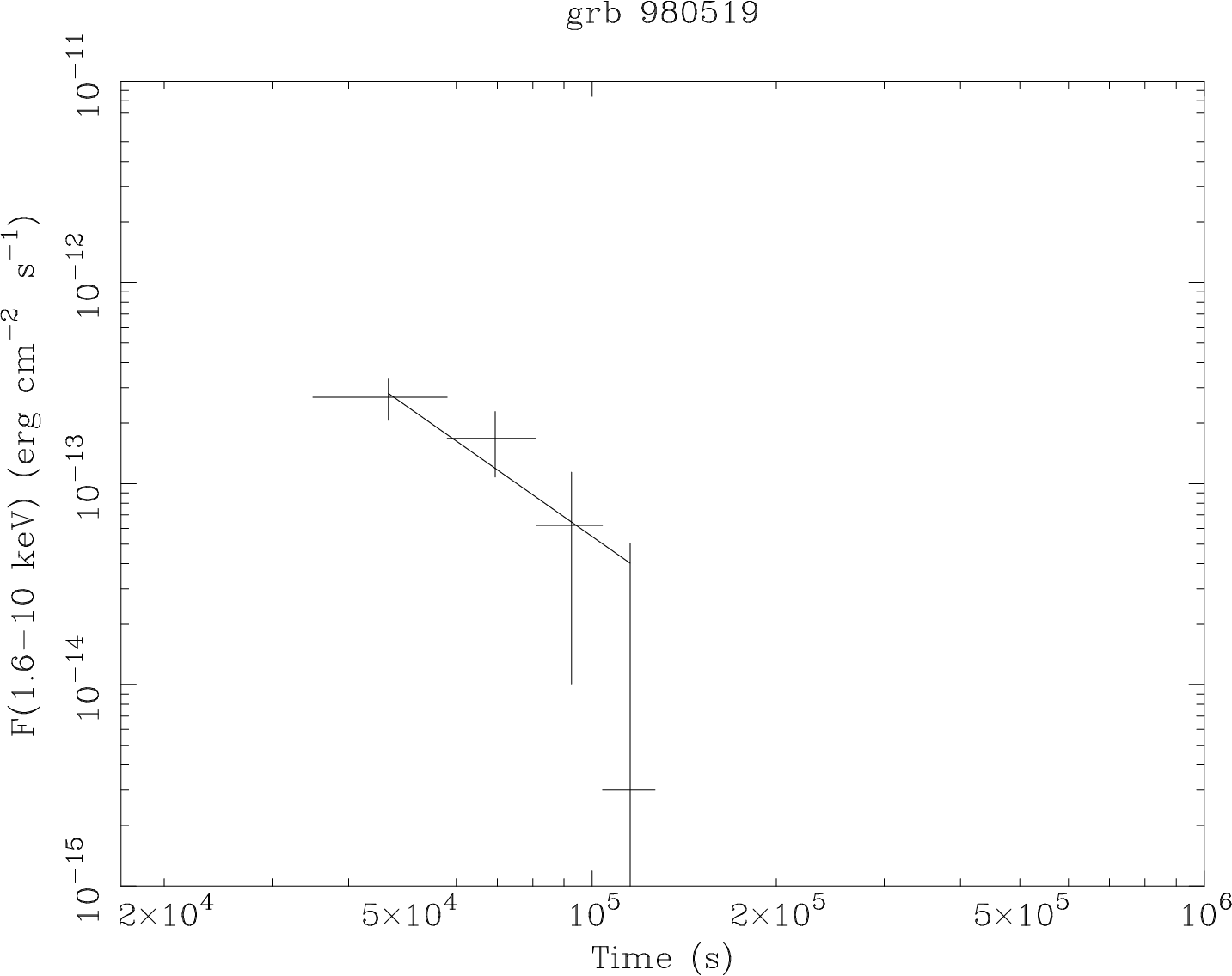}\vspace{1cm}

\includegraphics[width=6cm,angle=270]{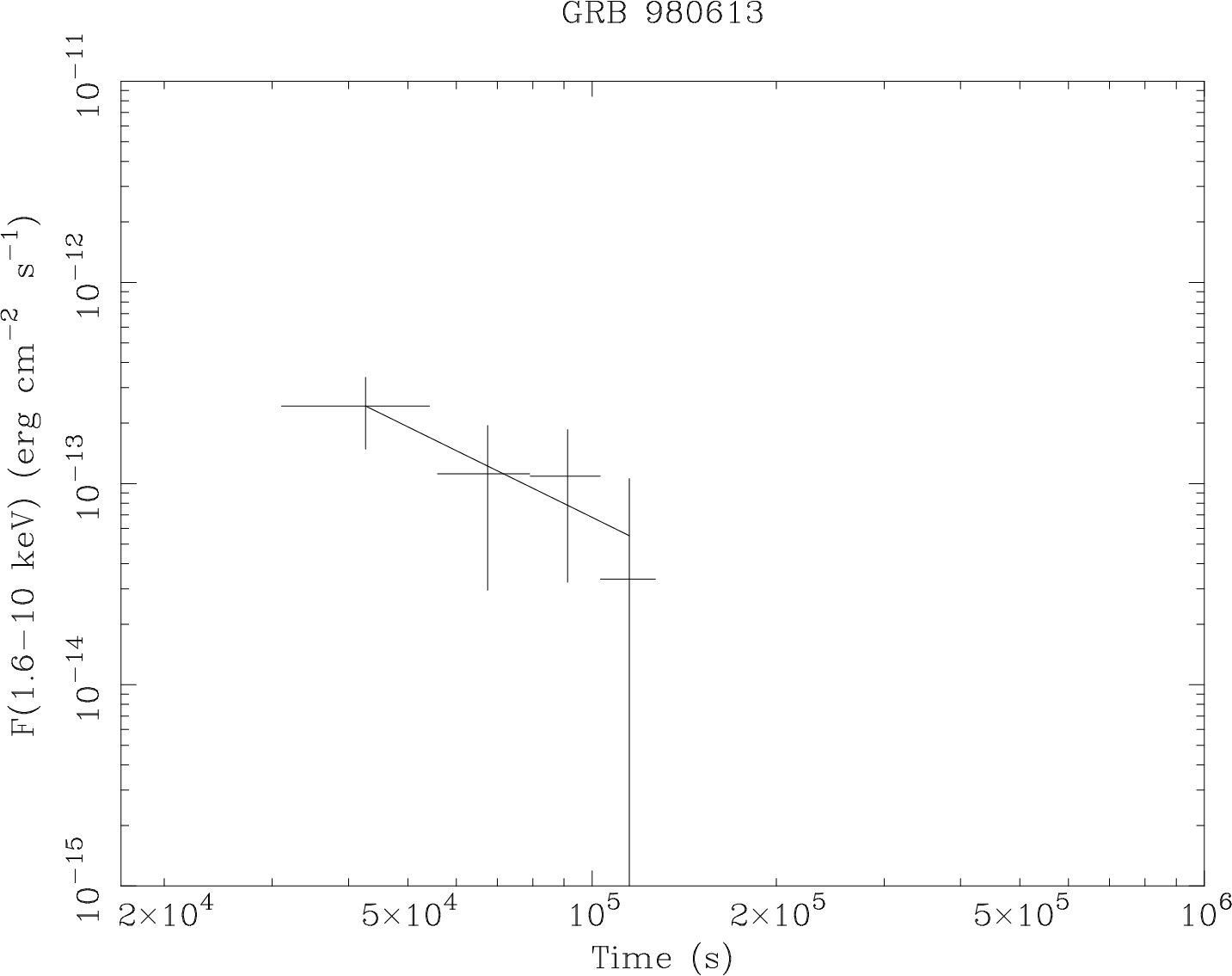}\hspace{1cm}
\includegraphics[width=6cm,angle=270]{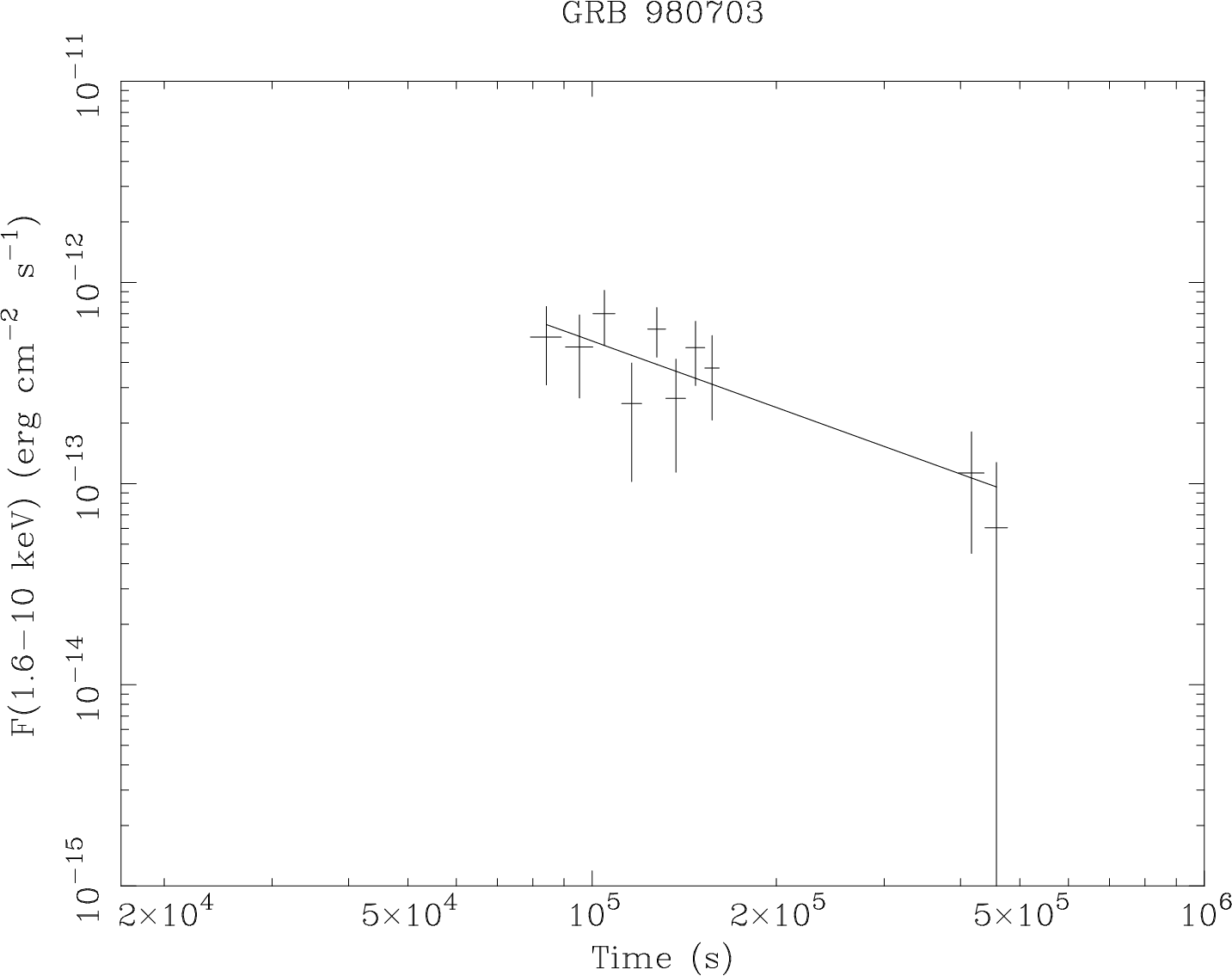}\vspace{1cm}
\caption{Continued.}
\end{figure*}

\setcounter{figure}{11}
\begin{figure*}
\centering
\includegraphics[width=6cm,angle=270]{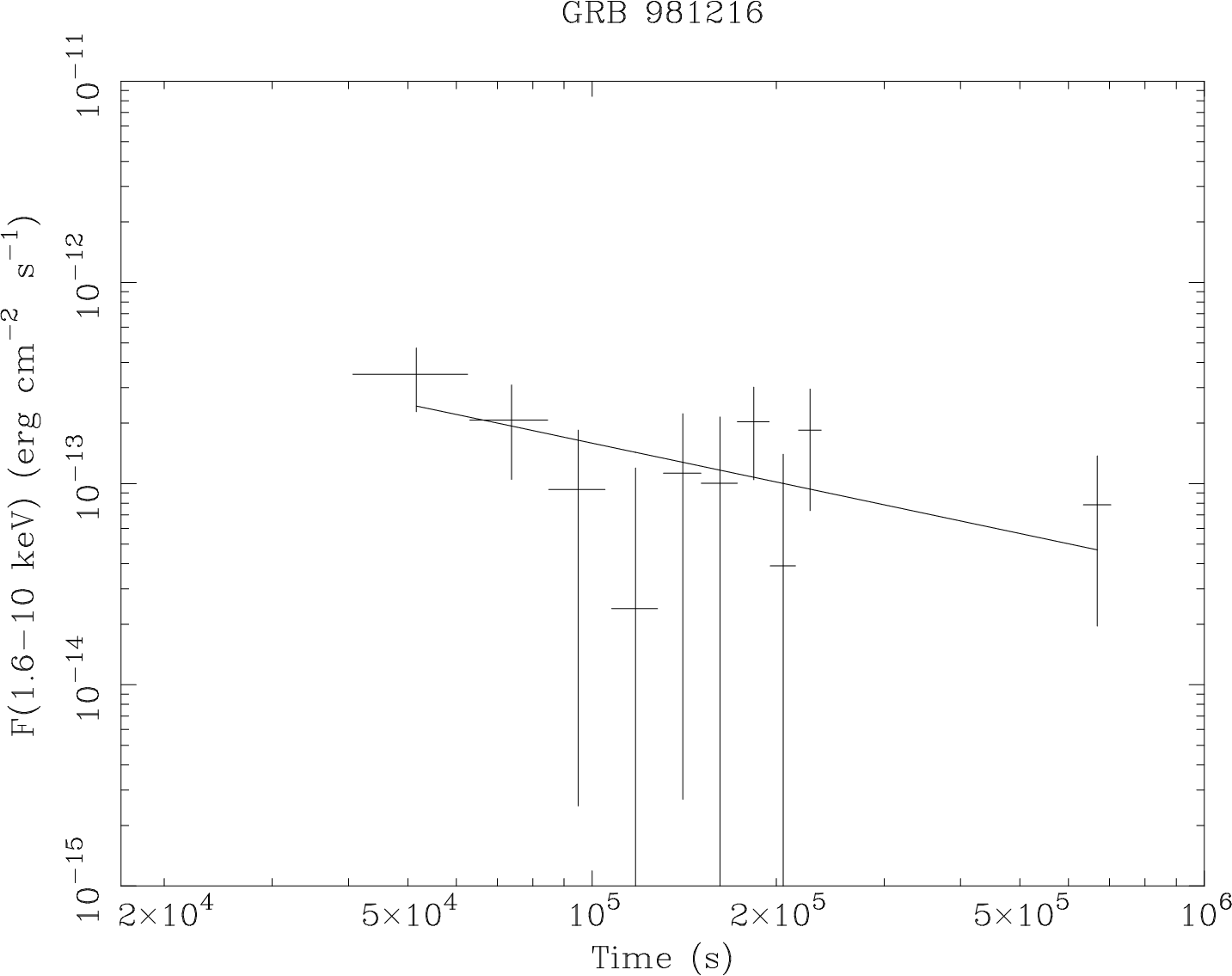}\hspace{1cm}
\includegraphics[width=6cm,angle=270]{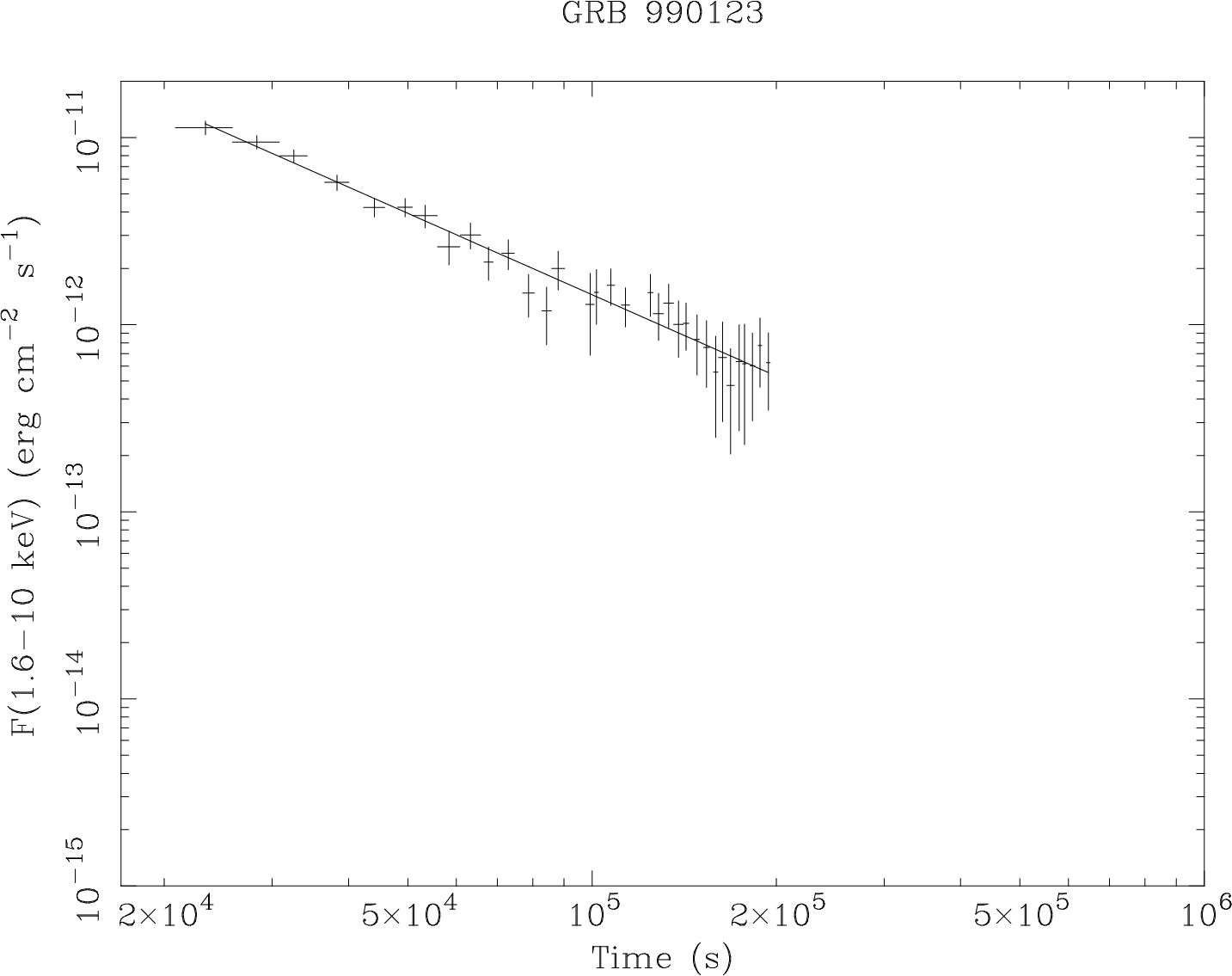}\vspace{1cm}

\includegraphics[width=6cm,angle=270]{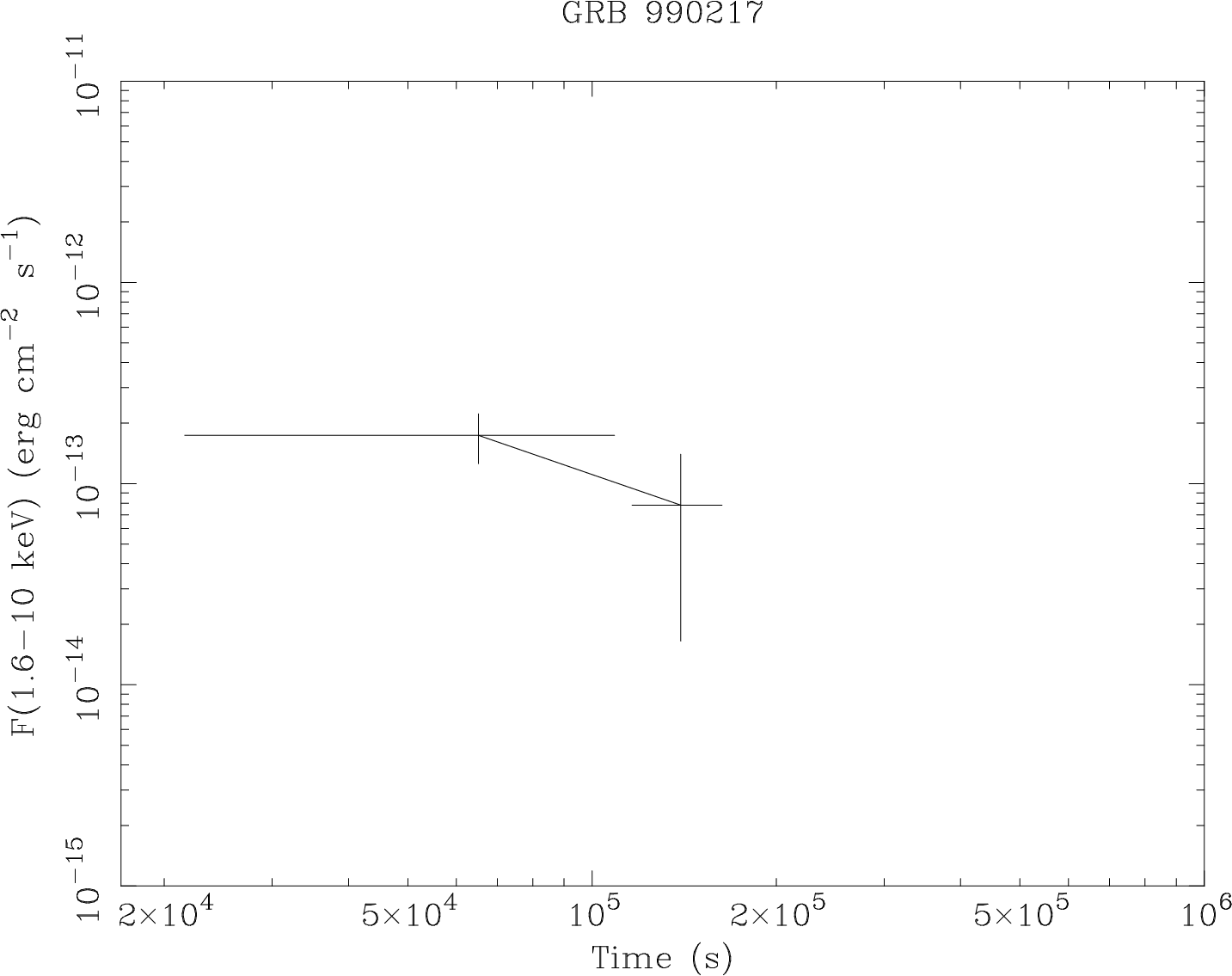}\hspace{1cm}
\includegraphics[width=6cm,angle=270]{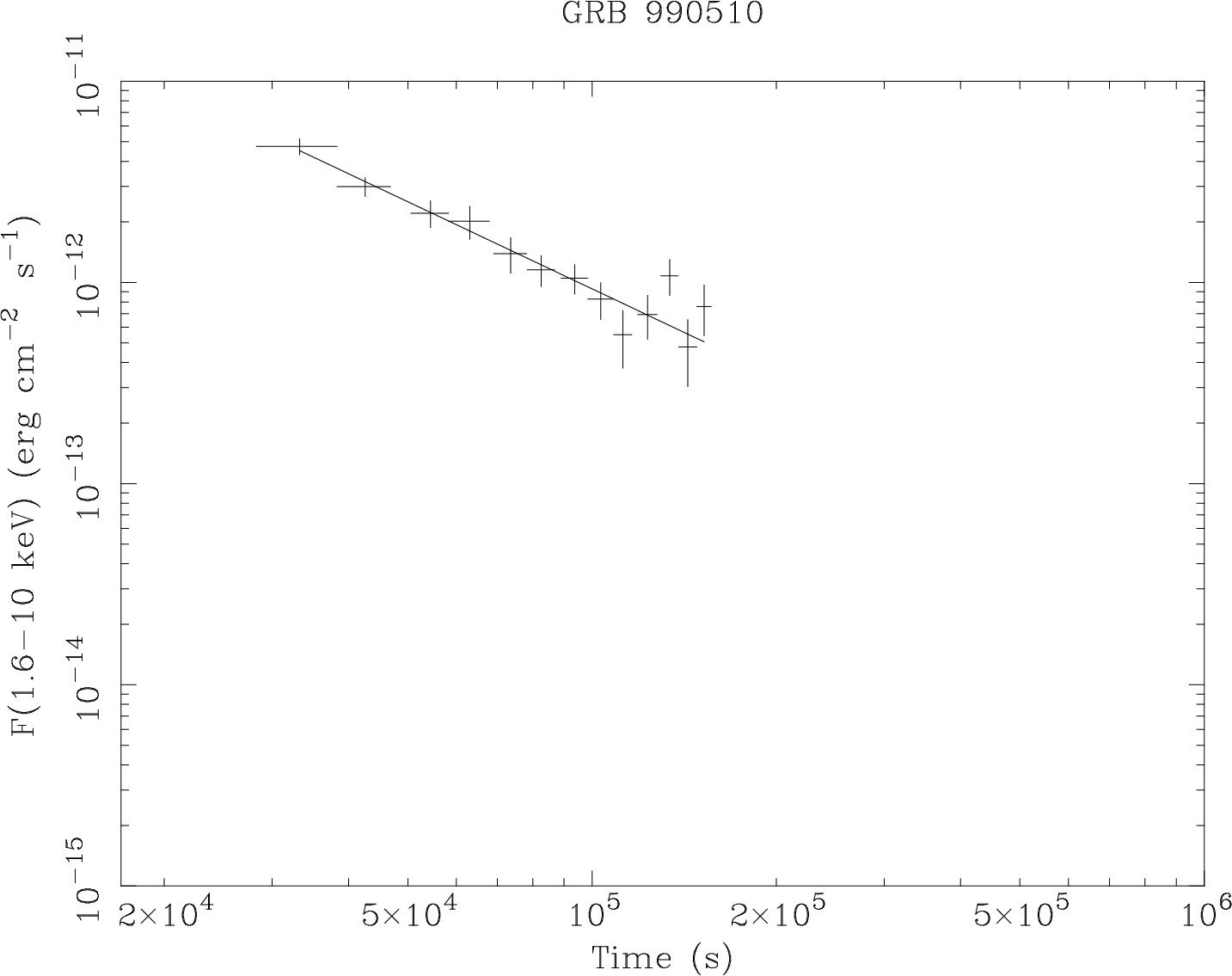}\vspace{1cm}

\includegraphics[width=6cm,angle=270]{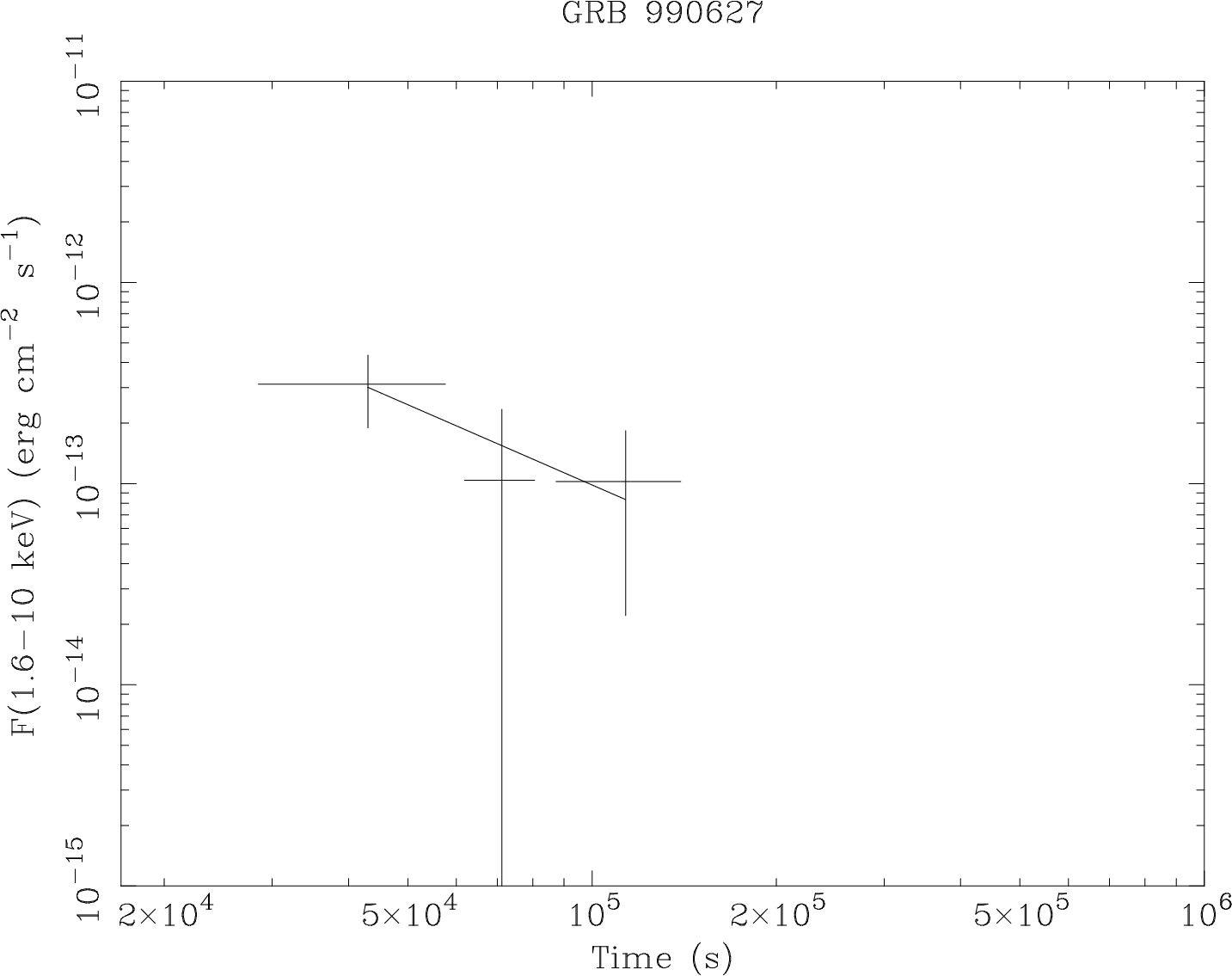}\hspace{1cm}
\includegraphics[width=6cm,angle=270]{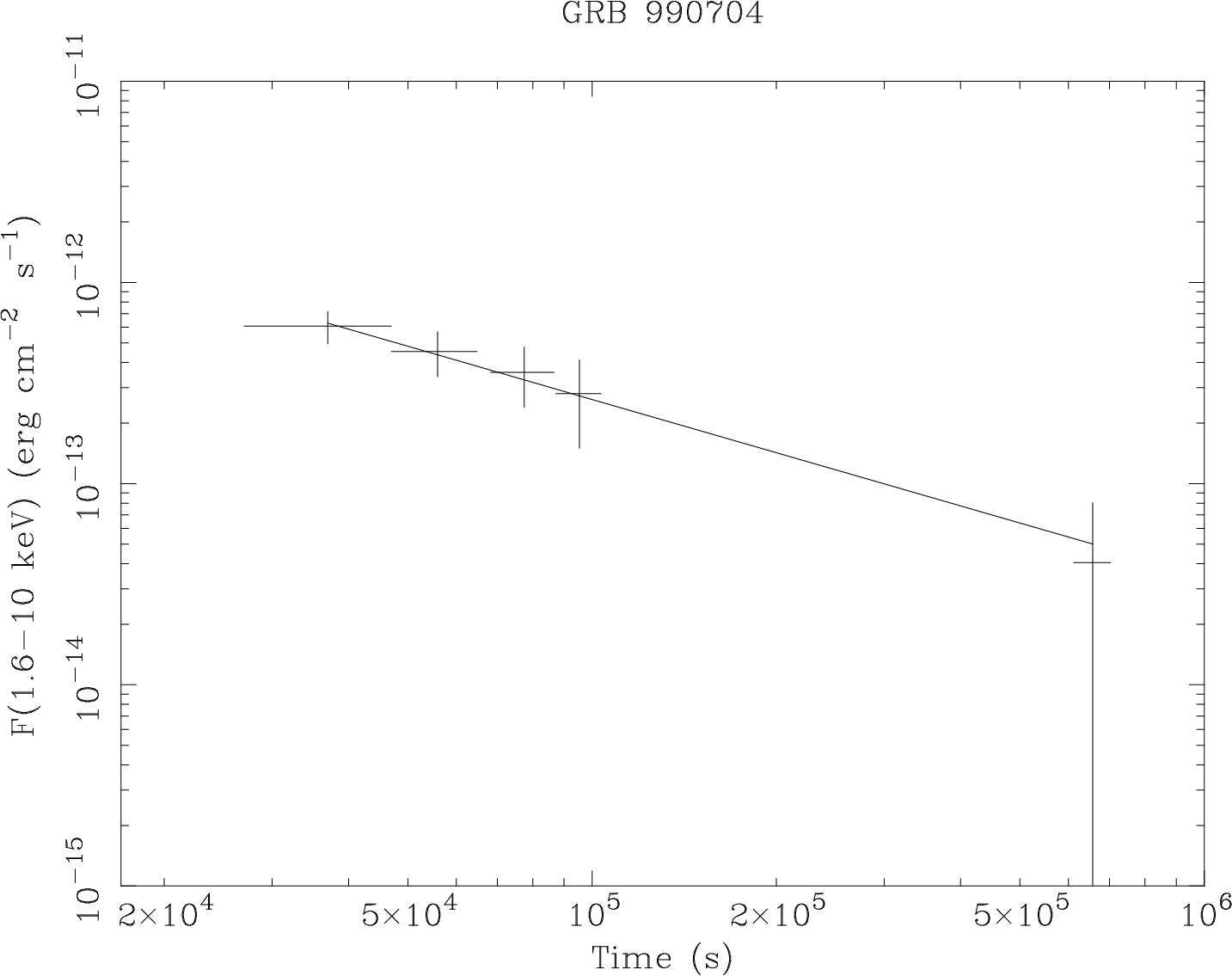}\vspace{1cm}
\caption{Continued.}
\end{figure*}

\setcounter{figure}{11}
\begin{figure*}
\centering
\vspace{1cm}
\includegraphics[width=6cm,angle=270]{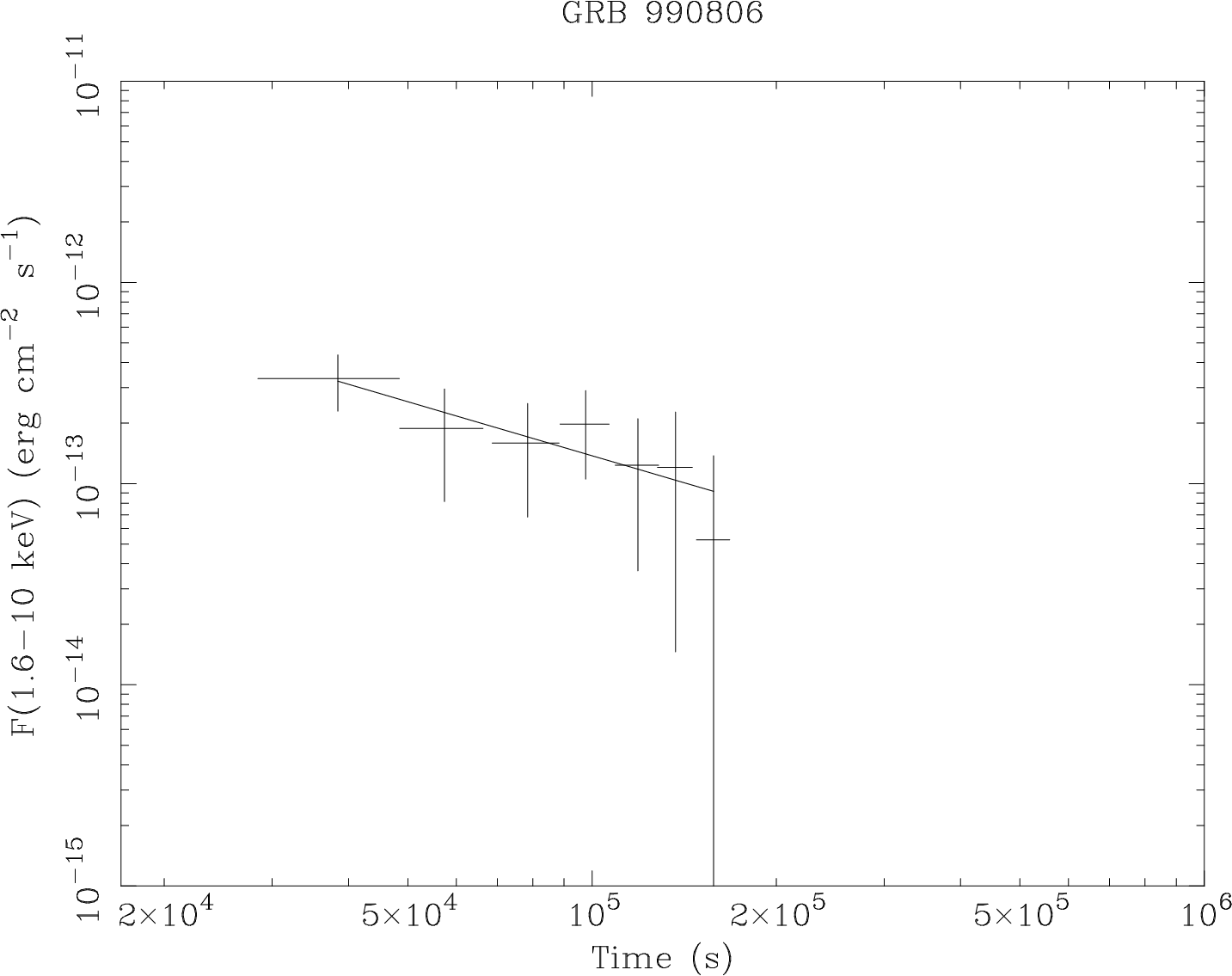}\hspace{1cm}
\includegraphics[width=6cm,angle=270]{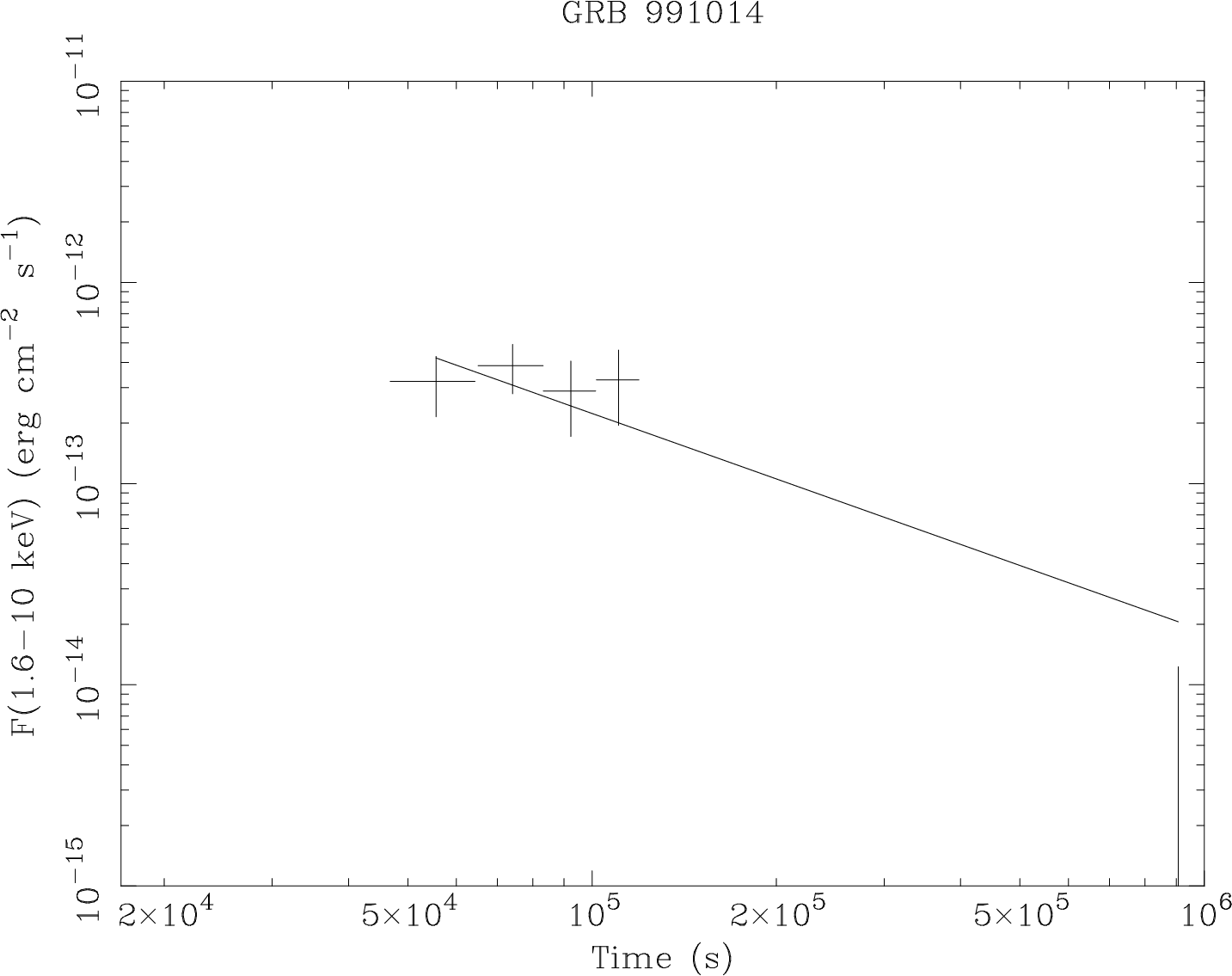}\vspace{1cm}

\includegraphics[width=6cm,angle=270]{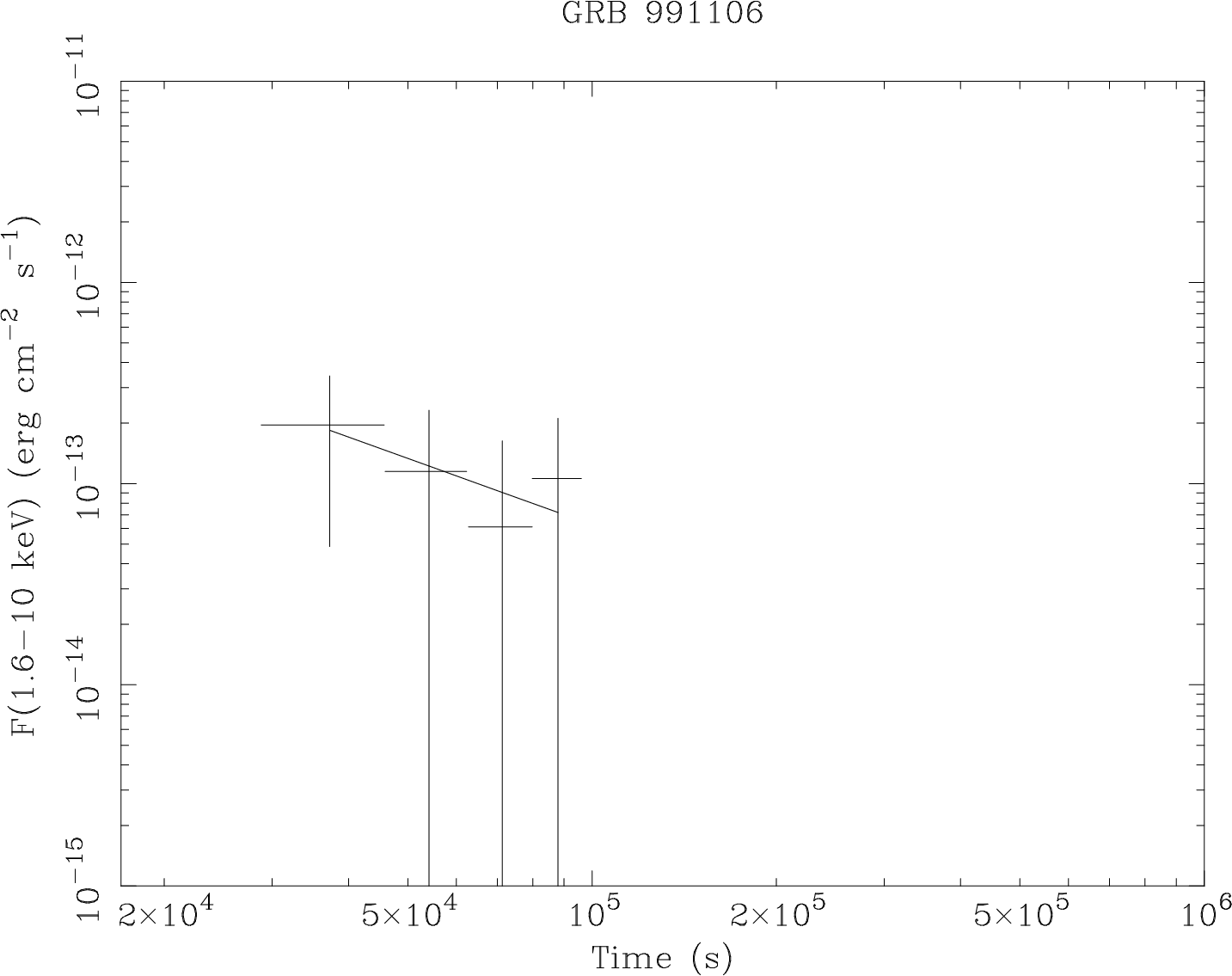}\hspace{1cm}
\includegraphics[width=6cm,angle=270]{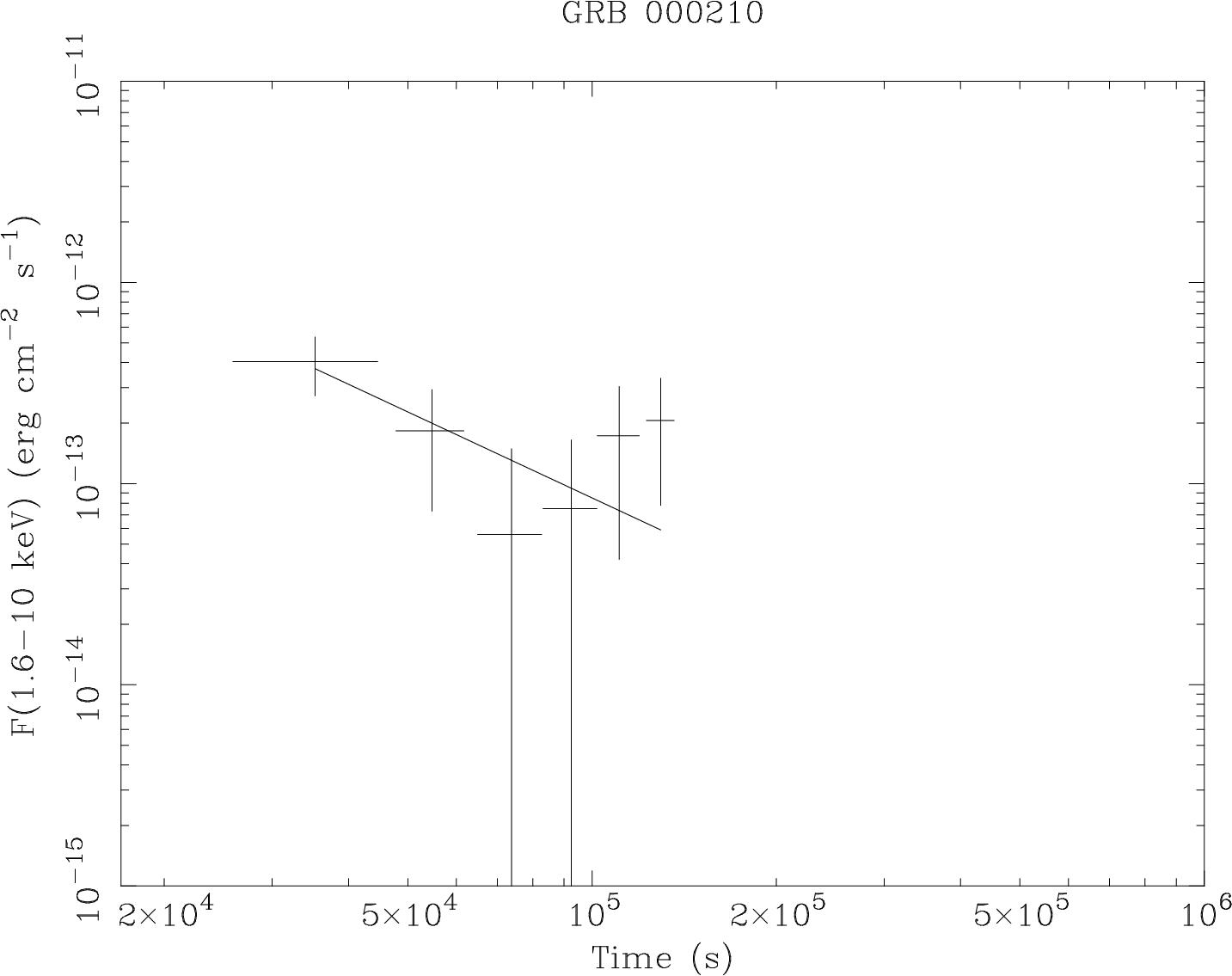}\vspace{1cm}

\includegraphics[width=6cm,angle=270]{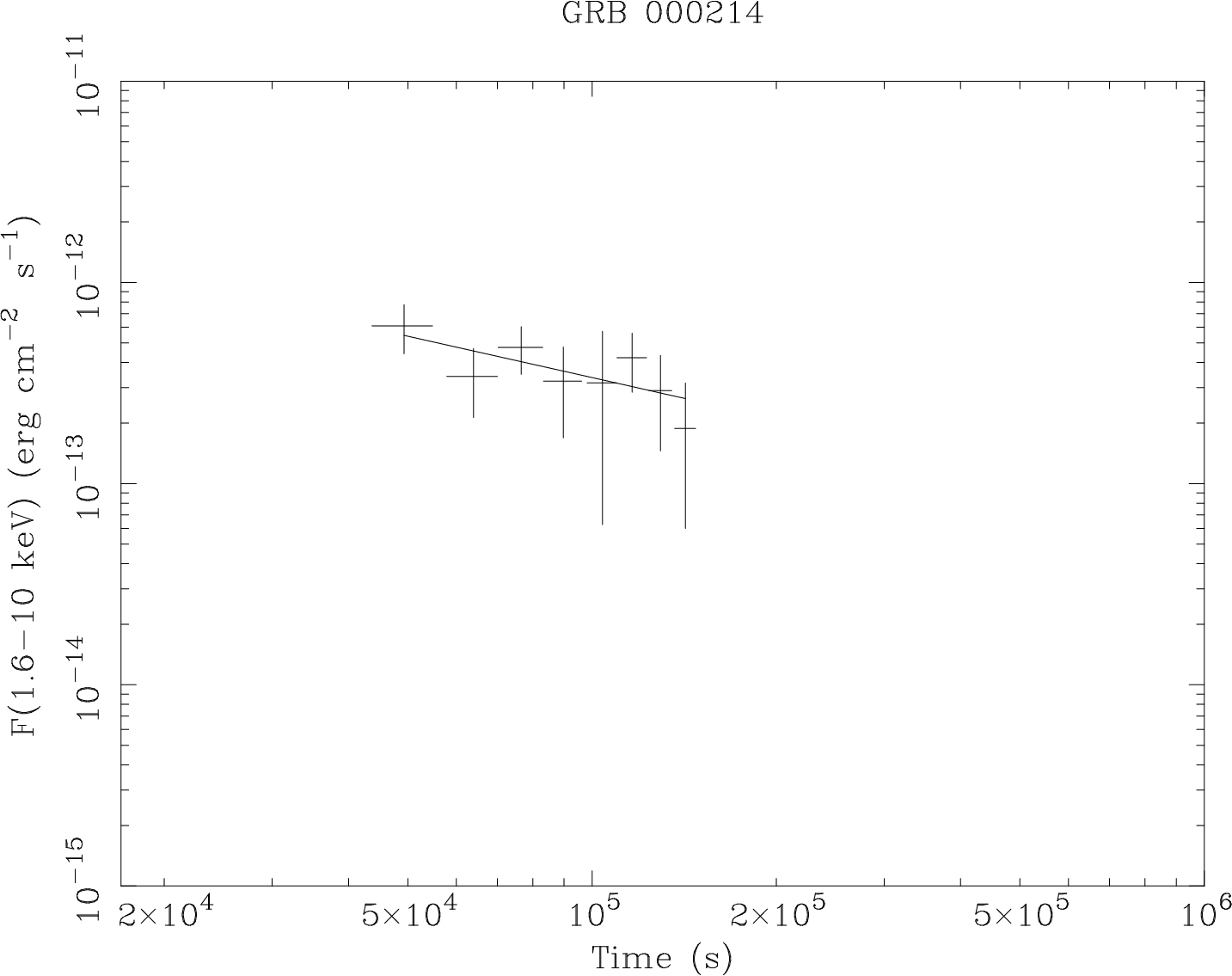}\hspace{1cm}
\includegraphics[width=6cm,angle=270]{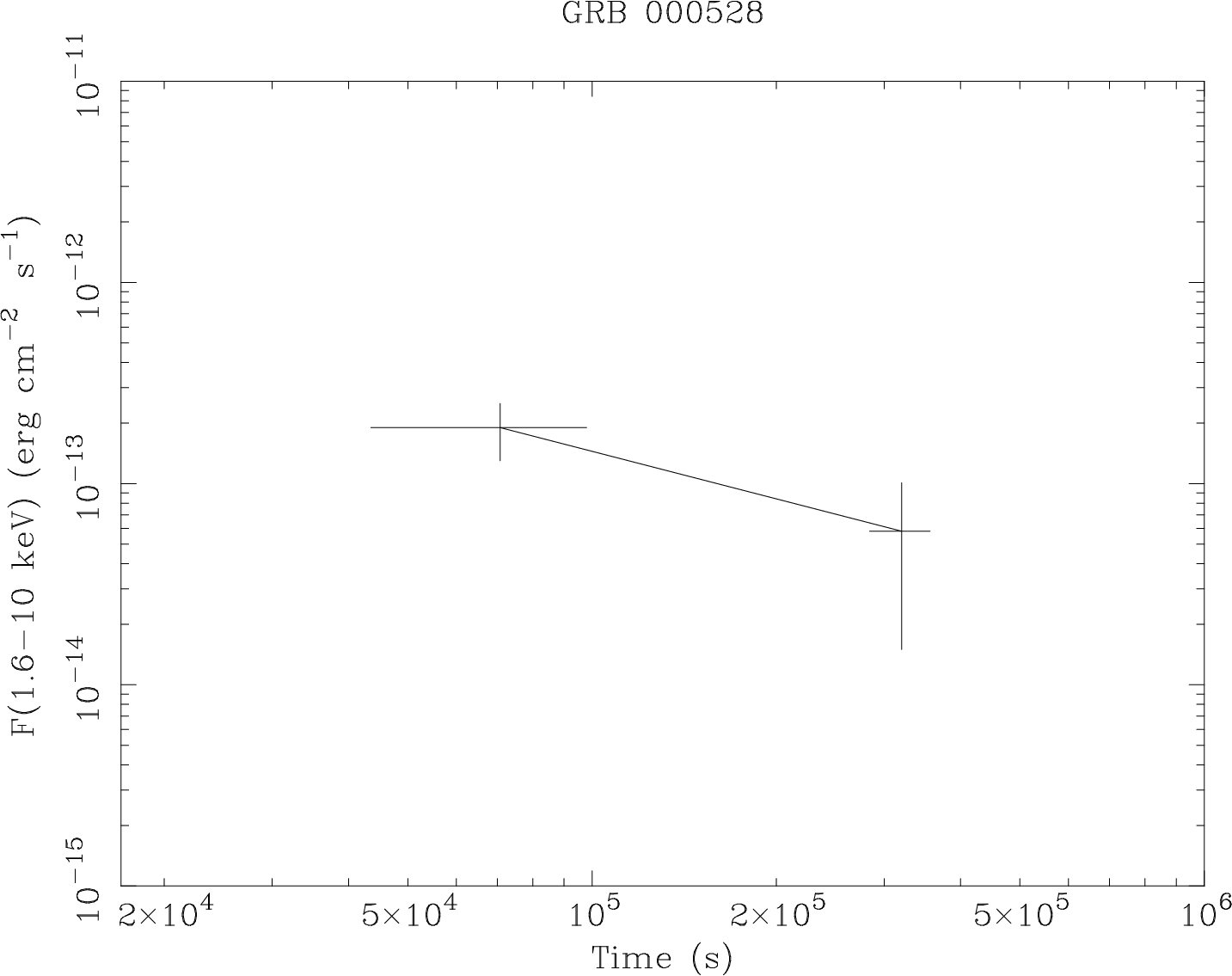}\vspace{1cm}
\caption{Continued.}
\end{figure*}

\setcounter{figure}{11}
\begin{figure*}
\centering
\vspace{1cm}
\includegraphics[width=6cm,angle=270]{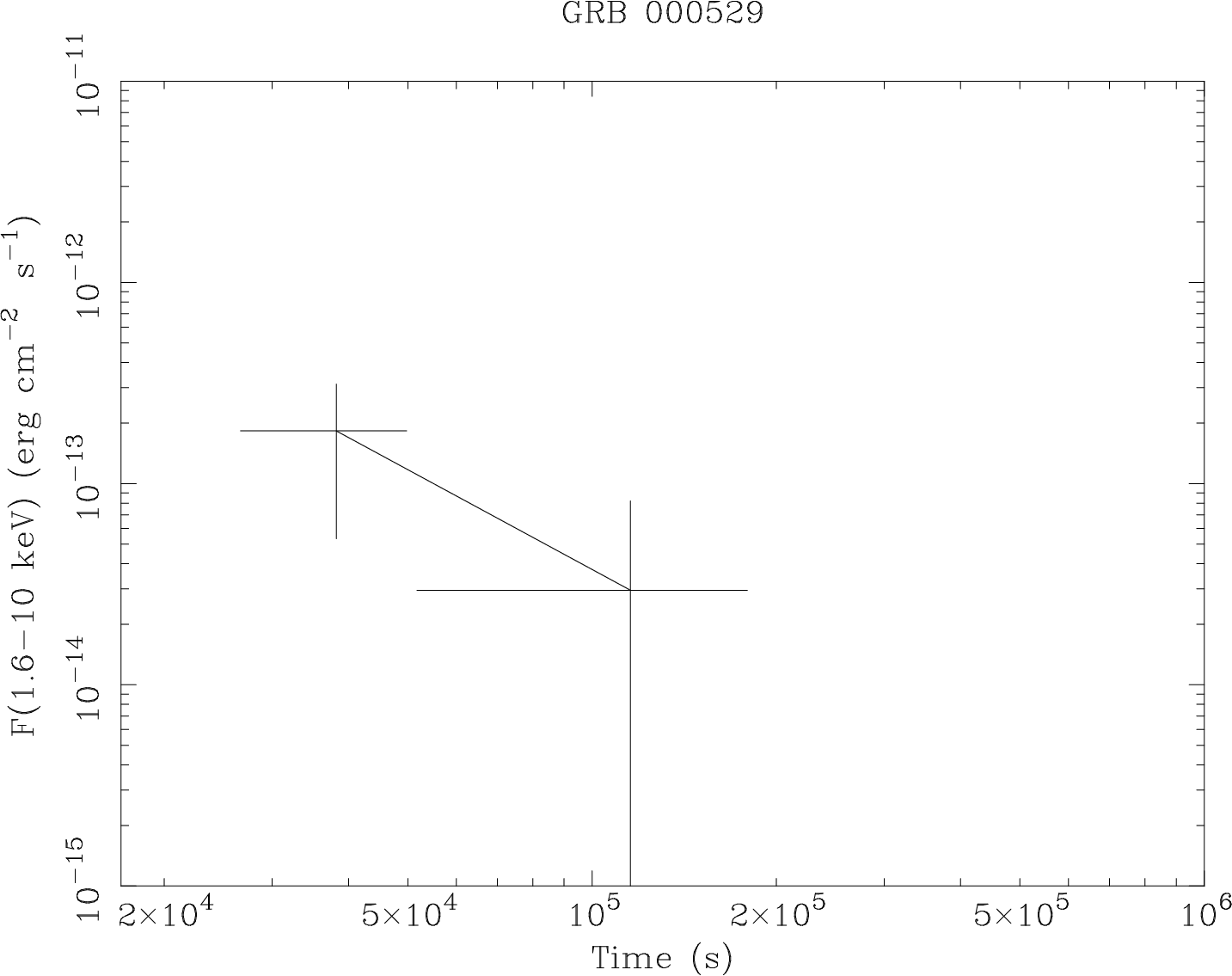}\hspace{1cm}
\includegraphics[width=6cm,angle=270]{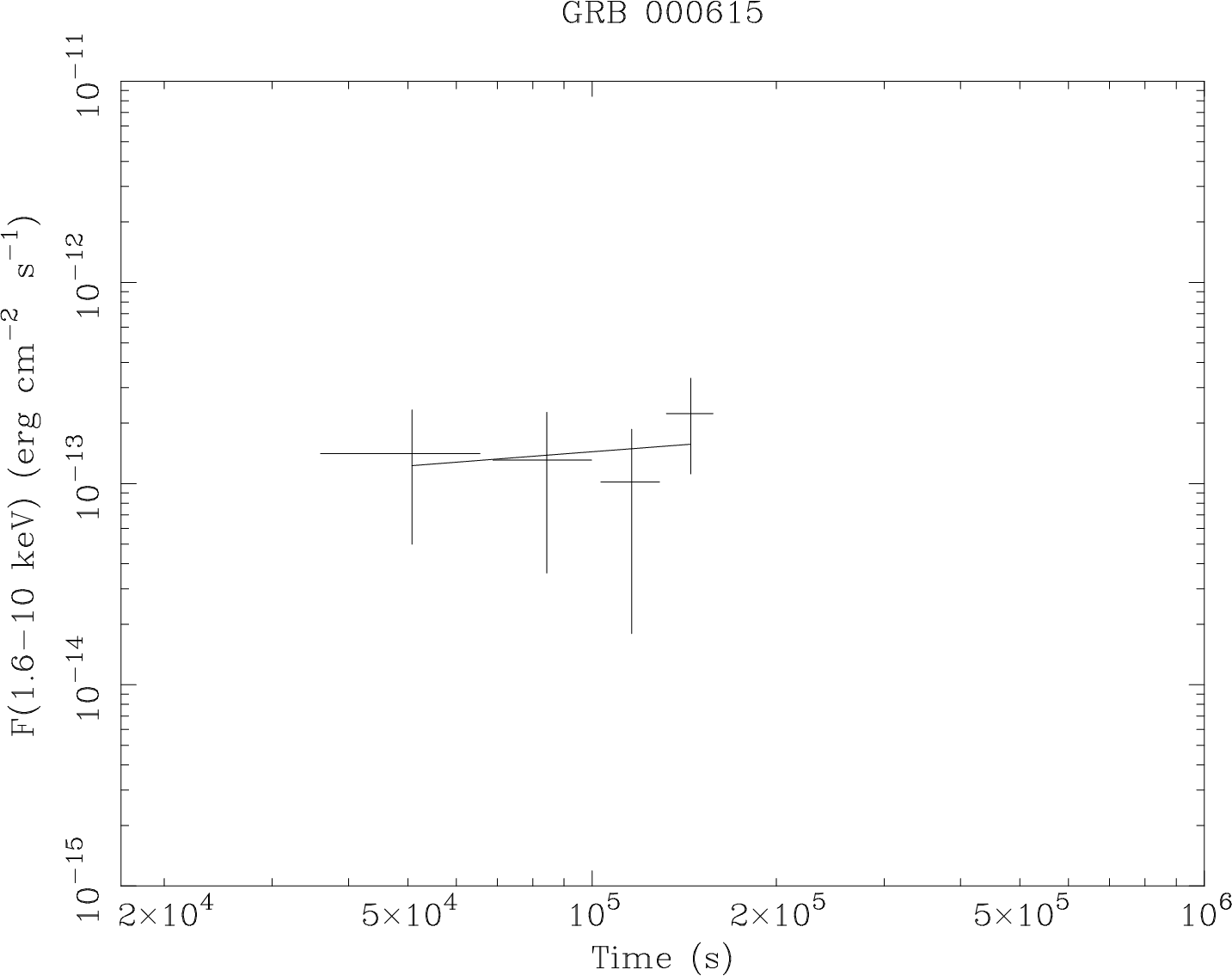}\vspace{1cm}

\includegraphics[width=6cm,angle=270]{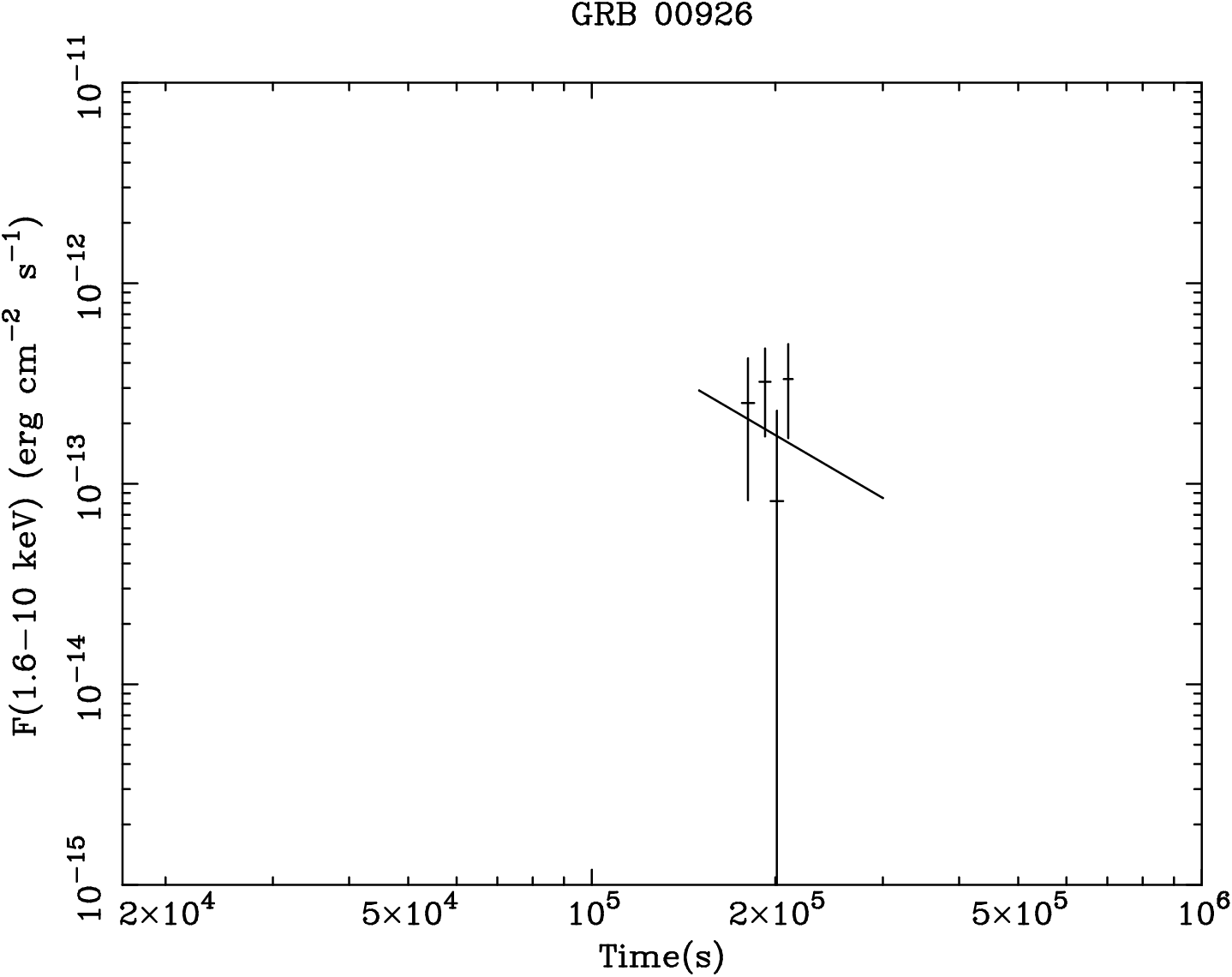}\hspace{1cm}
\includegraphics[width=6cm,angle=270]{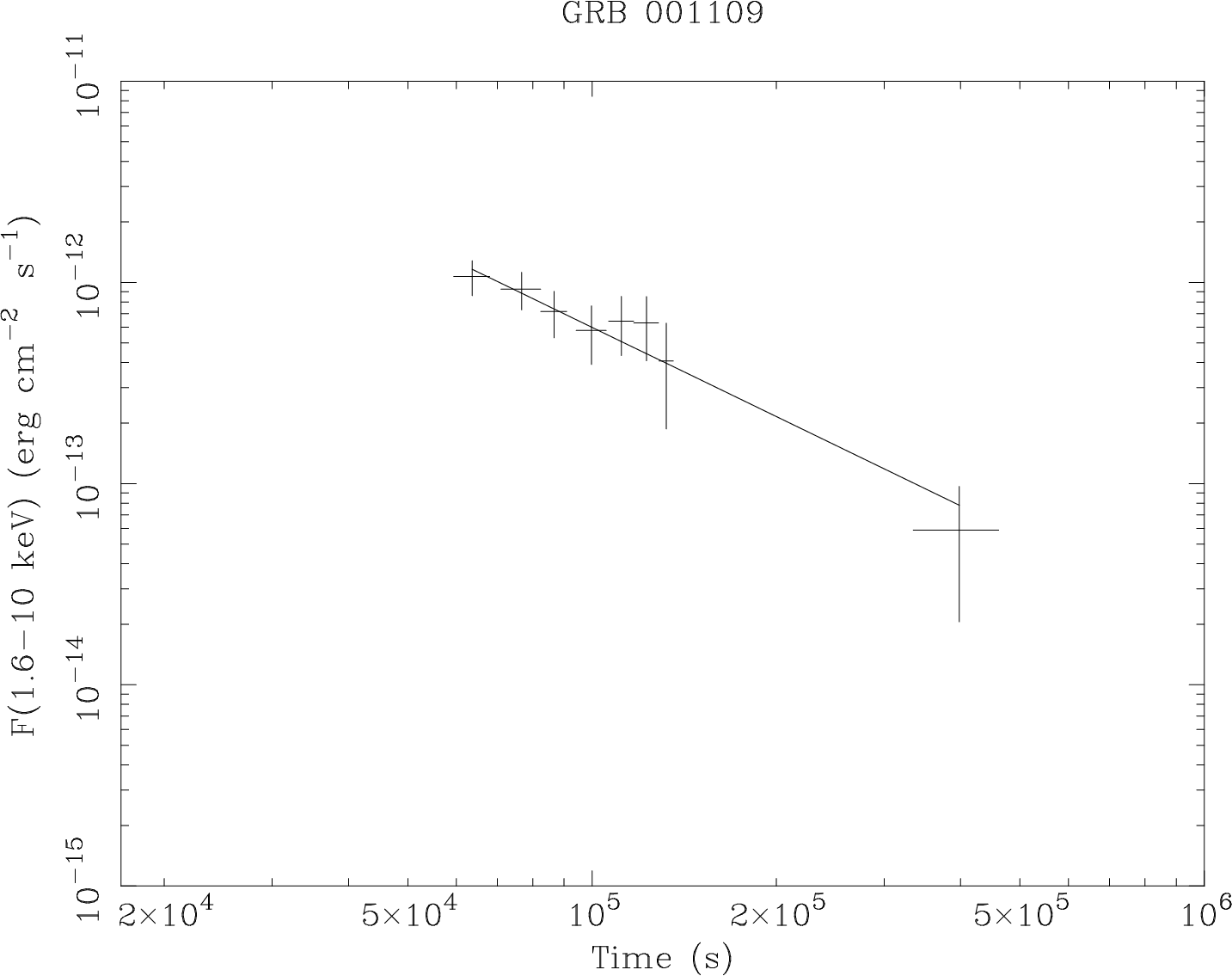}\vspace{1cm}

\includegraphics[width=6cm,angle=270]{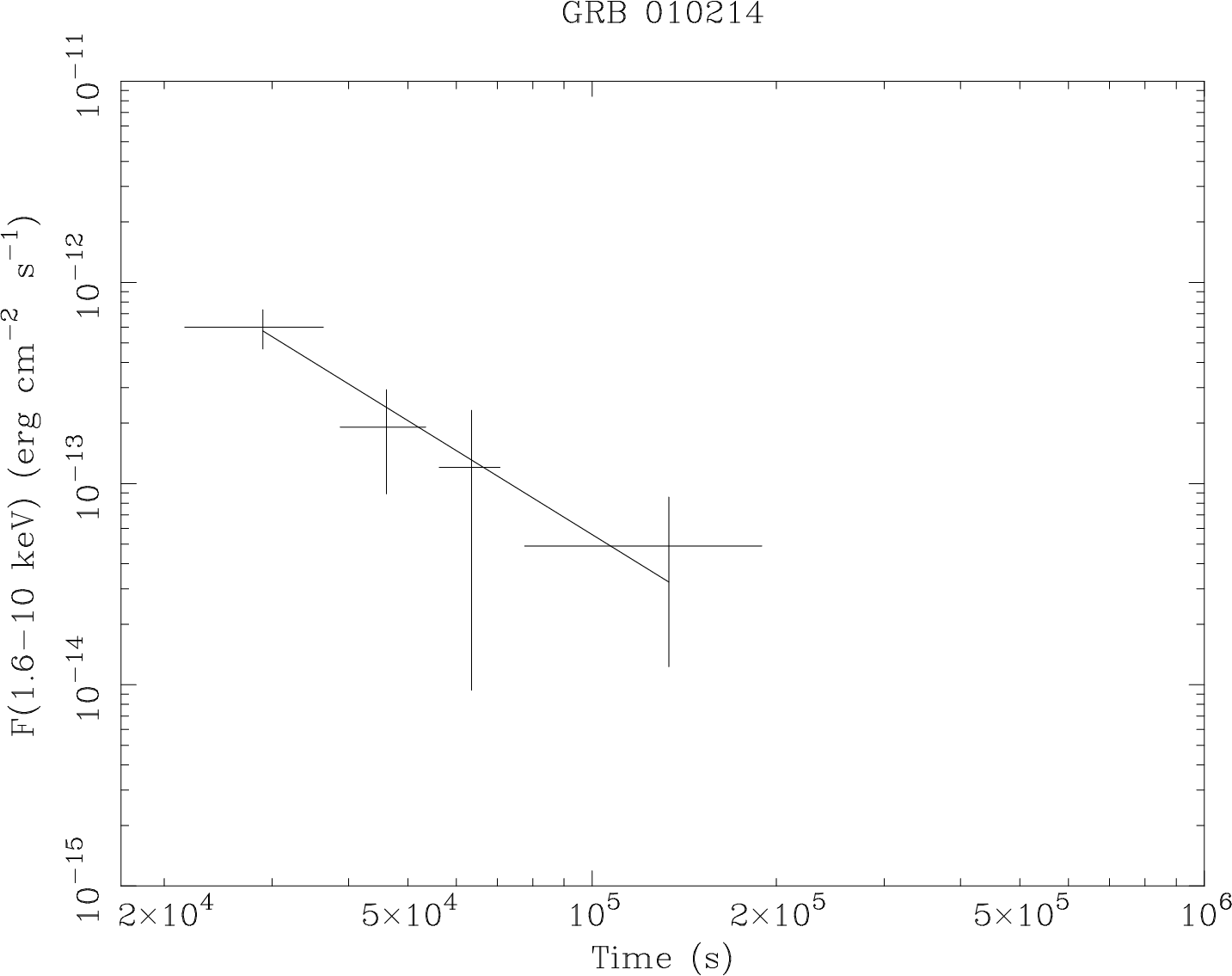}\hspace{1cm}
\includegraphics[width=6cm,angle=270]{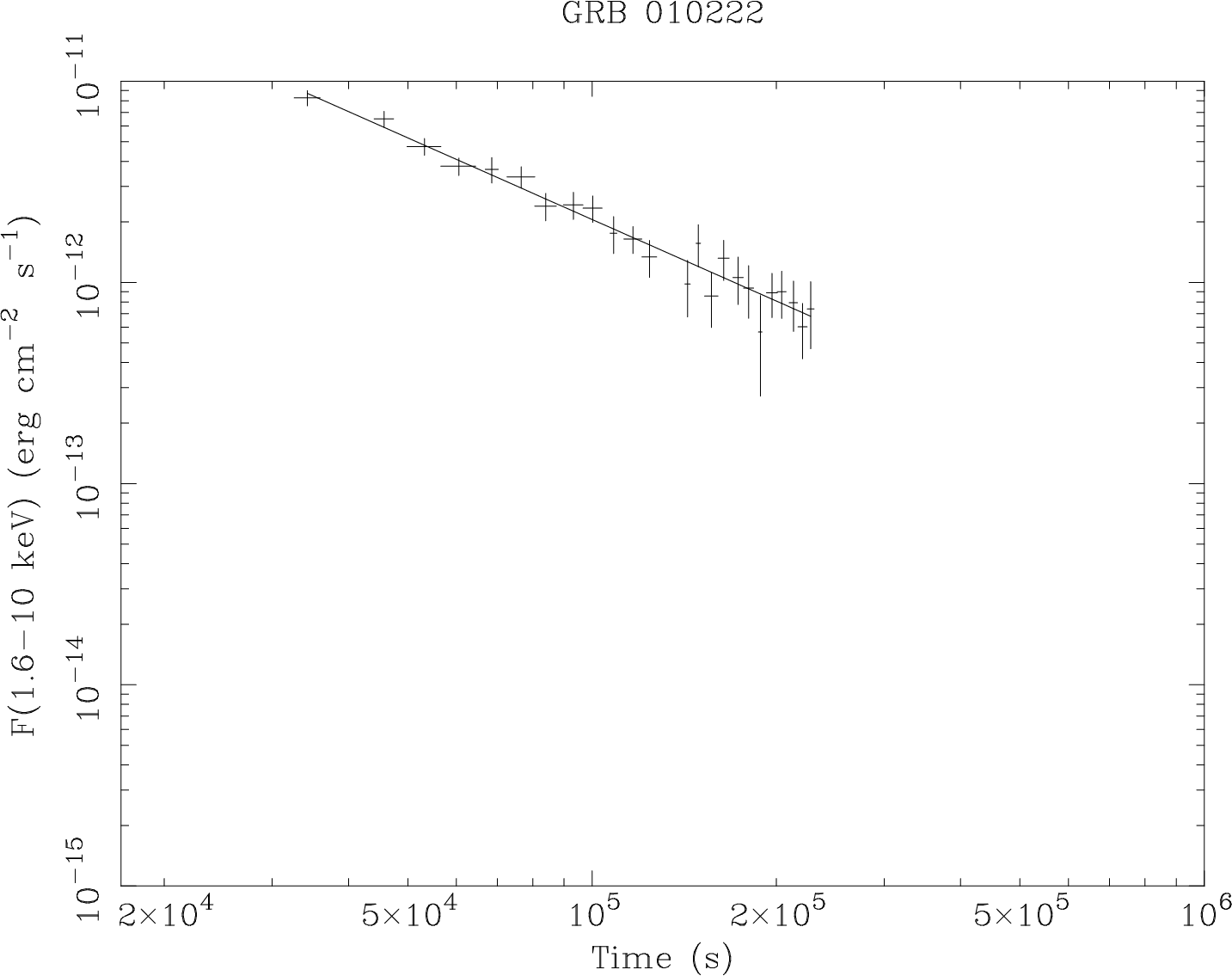}\vspace{1cm}

\caption{Continued.}
\end{figure*}

\setcounter{figure}{11}
\begin{figure*}
\centering
\vspace{1cm}
\includegraphics[width=6cm,angle=270]{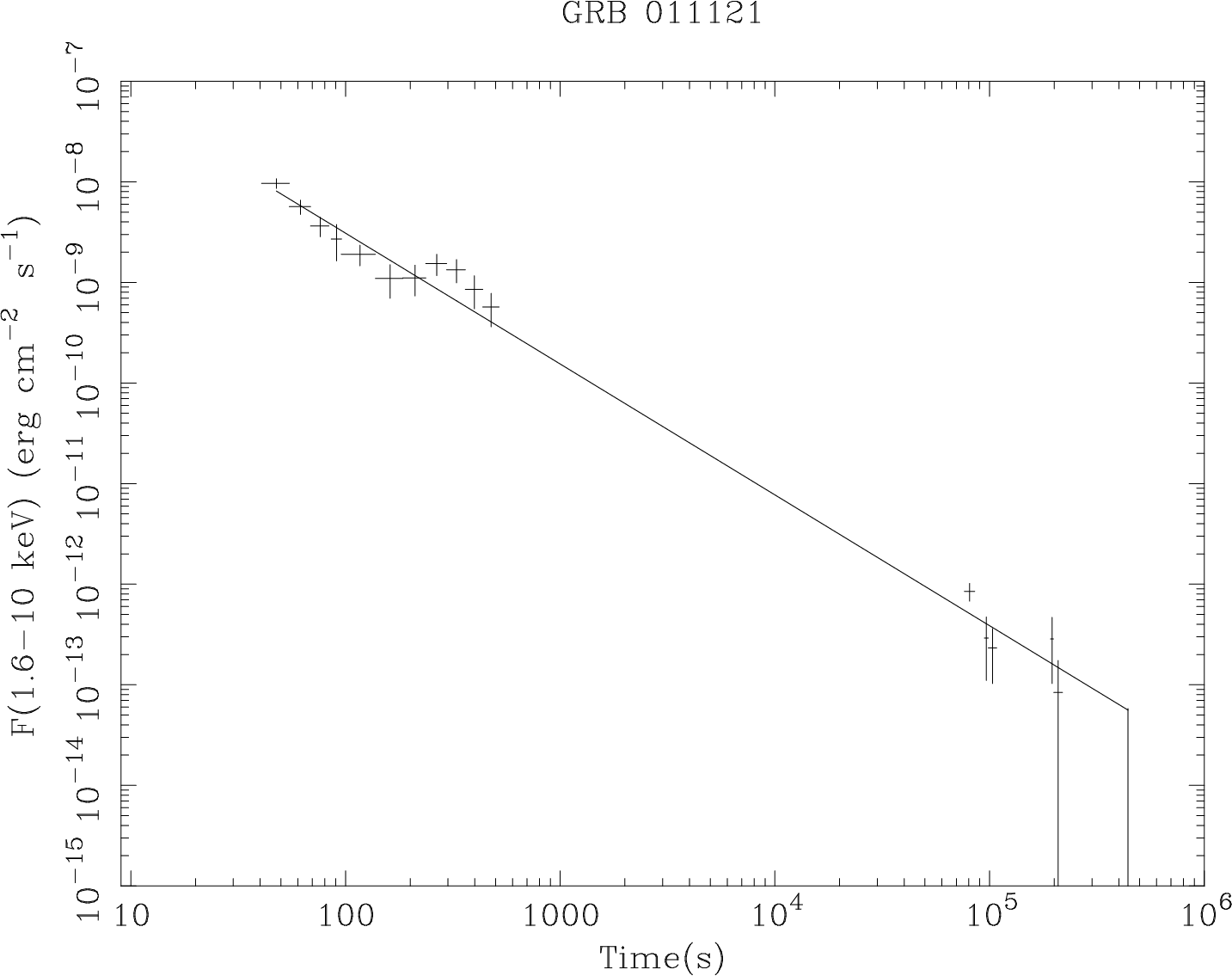}\hspace{1cm}
\includegraphics[width=6cm,angle=270]{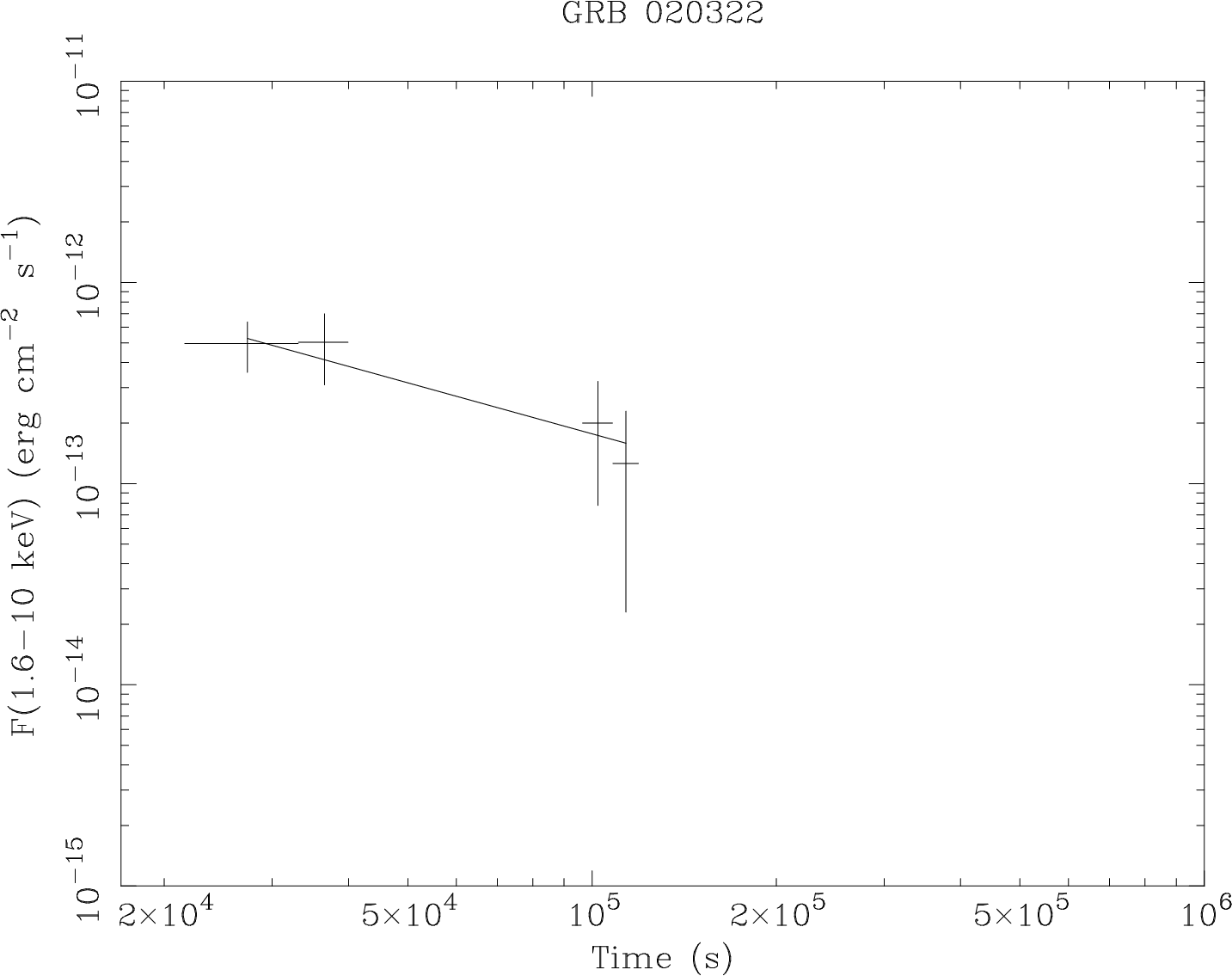}\vspace{1cm}

\includegraphics[width=6cm,angle=270]{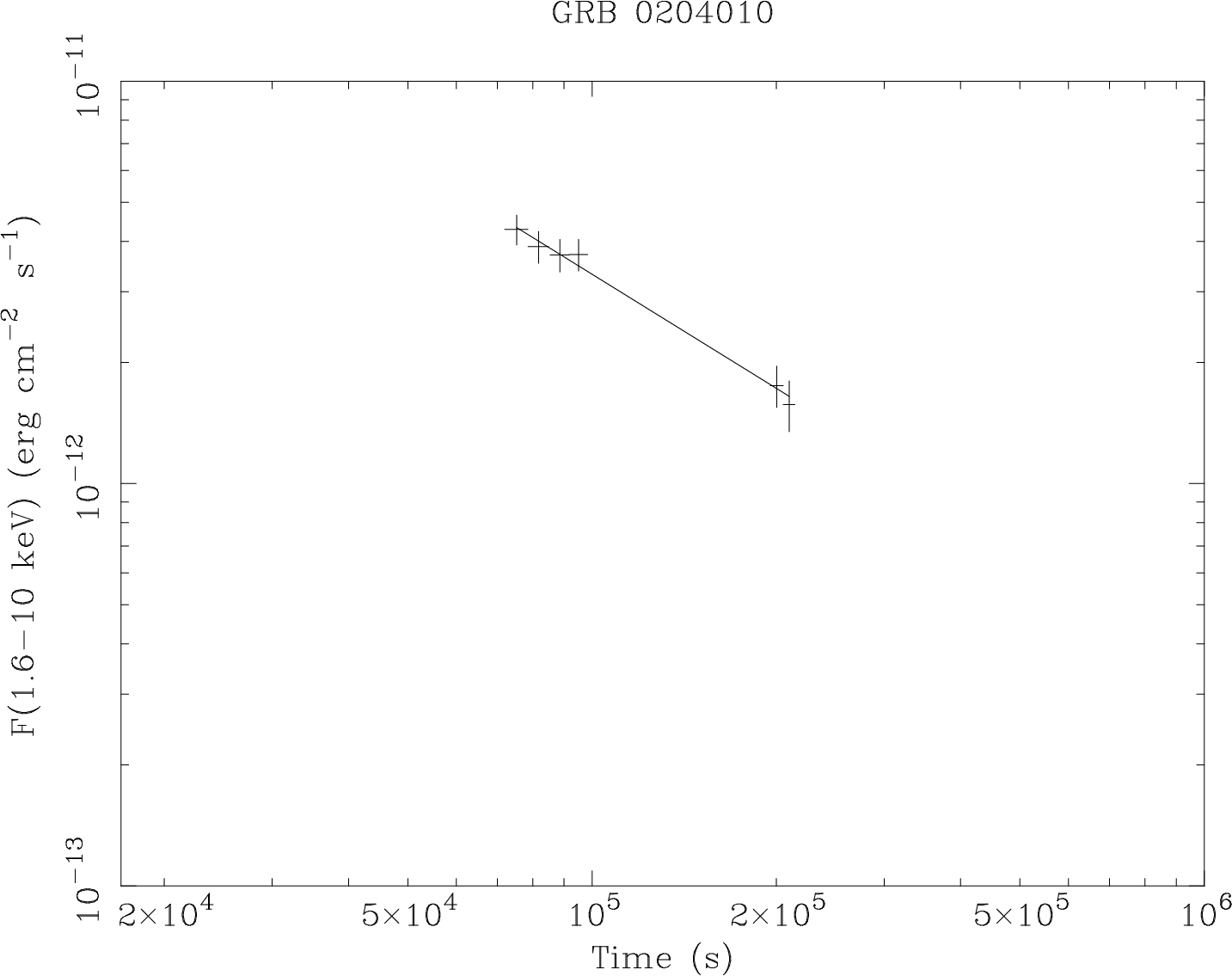}\hspace{1cm}
\includegraphics[width=6cm,angle=270]{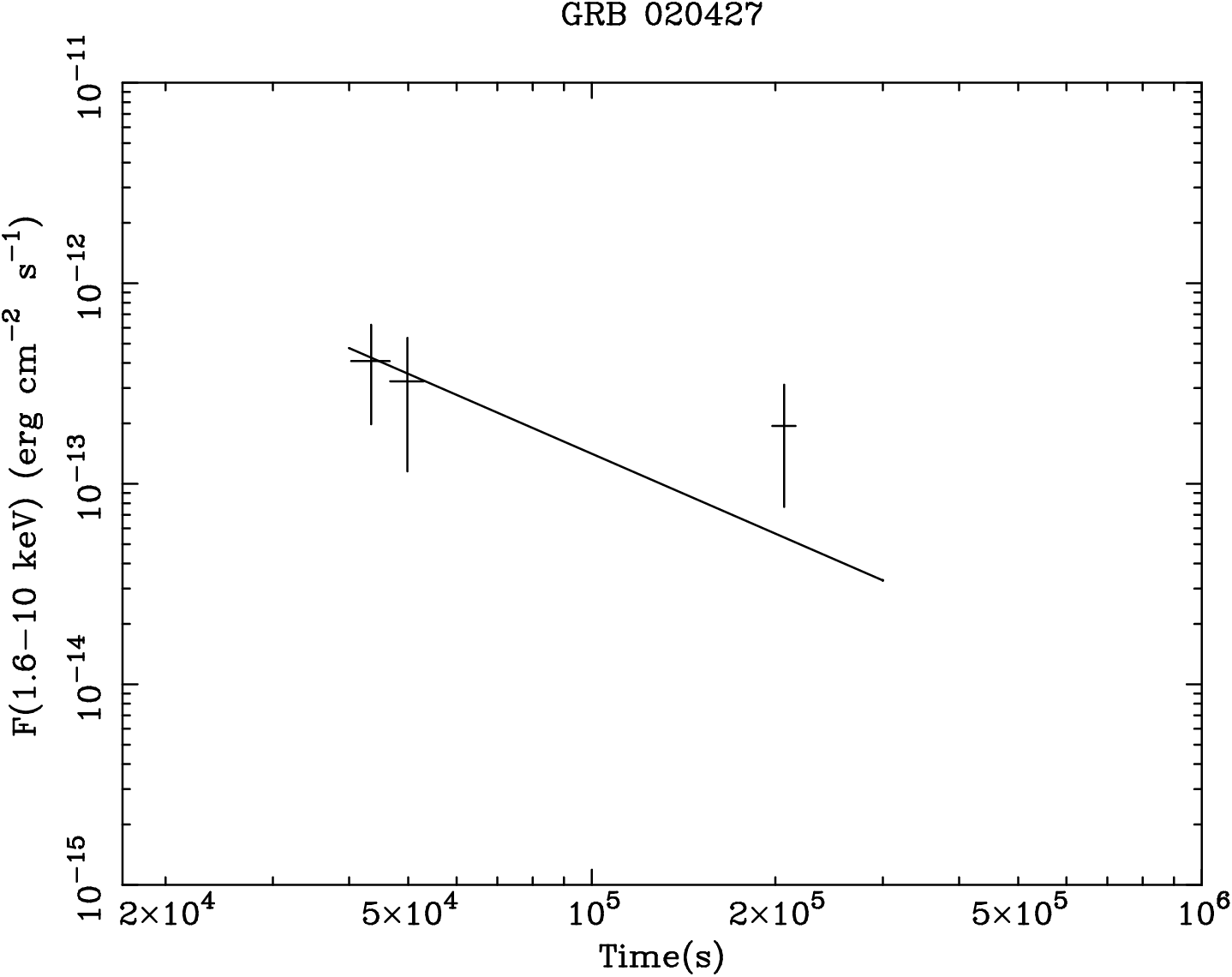}\vspace{1cm}
\caption{Continued.}
\end{figure*}


\begin{thebibliography}{}
\bibitem[Amati et al. (1998)]{ama98} Amati L., Frontera F., Costa E., Feroci, M., 1998, GCN \#146
\bibitem[Amati et al. (1999)]{ama99} Amati L., 1999, private comunication
\bibitem[Amati et al. (2002)]{ama02a} Amati L., Frontera F., Tavani M. et al. 2002, \aap, 390, 81
\bibitem[Amati et al. (2003)]{ama02b} Amati L., Frontera F., Castro-Ceron J.M. et al,  2003, Prooceedings of ``GRB and Afterglow Astronomy 2001'',  AIP Conference Proceedings, 662, 387
\bibitem[Amati et al. (2004)]{ama04} Amati L., Frontera F., in't Zand J. et al. 2004, \aap, 426, 415
\bibitem[Antonelli et al. (1999)]{ant99} Antonelli L.A., Fiore F., Amati L. et al. 1999, \aaps, 138,435
\bibitem[Antonelli et al. (2000)]{ant00} Antonelli, L. A., Piro, L., Vietri, M. et al. 2000, ApJ, 545L, 39
\bibitem[Berger et al. (2003)]{ber03} Berger E., Kulkarni S.R. \& Frail D.A., 2003, \apj, 590, 379
\bibitem[Bloom et al. (1998)]{blo99} Bloom J.S. Frail D.A., Kulkarni S.R. et al., 1998, \apj, 508, L21
\bibitem[Bloom et al. (2002)]{blo02} Bloom J.S., Mirabal N., Helpern J.P., Fox, D.W., Lopes, P.A.A., 2002, GCN \#1296
\bibitem[Boella et al. (1997)]{boe97} Boella G., Butler, R.C., Perola, G.C., et al., 1997, \aaps, 122, 299
\bibitem[Chevalier \& Li (1999)]{che99} Chevalier, R.A., \& Li, Z.Y., 1999, \apj, 520, L29
\bibitem[Chevalier et al. (2004)]{che04} Chevalier, R.A., Li, Z.Y., \& Fransson, C., 2004, \apj, 606, 369
\bibitem[Chincarini et al. (2005)]{chi05} Chincarini, G., Moretti, A., Romano, P., et al., 2005, submitted to \apj, astro-ph/0506453
\bibitem[Cornelisse et al. (2002)]{cor02} Cornelisse R., Verbunt F, in 't Zand J., et al., 2002, \aap, 392, 885
\bibitem[Costa et al. (1997)]{cos97} Costa E., Frontera F., Heise J., et al. 1997, Nature, 387, 783
\bibitem[Dal Fiume D. et al. (2000)]{dal00} Dal Fiume D., Amati, L., Antonelli, L. A., 2000, \aap, 355, 454
\bibitem[De Pasquale et al. (2003)]{dep03} De Pasquale M., Piro L., Perna R., et al. 2003, \apj, 592, 1018
\bibitem[Diercks et al. (1998)]{die98} Diercks A.H., Deutsch E.W., Castander F.J., et al. 1998, \apj, 503, L105
\bibitem[Djorgovski et al. (2001)]{djo01} Djorgovski, S.G., Frail, D. A.; Kulkarni, S. R., et al., 2001, \apj, 562, 654
\bibitem[Feroci et al. (1998)]{fer98} Feroci, M., Antonelli, L.A., Guainazzi, M., et al., 1998, \aap, 332, L29
\bibitem[Feroci et al. (2001)]{fer01} Feroci M., Antonelli, L. A., Soffitta, P., et al., 2001, \aap, 378, 441
\bibitem[Fiore et al. (1999)]{fio99} Fiore F., Guinazzi M. \& Grandi P., 1999, Handbook for \textit{BeppoSAX} NFI Spectral Analysis, ftp:\\\\www.sdc.asi.it/pub/sax/doc/software\_docs/saxabc\_v1.2.ps.gz
\bibitem[Fox et al. (2003)]{fox03} Fox, D.W., Price, P.A., Soderberg, A.M., et al., 2003, \apj, 586, L5
\bibitem[Frail et al. (1997)]{fra97} Frail D.A., Kulkarni S.R., Nicastro S.R., Feroci, M., Taylor, G. B., 1997, Nature, 389, 261
\bibitem[Frail et al. (2001)]{fra01} Frail D.A., Kulkarni S.R., Sari R., et al. \apj, 2001, 562, L55
\bibitem[Freedman \& Waxman (2001)]{fre01} Freedman D. \& Waxman E., 2001 ApJ, 547, 922
\bibitem[Frontera et al. (1997)]{fro97} Frontera, F., Costa, E., Dal Fiume, D., et al., 1997, \aaps, 122, 357
\bibitem[Frontera et al. (1998)]{fro98} Frontera F., Costa E., Dal Fiume D. et al, 1998, \apj, 493L, 67
\bibitem[Frontera et al. (2000a)]{fro00a} Frontera F., Amati L., Costa, E., et al., 2000a, \apjs, 127, 59
\bibitem[Frontera et al. (2000b)]{fro00b} Frontera F., Antonelli L. A., Amati L., et al. 2000b, \apj, 540, 697
\bibitem[Frontera et al. (2004)]{fro04} Frontera F., 2004, Proceedings of ``GRBs in the afterglow Era 2002'', ASP conference series, 312, 3
\bibitem[Frontera et al. (2003)]{fro05} Frontera F., 2003, Lecture Notes in Physics, 598, p.317 (astro-ph/0406579)
\bibitem[Fynbo et al. (2001)]{fyn01} Fynbo, J.U., Jensen, B.L., Gorosabel, J., et al., 2001, \aap, 369, 373
\bibitem[Galama et al. (1998)]{gal98} Galama T.J., Groot, P.J., van Paradijis J., et al., 1998, \apj, 497, L13
\bibitem[Gandolfi et al. (1999)]{gan99} Gandolfi, G., Soffitta, P., Heise, J., et al., 1999, GCN \#448
\bibitem[Gehrels et al. (2005)]{geh05} Gehrels, N., Chincarini, G., Giommi, P., et al., 2005, \apj, 611, 1005
\bibitem[Gendre et al. (2005)]{gen05} Gendre, B., Corsi, A., \& Piro, L, 2005 submited to \aap
\bibitem[Giommi et al. (2000)]{gio00} Giommi, P., Perri, M., \& Fiore, F., 2000, \aap, 362, 799
\bibitem[Greiner et al. (2002)]{gre02} Greiner J, Thiele U., Klose S., Castro-Tirado, A.J., 2002, GCN \#1298
\bibitem[Guidorzi et al. (2000)]{gui00} Guidorzi C., Montanari E., Frontera F., et al., 2000, GCN \#675
\bibitem[Guidorzi et al. (2003)]{gui03} Guidorzi C., Frontera, F., Montanari, E., et al., 2003, \aap, 401, 491
\bibitem[Harrison et al. (1999)]{har99} Harrison F.A., Bloom J.S., Frail D.A., et al., 1999, \apj, 523, L121
\bibitem[Heise et al. (2002)]{hei03} Heise J., in 't Zand J., Kippen M. et al. 2002, Proceedings of the 2000 Rome Workshop on "Gamma Ray Burst in the Afterglow Era", AIP, 229
\bibitem[Hjorth et al. (2002)]{hjo02} Hjorth J., Thomsen, B., Nielsen, S.R., et al., 2002, \apj, 576, 113
\bibitem[Hjorth et al. (2003)]{hjo03} Hjorth J., Sollerman, J., M\o ller, P., et al. 2003, Nature, 423, 847
\bibitem[Hurley et al. (2000)]{hur00} Hurley K., Mazets E., Golenetskii S., et al. 2000, GCN 801
\bibitem[Jacobsson et al. (2003)]{jac03} Jakobsson, P., Hjorth, J., Fynbo, J.P.U., et al., 2003, \aap, 408, 941
\bibitem[Jacobsson et al. (2004)]{jac04} Jakobsson, P., Hjorth, J., Fynbo, J.P.U., et al., 2004, \apj, 617, L21
\bibitem[Jager et al. (1997)]{jag97} Jager, R., Mels, W.A., Brinkman, A.C., et al., 1997, \aaps, 125 557
\bibitem[Jaunsen et al. (2001)]{jau01} Jaunsen A.O., Hjorth J., Bj\"ornsson, G., et al., 2001, \apj, 546, 127
\bibitem[Klebesadel et al. (1973)]{kle73} Klebesadel, R.W., Strong, I.B., \& Olson, R.A., 1973, \apj, 182, L85
\bibitem[Kobayashi et al. (1997)]{kob97} Kobayashi, S., Piran, T., Sari, R., 1997, \apj, 490, 92
\bibitem[Kulkarni et al. (1999)]{kul99} Kulkarni, S.R., Djorgovski, S.G., Odewahn, S.C., et al, 1999, Nature, 398, 389
\bibitem[Kuulkers et al. (2000)]{kuu00} Kuulkers, E., Antonelli, L.A., Kuiper, L, et al., 2000, \apj, 538, 638
\bibitem[Kumar \& Piran (2000)]{kup00} Kumar, P., \& Piran, T., 2000, \apj, 535 152
\bibitem[Lamb \& Reichart (2000)]{lam00} Lamb, D. \& Reichart, E., 2000, \apj, 536, 1
\bibitem[Lazzati et al. (2002)]{laz02} Lazzati, D., Covino, S., \& Ghisellini, G., 2002, \mnras, 330, 583
\bibitem[Levine et al. (1998)]{lev98} Levine, A., Morgan, E., \& Muno, M., 1998, IAUC 6966
\bibitem[Maccacaro et al. (1988)]{mac88} Maccacaro T, Gioia I.M., Wolter A. et al., 1988, \apj, 326, 680
\bibitem[Maiorano et al. (2005)]{mai05} Maiorano E., Masetti N., Palazzi E. et al. 2005, \aap, in press (astro-ph/0504602)
\bibitem[Masetti et al. (1998)]{mas98} Masetti N., Bartolini C., Guarnieri A., \& Piccioni, A., 1998, Proceedings of the Active X-ray Sky symposium 1997, Editors L. Scarsi, H. Bradt, P. Giommi, and F. Fiore, p.674
\bibitem[Masetti et al. (2001)]{mas01} Masetti N., Palazzi E., PianE. et al., 2001, A\&A, 374, 382
\bibitem[Meszaros \& Rees (1997)]{mes97} Meszaros, P, \& Rees, M.J., 1997, \apj, 476, 232
\bibitem[Metzger et al. (1997)]{met97} Metzger, M.R., Djorgovski, S.G., Kulkarni, S.R., et al., 1997, Nature, 387, 879
\bibitem[Montanari et al. (2002)]{mon01} Montanari, E., Amati, L., Frontera, F., et al., 2002, Proceedings of ``2nd Rome Workshop on Gamma-Ray Burst in the afterglow Era'', 195
\bibitem[Muller et al. (1999b)]{mul99b} Muller, J.M, Costa, E., Gandolfi, G., et al., 1999b, IAUC 7211
\bibitem[Nicastro et al. (1998)]{nic98v} Nicastro, L. Amati, L. Antonelli, L. A. et al., 1998, A\&A 338, L17
\bibitem[Nicastro et al. (1999)]{nic98} Nicastro L., Amati, L., Antonelli, L.A., et al., 1999, \aaps, 138, 437
\bibitem[Nicastro et al. (2001)]{nic01} Nicastro L., Cusumano G., Antonelli L.A., et al. 2001, Proceedings of ``GRBs in the afterglow Era 2000'', Eds Enrico Costa, Filippo Frontera, and Jens Hjorth, 198
\bibitem[Nicastro et al. (2004)]{nic04} Nicastro L., in't Zand, J.; Amati, L., et al. 2004, \aap, 427, 445
\bibitem[Panaitescu et al. (1998)]{pan98} Panaitescu, A., Meszaros, P., \& Rees, M.J., 1998, \apj, 503, 314
\bibitem[Panaitescu \& Kuman (2002)]{pan01} Panaitescu, A., \& Kumar, P., 2002, \apj, 571, 779
\bibitem[Parmar et al. (1997)]{par97} Parmar, A.N., Martin, D.D.E., Bavdaz, M., et al. 1997, \aaps, 122, 309
\bibitem[Pian et al. (1999)]{pia99} Pian E. Amati, L., Antonelli, L. A., et al., 1999, \aaps, 138, 463
\bibitem[Pian et al. (2000)]{pia00} Pian, E., Amati L., Antonelli, L. A. et al., 2000, \apj, 536, 778
\bibitem[Piran et al. (2001)]{pira01} Piran, T., Kumar, P., Panaitescu, A., \& Piro, L., 2001, \apj, 560, L167
\bibitem[Piro (1995)]{piro95} Piro, L., 1995, ``SAX Observer Handbook'', Agenzia Spaziale Italiana, |c1995, Issue 1.0, edited by Piro, L.
\bibitem[Piro et al. (1998)]{piro98} Piro, L., Heise, J., Jager, R., et al., 1998, \aap, 329, 906
\bibitem[Piro et al. (1998b)]{piro98b} Piro, L., Amati, L., Antonelli, L.A. et al., 1998b, \aap, 331, L41
\bibitem[Piro et al  (1999)]{piro99} Piro L., Costa, E., Feroci, M. et al., 1999, 514, L73
\bibitem[Piro et al. (2001)]{piro01} Piro, L., Garmire, G., Garcia, M., et al., 2001, \apj, 558, 442
\bibitem[Piro et al. (2002)]{piro02} Piro, L., Frail, D.A., Gorosabel, J., et al. 2002, \apj, 577, 680
\bibitem[Piro (2004)]{piro04} Piro, L., 2004,  Proceedings of ``GRBs in the afterglow Era 2002'', ASP conference series, p. 149
\bibitem[Piro et al. (2005)]{piro05} Piro, L., De Pasquale, M., Soffita, P., et al., 2005, \apj, 623, 314
\bibitem[Price et al. (2002)]{pri02} Price, P., Berger, E., Reichart, D.E., et al., 2002, \apj 572, L51
\bibitem[Ramirez-Ruiz et al. (2001)]{ram01} Ramirez-Ruiz, E., Dray, L.M., Madau, P., Tout, C.A., 2001, \mnras, 327, 829
\bibitem[Rees \& Meszaros (1992)]{ree92} Rees, M.J., \& Meszaros, P., 1992, \mnras, 258, 41
\bibitem[Reichart et al. (1999)]{rei99} Reichart, D.E., Lamb, D.Q, Metzger, M.R., et al., 1999, \apj, 517, 692
\bibitem[Rhoads (1997)]{rho97} Rhoads, J.E., 1997, \apj, 487, L1
\bibitem[Rol et al. (2005)]{rol05} Rol, E., Wijers, R.A.M.J., Kouveliotou, C., Kaper, L., Kaneko, Y., \apj, 624, 868
\bibitem[Sari et al. (1998)]{sar98} Sari, R., Piran, T., \& Narayan, N., 1998, \apj, 497, L17
\bibitem[Sari et al. (1999)]{sar99} Sari, R., Piran, T., \& Helpern, J.P., 1999, \apj, 519, L17
\bibitem[Soffitta et al. (2002)]{sof02} Soffitta P., Amati L., Antonelli L.A., et al., 2002, Proceedings of ``GRB in the Afterglow Era 2000'', p.201
\bibitem[Stanek et al. (2003)]{sta03} Stanek, K.Z., Matheson, T., Garnavich, P. M., et al., 2003, \apj, 591, L17
\bibitem[Stratta et al. (2004)]{str04} Stratta, G., Fiore, F., Antonelli, L.A., et al., 2004, \apj, 608, 846
\bibitem[Tassone et al. (1999)]{tas99} Tassone, G., in 't Zand, J., Frontera, F., \& Gandolfi, G., 1999, IAUC \#7281
\bibitem[van Paradijis et al. (1997)]{van97} van Paradijs, J., Groot, P.J., Galama, T.J., et al., 1997, Nature, 386, 686
\bibitem[Vreeswijk et al. (1999)]{vre99} Vreeswijk, P. M., Galama, T. J., Owens, A., 1999, \apj, 528, 171
\bibitem[Woosley (1993)]{woo93} Woosley, S., 1993, \apj, 405, 273
\bibitem[in't Zand et al. (1998)]{zan98} in 't Zand J., Amati, L., Antonelli, L. A. et al.,  \apj, 505L 119
\bibitem[in't Zand et al. (1999)]{zan99} in't Zand J., Heise, J., van Paradijs, J., \& Fenimore, E. E., 1999, \apj, 516, L57
\bibitem[in't Zand et al. (2000a)]{zan00a} in't Zand J., Heise J., Kuulkers E. et al. 2000a, GCN \#677
\bibitem[in't Zand et al. (2000b)]{zan00b} in't Zand J., Kuiper L., Amati L., et al. 2000b, \apj, 545, 266
\bibitem[in't Zand et al. (2001)]{zan01} in't Zand J., Kuiper L., Amati, L. et al.,  2001, \apj 559, 710
\bibitem[in't Zand et al. (2004)]{zand03} in't Zand J., Kuiper L., Heise J. et al. 2004, Proceedings of ``GRBs in the afterglow Era 2002'', ASP Conference Series, p. 209
\end{thebibliography}
\end{document}